\documentclass[%
 reprint,
 superscriptaddress,
 amsmath,amssymb,
 aps,
 twocols
]{revtex4-1}

\usepackage{graphicx}
\usepackage{dcolumn}
\usepackage{bm}
\usepackage{color}

\newcommand{\VECM}{\bm M}
\newcommand{\VECr}{\bm r}
\newcommand{\VECR}{\bm R}

\newcommand{\MATD}{\text{\sffamily\bfseries D}}
\newcommand{\MATG}{\text{\sffamily\bfseries G}}
\newcommand{\MATH}{\text{\sffamily\bfseries H}}

\newcommand{\MATS}{\text{\sffamily\bfseries S}}

\newcommand{\romd}{\text{d}}
\newcommand{\rome}{\text{e}}

\definecolor{darkgreen}{rgb}{0,0.5,0}
\definecolor{purple}{rgb}{0.543,0,0.543}

\newcommand{\ake}{{\it Ake}}
\newcommand{\add}{{\it add}}

\begin{document}


\title{Optimal coarse-grained site selection in elastic network models of biomolecules}


\author{Patrick Diggins IV}
\affiliation{Department of Physics, Carnegie Mellon University, Pittsburgh, PA 15213, USA}

\author{Changjiang Liu}
\affiliation{Department of Physics, Carnegie Mellon University, Pittsburgh, PA 15213, USA}
\affiliation{Department of Biophysics, University of Michigan, Ann Arbor, MI 48109, USA}

\author{Markus Deserno}
\affiliation{Department of Physics, Carnegie Mellon University, Pittsburgh, PA 14213, USA}

\author{Raffaello Potestio}
\email{raffaello.potestio@unitn.it}
\affiliation{Physics Department, University of Trento, via Sommarive, 14 I-38123 Trento, Italy}
\affiliation{INFN-TIFPA, Trento Institute for Fundamental Physics and Applications, I-38123 Trento, Italy}


\date{\today}

\begin{abstract}
Elastic network models, simple structure-based representations of biomolecules where atoms interact {\it via} short-range harmonic potentials, provide great insight into a molecule's internal dynamics and mechanical properties at extremely low computational cost. Their efficiency and effectiveness have made them a pivotal instrument in the computer-aided study of proteins and, since a few years, also of nucleic acids. In general, the coarse-grained sites, i.e.\ those effective force centres onto which the all-atom structure is mapped, are constructed based on intuitive rules: a typical choice for proteins is to retain only the C$_\alpha$ atoms of each amino acid. However, a mapping strategy relying only on the atom type and not the local properties of its embedding can be suboptimal compared to a more careful selection. Here we present a strategy in which the subset of atoms, each of which is mapped onto a unique coarse-grained site of the model, is selected in a stochastic search aimed at optimising a cost function. The latter is taken to be a simple measure of the consistency between the harmonic approximation of an elastic network model and the harmonic model obtained through exact integration of the discarded degrees of freedom. The method is applied to two representatives of structurally very different types of biomolecules: the protein Adenylate kinase and the RNA molecule adenine riboswitch. Our analysis quantifies the substantial impact that an al\-go\-rithm-dri\-ven selection of coarse-grained sites can have on a model's properties.\\

This manuscript was published on the Journal of Chemical Theory and Computation. DOI: 10.1021/acs.jctc.8b00654
\end{abstract}


\maketitle

\section{Introduction}

Our current understanding of biological processes at the molecular level has benefited greatly from computer simulations and \emph{in silico} studies. Computational models of fundamental molecules and molecular assemblies, such as proteins, nucleic acids, or lipid bilayers, allow us to observe and quantitatively investigate them under a broad range of physical conditions, and at a level of resolution usually inaccessible to experiments.

Since the first pioneering simulations of simple model systems \cite{alder1957} and biological molecules \cite{Levitt1975}, computational models have enjoyed a steady increase in force field accuracy, system sizes, and accessible time scales. State-of-the-art simulations, especially those performed through purposefully constructed machines such as ANTON \cite{Shaw2009}, attain durations compatible with the folding time of small proteins \cite{Piana2012,Piana2013}, while systems composed of millions of atoms can be studied on more standard supercomputing machines \cite{Freddolino2006,Bock2013}.

However, many situations remain where investigating fully atomistic models of bio\-mo\-le\-cu\-les is neither a \emph{viable option}, nor in fact an \emph{adequate strategy}. It is uncontested that the sizes and time scales involved in many exciting problems still substantially exceed the typical computational power accessible to a majority of research groups. However, even ignoring this aspect, we should recall that from an epistemological point of view an all-atom treatment might not only be impractical or impossible {\it tout court}, but explicitly undesirable: a ``complete'' representation of some complex system will \emph{of course} exhibit all the emergent behaviour it is capable of displaying; but if a much simpler representation captures the same phenomenology, this offers novel and often deep explanatory insight into the underlying mechanisms and helps to distill causations that otherwise remain opaque. Good models are \emph{necessarily} highly simplified versions of the systems, for the same reason that useful maps are highly simplified versions of reality \cite{borges2002exactitude}.

These two principles ---efficiency and simplicity--- have inspired the development of \emph{coarse-grained} (CG) models \cite{Takada2012,Noid2013,Saunders2013,Potestio2014}, which demagnify the atomistic resolution of a molecule by combining several atoms or entire chemical groups into effective degrees of freedom (called ``interaction sites'' or coarse-grained ``beads'') that are subject to suitably chosen effective interaction potentials. It is worth recalling that classical atomistic force fields are \emph{also} coarse-grained: they have removed the electrons---and all the quantum mechanics that goes with them---and replaced them by effective interactions: strong short-range repulsions arising from the Pauli principle, long-range van-der-Waals attractions to account for correlated charge fluctuations, and Coulomb interactions for the case where a local unit is not entirely charge neutral. Doing this is neither loss-less nor unique, which explains why more than one atomistic force field exists.

The spectrum of CG models developed during the past few decades spans from particle-based models \cite{Ueda1978,Go1978,Gohlke2006,Potestio2009,Tozzini2010,Noid2013,Polles2013,Najafi2015}, where each bead is taken to represent groups of atoms (from parts of a side chain over single amino acids up to entire proteins), all the way up to continuum descriptions employed in the study of very large or mesoscale systems such as viruses \cite{Gibbons2007,Gibbons2008,Roos2010,Aggarwal2016}, membranes \cite{marrink2004coarse, izvekov2005multiscale, venturoli2006mesoscopic, muller2006biological, murtola2009multiscale, Deserno2009}, or even cells \cite{alber2003cellular, izaguirre2004compucell, shirinifard20093d}.

A particular flavour of CG modelling, which is widely used, is the so-called \emph{Elastic Network Model} (ENM) \cite{Tirion1996,Bahar1997,Hinsen1998,Atilgan2001,Delarue2002,Micheletti2004}. This group encompasses a class of particle-based representations of biomolecules in which the gamut of realistic interactions is replaced by harmonic springs. ENMs have gained widespread attention following the pioneering work of Tirion \cite{Tirion1996}, who demonstrated that an all-atom model of a protein, whose detailed force field has been replaced by local springs, all of the same strength, can reproduce the protein's low-energy vibrational spectrum with astounding faithfulness. Observe that since a normal mode analysis of a harmonic system can be performed analytically, we do not even have to run a simulation to get the answer. In subsequent developments, ENMs of even lower resolution have been studied, keeping only one or two atoms per residue \cite{Bahar1997,Hinsen1998,Atilgan2001,Delarue2002,Micheletti2004}. These CG models have proven extremely useful in characterising the collective motions of proteins, which matters because these low energy conformational fluctuations often relate directly to a protein's function \cite{Amadei1993,Pontiggia2007,Potestio2009,Hensen2012,Nussinov2014,Wei2016}.

The construction of CG ENMs is carried out starting from a reference conformation (typically the native structure, as determined from crystallography), of which only the C$_\alpha$ atoms are retained. Springs are then placed between those C$_\alpha$ atoms falling within a predetermined cutoff distance. More detailed models exist \cite{Micheletti2004}, which include also interaction centers representative of the side chains; their position in space, however, is uniquely determined by that of the C$_\alpha$ atoms, thus maintaining the same number of degrees of freedom as the former models. This strategy, in all its many variants, constitutes a simple rule to define a versatile and computationally efficient model of the protein.

Nonetheless, the question remains if the specific choice of the degrees of freedom retained in ENMs---for instance, the $\alpha$ carbons---is in any way optimal. In fact, one may reasonably expect that a different selection of atoms as CG sites, performed so as to maximise the consistency between the reference system and the resulting CG model, could outperform a strategy that entails no system specificity. Various authors have already shown that the \emph{number} as well as the {\it distribution} of CG sites can be adjusted in order to optimise the balance between efficiency and accuracy. Gohlke and Thorpe \cite{Gohlke2006}, for example, have suggested that particularly rigid subregions of a protein represent a most natural notion of large-scale, variable-sized coarse-grained groups. This concept has been employed by Zhang and coworkers \cite{ZHANG_BJ_2008,ZHANG_BJ_2009,ZHANG_JCTC_2010} as well as by Potestio and coworkers \cite{Potestio2009,aleksiev2009,Polles2013} to develop optimisation schemes aimed at identifying these quasi-rigid domains in proteins, either by exploring various mappings with fixed number of CG sites, or searching for the best CG site number and distribution. Sinitskiy and collaborators \cite{Sinitskiy2012} have built on the work by Zhang {\it et al.} to single out an optimal number of CG sites to be employed in a simplified representation of the system. More recently, the study of Foley and coworkers \cite{Foley2015} has shed further light on this latter aspect by making use of the notion of relative entropy \cite{Shell2008} to quantify the balance between the simplification of a CG model and its information content.

Refining the \emph{mapping} of CG sites should thus further improve a model's quality; of course, if the latter required us to actually \emph{simulate} the original system (for instance in order to learn more about the mode spectrum), we would lose one of the key redeeming virtues of the whole approach---the fact that we can get a good proxy for the low energy fluctuations without ever running an atomistically detailed simulation.

In the present work we propose a simulation-free strategy for improving the construction of an ENM, which amounts to selecting the CG beads via an algorithmic optimisation procedure. This procedure in turn relies on an intermediate step, in which the number of atoms in an existing ENM is reduced by performing a partial trace over ``undesired'' degrees of freedom in the system's partition function. Performing such a partial trace has been proposed before \cite{Hinsen2000,Carnevale2007,Zen2007}; its chief attraction lies in the fact that harmonic partition functions can be computed \emph{exactly}. However, there is a snag, and in the present context it is an important one: an ENM, while entirely consisting of harmonic springs, is \emph{not} harmonic in the coordinates over which we wish to integrate (that is, the \emph{Cartesian displacements from a reference conformation}), rather only in the \emph{distances} (a distinction which sometimes seems to be missed). Hence, it first needs to be harmonically expanded in these coordinates, a model that for clarity we dub here hENM. Unfortunately, though, a CG-hENM obtained by performing a partial trace over some of its parent's coordinates no longer corresponds to a CG-ENM of which it would be the harmonic expansion. This results in artefacts at the ENM level despite the exact transformation at the hENM level.

The key idea of our paper is to show that this admittedly annoying artefact, which to our knowledge has not been previously realised, can be exploited to optimise the modelling: in fact, we propose to \emph{choose the CG sites such as to minimise the corresponding mapping error}. We construct a quantity that serves as a proxy for this error, and employ it to construct models which outperform, in terms of this and other observables, models built on more conventional approaches. We illustrate the properties of this new method by explicitly applying it to two examples: (\emph{i}) a small protein (Adenylate kinase) and (\emph{ii}) an RNA molecule (adenine riboswitch).

\section{Theory}

\subsection{Overview of Elastic Network Models}

Elastic network models for proteins have been first introduced by Tirion \cite{Tirion1996} as a simplified approximation of an atomistic force field. The assumption underlying this approach is that the small-amplitude, low-energy, and collective vibrations of proteins emerge from the concurrent action of a large number of interactions, whose specific functional form and strength is rendered unimportant by the central limit theorem. The complex and accurate potential of a realistic model---including bonds, angles, van der Waals forces, and electrostatic interactions---is thus replaced by an effective potential of the form
\begin{equation}\label{tirion1}
V_{\rm ENM}^{\rm AT}(\{\VECr_i\}) = \frac{1}{2}K \sum_{i<j}  C_{ij} \left(r_{ij} - r^0_{ij}\right)^2 \ .
\end{equation}

Here, $r_{ij}$ is the \emph{scalar distance} between particles $i$ and $j$, calculated as the magnitude of the distance \emph{vector} $\VECr_{ij} \equiv \VECr_i - \VECr_j$. The superscript $0$ indicates the same quantity, but evaluated in the ground state (reference) structure, obtained for instance from X-ray crystallography. Only two model parameters remain: first, the elastic strength (``spring constant'') $K$; and second, the cutoff distance $R_{\rm c}$ within which two atoms must be located in the reference structure in order to interact. This cutoff enters the definition of the \emph{contact matrix}
\begin{equation}
C_{ij} = \left\{\begin{array}{c@{\;\;\;\;}c} 1 & \mbox{if } r^0_{ij} \leq R_{\rm c} \\[0.5em] 0 & \mbox{otherwise} \end{array} \right. \ .
\end{equation}

It is important to realize that the potential energy function (\ref{tirion1}) is \emph{not} quadratic in the actual coordinates $\VECr_i$, despite consisting entirely of harmonic springs, because calculating the distance $r_{ij}=|\VECr_{ij}|$ involves taking a square root. However, we can expand (\ref{tirion1}) quadratically in the displacements $\Delta\VECr_i=\VECr_i-\VECr^0_i$ away from the reference structure, which---up to an irrelevant constant---leads to
\begin{equation}\label{tirion2}
V_{\rm hENM}^{\rm AT}(\{\VECr_i\}) = \frac{1}{2} \sum_{k,l} \Delta\VECr_k^\dagger\,\MATH_{kl}\,\Delta\VECr_l \ .
\end{equation}

Here, the \emph{Hessian matrix} $\MATH_{kl}$ is given by
\begin{subequations}
\label{eq:Hessian-1-all}
\begin{align}
\MATH_{kl} &= \frac{\partial^2 V_{\rm ENM}^{\rm AT}(\{\VECr_i\})}{\partial\VECr_k\,\partial\VECr_l}\bigg|_{\{\VECr_i^{0}\}} \label{eq:Hessian-1-1} \\[0.5em]
&= - \MATD_{kl} + \delta_{kl}\sum_j \MATD_{kj} \label{eq:Hessian-1-2} \\
&=
\left\{\begin{array}{c@{\;\;,\;\;}c}
-\MATD_{kl} & k\ne l \\[0.5em] \sum_j^{j\ne k} \MATD_{kj} & k= l
\end{array}\right. \ , \label{eq:Hessian-1-3}
\end{align}
\end{subequations}
where the ``elastic dyad'' $\MATD_{kl}$ is defined by
\begin{equation}
\MATD_{kl} = K\,C_{kl}\, (\hat\VECr^0_{kl}\otimes\hat\VECr_{kl}^0) \ ,
\label{eq:Dkl}
\end{equation}
and $\hat\VECr_{kl}^0=\VECr_{kl}^0/r_{kl}^0$ is the unit vector pointing from the site $l$ to the (different) site $k$ (in the reference state), such that $(\hat\VECr^0_{kl}\otimes\hat\VECr_{kl}^0)$ is the projector onto the line between them.

Several comments are in order:
\begin{enumerate}
\item Each element of the Hessian matrix is in fact a $3\times 3$ \emph{sub-matrix}, due to the occurrence of the dyads. This is necessary because the displacements $\Delta\VECr_k$ and $\Delta\VECr_l$ in Eqn.~(\ref{tirion2}) are themselves vectorial.
\item For any pair $k\ne l$, the Hessian sub-matrix is simply the negative of the elastic dyad, and as such it is a $3\times $3 matrix which has exactly one non-zero eigenvalue, which corresponds to the (negative of the) spring constant $K$, and whose eigenvector aligns with the bond between $k$ and $l$.
\item The second term in (\ref{eq:Hessian-1-2}) ensures that the sum over the elements in any row or any column of $\MATH_{kl}$ vanishes. This removes the contribution of pure translations to the energy---a physically pleasing outcome that has not been imposed by hand but is a natural consequence of the fact that the elastic energy (\ref{tirion1}) is a sum of terms that depend only on the \emph{difference} between pairs of coordinates.
\item Taken together, we recognize $\MATH_{kl}$ as a generalized Kirchhoff matrix.
\end{enumerate}

What makes the quadratic expansion (\ref{tirion2}) of the ENM (\ref{tirion1}) so attractive is that it is exactly solvable---in the sense that we can exactly calculate its correlation matrix in the canonical state,
\begin{equation}
\langle\Delta \VECr_k\otimes \Delta \VECr_l\rangle=
k_{\rm B}T\,(\MATH^{-1})_{kl} \ ,
\label{eq:exact-correlations}
\end{equation}
where $k_{\rm B}$ is Boltzmann's constant and $T$ the temperature. To clarify the notation: if we view the Hessian as a $3N\times 3N$ matrix, subdivided into $3\times 3$ blocks for the $(x,y,z)$ components of the position variations of particle $k$ and $l$, then the right hand side of Eqn.~(\ref{eq:exact-correlations}) contains the inverse of the \emph{entire} $3N\times 3N$ matrix, which subsequently gets re-parceled into sub-blocks.

For historical reasons, the elastic network model described in Eqn.~(\ref{tirion1}) is dubbed \emph{anisotropic} ENM (or ANM for short), because the energy cost associated with the displacement of an atom depends on its direction: for a given $i$-$j$-bond, no energy is required to move atom $i$ in the direction perpendicular to $\VECr_{ij}$, only displacements parallel to it affect the energy. This distinction is not present in the so-called \emph{Gaussian} ENM (or GNM) \cite{Bahar1997,Doruker2000,Cui2005}, where the pairwise interaction is proportional to the squared \emph{vectorial} displacement $(\Delta\VECr_i - \Delta \VECr_j)^2$ and, therefore, a given displacement will increase the energy by the same amount irrespective of the direction in which it is performed. In the following, we will focus on ANMs and hANMs only.

We conclude this section by introducing a further distinction between classes of matrices $\MATH_{kl}$ that can be employed to build a network of the general form (\ref{tirion2}), and those that can be expressed according to Eqs. (\ref{eq:Hessian-1-all},\ref{eq:Dkl}). The latter are a subclass of the former, more general class that can be dubbed ``quadratic displacement networks'', or QDN. Quadratically expanding an ENM leads to an hENM, a special case of a QDN. Also, \emph{all QDNs can be coarse-grained exactly}. However, if a QDN happens to belong to the special subclass of hENMs, \emph{it generally loses that property upon coarse-graining}.

\subsection{The issue of mapping in ENMs}

Approaches to coarse-graining fall into two major categories: \emph{bottom-up} \cite{Muller-Plathe2002,Reith2003,Izvekov2006,Spyriouni2007,Noid2008,Shell2008,Shell2012,Noid2013b,Potestio2014} and \emph{top-down} \cite{Tirion1996,Bahar1997,Ueda1978} methods. Those belonging to the first class assume the existence of a higher-resolution ``reference'' model from which they construct a simplified representation via a set of systematic rules. In contrast, those belonging to the second class postulate empirical models suggested by generic physical principles, without insisting on a microscopic underpinning. Their parameters, however, may get further refined by higher level knowledge (e.g.\ known structure or thermodynamic properties) that could for instance be obtained from experiment.

Classical (h)ENMs \cite{Tirion1996,Bahar1997,Hinsen1998,Atilgan2001,Delarue2002,Micheletti2004,Polles2013} are representatives of this second class, in that the interactions among the CG sites are parametrized based on a reference structure, but without incorporating any more accurate knowledge of the real forces acting between the atoms. One could of course do the latter, for instance by combining the crystal structure with an atomistic force field, evaluate the interactions, and thereby systematically improve the spring constants \cite{phdPontiggia,Globisch2013}, but this is much less common. However, once we construct lower resolution ENMs, we have the choice to either follow the same top-down strategy as used for more finely resolved ENMs, or to systematically derive lower resolution ENMs in a bottom-up fashion, using finely-resolved ENMs as the reference. The latter is the topic of the present paper.

To construct a low-resolution ENM, we need to do two things: first, agree on a smaller set of new degrees of freedom; and second, define effective interactions between them. The usual way to formalize the first step is to establish a \emph{mapping} \cite{Noid2013} between atoms of the high resolution description and the smaller number of CG sites of the lower resolution model. This mapping can be expressed as vector-valued functions $\VECM_I(\{\VECr_i\})$ which specify the (typically Cartesian) coordinates $\VECR_I$ of the CG sites in terms of the set $\{\VECr_i\}$ of high resolution coordinates: $\VECR_I = \VECM_I(\{\VECr_i\})$. These mappings are almost invariably linear \cite{Noid2013b}, and the most common choices are ($i$) the definition of cen\-ter-of-mass coordinates of the set of atoms grouped together and ($ii$) the reduction to one particular coordinate from that set. It is generally understood that the choice of mapping affects the quality of the resulting CG model, but systematic studies for how to optimise this step have only been undertaken quite recently \cite{Potestio2009,Sinitskiy2012,Polles2013,Foley2015}

When constructing CG-ENMs, the most common choice for a mapping is to remove all atoms of a given residue except for their $\alpha$-carbon. This reduces the number of interaction sites to that of amino acids and leads to a (quasi) uniform mass distribution along the backbone. A less frequent strategy is to keep the C$_\alpha$ as well as, from each non-glycine residue, a second site representative of the side chain, thus approximately doubling the number of interaction sites with respect to C$_\alpha$--only models.

Once the mapping has been established, interactions must be defined, which are typically of the form (\ref{tirion1}), possibly with bond-specific spring constants:
\begin{equation}\label{CG-ENM-1}
V_{\rm ENM}^{\rm CG}(\{\VECR_I\}) = \frac{1}{2} \sum_{I<J}  K_{IJ} \left(R_{IJ} - R^0_{IJ}\right)^2 \ ,
\end{equation}
where $K_{IJ}$ is the spring constant between sites $I$ and $J$; if there is no spring between two sites, we simply set $K_{IJ}=0$. Once again, this model can be quadratically expanded in the $\Delta\VECR_I$, just as we did for the more finely resolved model (\ref{tirion1}), leading to
\begin{equation}\label{enm1}
V_{\rm hENM}^{\rm CG}(\{\VECR_I\}) = \frac{1}{2}\sum_{K,L} \Delta \VECR_K^\dagger \MATH_{KL}^{\rm CG} \Delta \VECR_L \ ,
\end{equation}
where the Hessian $\MATH_{KL}^{\rm CG}$ is constructed analogously to Eqns.~(\ref{eq:Hessian-1-all},\ref{eq:Dkl}), except for the additional obvious replacement $KC_{kl}\rightarrow K_{KL}$. This model can again be solved analytically by virtue of being quadratic, leading to the full spectrum of CG eigenmodes of the dynamics. We note, in passing, that what we refer to with the term {\it dynamics} is to be intended as the equilibrium fluctuations of the system, and not the time evolution of its conformation. We will employ the term {\it dynamics} with this meaning throughout the manuscript.

At this point, an intriguing idea might suggest itself: the systematic construction of CG models, in one way or the other, tries to capture as much thermodynamic properties as possible from its more finely resolved reference. The quality with which this is doable is limited, trivially, by the fact that the CG model has a lower resolution; and more practically, by the fact that we usually cannot calculate the full thermodynamic information of the finely resolved model. However, in this case our underlying model consists of harmonic springs, and its quadratic expansion is exactly solvable. Can we exploit this property and \emph{analytically} calculate the optimal CG model, without the need to perform simulations to approximately track thermodynamic information, as we would do in other more complex cases? The systematic and semi-analytic reduction of degrees of freedom in ENMs has been attempted \cite{Hinsen2000,Carnevale2006,Zhou2008}, however always retaining the structure of a simplified (CG) model that is quadratic in the Cartesian displacements, i.e.\ of the form (\ref{enm1}). Here, we will show that a break exists in the continuity of the connections between different models; more precisely, \emph{we can analytically link model (\ref{tirion2}) and (\ref{enm1}), but not model (\ref{tirion1}) and (\ref{CG-ENM-1})}. The reason is subtle, and the result might at first sight be annoying; however, we will argue that it permits us to make significant headway on the first and understudied coarse-graining question: how to pick good CG sites.

\subsection{Coarse-graining an hENM}

A powerful way to conceptualise coarse-graining is to view it as a mapping of the canonical state of a microscopic system into a smaller phase space via the transformation theorem for probability densities \cite{Noid2008}. Having established the connection $\VECR_I=\VECM_I(\{\VECr_i\})$, one writes the canonical partition function in the degrees of freedom $\{\VECr_i\}$ and encodes the mapping by including the additional delta function $\delta(\VECR_I-\VECM_I(\{\VECr_i\}))$, thereby arriving at an equivalent canonical partition function which now depends on the $\{\VECR_I\}$; its logarithm, multiplied by $-k_{\rm B}T$, equals the potential of mean force in the coarse-grained coordinates.

In our case the situation is even simpler, because the linear mapping we have in mind picks a subset of degrees of freedom from the fine-grained level, in which case one merely has to perform a partial trace over all the degrees of freedom one wishes to eliminate. Specifically, let us assume that we can subdivide the total set of degrees of freedom into a subset $A$ that will be kept and a subset $B$ that will be removed:
\begin{equation}
\{\VECr_i\} = \{\VECr_i\}_A \cup \{\VECr_i\}_B \ .
\end{equation}
Starting out with a linearised ENM (thus an hENM) of the form (\ref{tirion2}), we can derive its coarse-grained version as follows:
\begin{equation}
\rome^{-\beta V_{\rm hENM}^{\rm CG}(\{\VECr_i\}_A)}
=
\!\int\!\romd\{\VECr_i\}_B\;\rome^{-\beta V_{\rm hENM}^{\rm AT}(\{\VECr_i\}_A,\{\VECr_i\}_B)} \ ,
\label{eq:CG-partial-trace}
\end{equation}
where for simplicity we ignore the momenta, as well as normalisation factors, as they will only contribute irrelevant constants to the new potential. Since the linearised ENM is quadratic in the $\{\VECr_i\}$, the right hand side of (\ref{eq:CG-partial-trace}) is a multi-dimensional Gaussian integral that can be performed exactly. As a consequence, we can write down a simple closed-form expression for the left hand side. If we order our degrees of freedom so that the Hessian of the microscopic system can be written in the following block form,
\begin{equation}\label{cgHess1}
\MATH = \left( \begin{array}{cc} 
\MATH_A & \MATG \\[0.4em]
\MATG^\dagger & \MATH_B 
\end{array} \right) \ ,
\end{equation}
the coarse-grained system will again be of Hessian form---see Eqn.~(\ref{enm1})---and its Hessian is explicitly given by \cite{Hinsen2000,Carnevale2006,Zhou2008}
\begin{equation}
\MATH^{\rm CG} = \MATH_A - \MATG\,\MATH_B^{-1}\,\MATG^\dagger \ . \label{cgHess3}
\end{equation}

Several things are worth noting here:
\begin{enumerate}
\item The calculation of the coarse-grained Hessian is non-iterative and computationally inexpensive: it only requires the inversion of a matrix.
\item The CG interactions $\MATH^{\rm CG}$ in the $A$-subset are not identical to the bare interactions $\MATH_A$: eliminated degrees of freedom leave a trace (no pun intended) in the effective Hamiltonian.
\item The new potentials are effectively \emph{free energies} of interaction (or so-called ``multibody potentials of mean force''). Curiously, they do not depend on temperature, even though the mapping equation (\ref{eq:CG-partial-trace}) explicitly does. This absence of a sta\-te-point-de\-pen\-den\-cy is unusual and generally \emph{not} true for this type of coarse-graining. It holds here because the microscopic Hamiltonian is quadratic. \item $\MATH^{\rm CG}$ might be temperature independent, but performing the partial trace in (\ref{eq:CG-partial-trace}) creates $T$-de\-pen\-dent \emph{prefactors}, which we ignored. This would matter if we cared about absolute free energies, not just effective potentials.
\item The effective Hessian in Eqn.~(\ref{cgHess3}) is generally \emph{not} of the form (\ref{eq:Hessian-1-all},\ref{eq:Dkl}) corresponding to a linearised ENM.
\end{enumerate}

The last point is extremely important, so let us elaborate. The most general form of a quadratic displacement network, or QDN as it was previously christened---Eqn.~(\ref{tirion2})--couples any two vector displacements $\Delta\VECr_i$ and $\Delta\VECr_j$ by a $3 \times 3$ sub-matrix $\MATH_{ij}$. The values of the $9$ sub-block elements are in principle not restricted by particular requisites: in fact, while the symmetry \emph{of the overall} $\MATH^{\rm CG}$ \emph{matrix} has to be enforced, as it grants the preservation of the action--reaction principle, this constraint does not necessarily hold for the single sub-blocks. This generality allows for different responses to the different displacements applied to pairs of residues in one order or another: that is to say that
\begin{equation}
{\bf u}^\dagger \MATH_{ij} {\bf v} \neq {\bf v}^\dagger \MATH_{ij} {\bf u}.
\end{equation}

When the $\MATH^{\rm CG}$ matrix is obtained by integrating a subset of degrees of freedom from a finer-grained Hamiltonian $\MATH$ (see Eq. \ref{cgHess3}), the sub-block matrix $\MATH^{\rm CG}_{IJ}$ does not need to be symmetric for $I \neq J$. Indeed, the off-diagonal $3 \times 3$ ``elements'' of this tensor emerge from the integration of several degrees of freedom, and entail the effect of the removed particles. Consequently, the $9$ sub-matrix elements can have arbitrary and independent values.
In contrast, the Hessians which arise from the linearisation of an ENM have the particular form (\ref{eq:Hessian-1-all},\ref{eq:Dkl}), in which the interaction between two (different) vector displacements is given by a dyad of the form $\Delta \VECR_{IJ}\otimes \Delta \VECR_{IJ}$. But dyads only have three degrees of freedom, since they can be fully specified by a vector $\Delta \VECR_{IJ}$.

This simple counting argument teaches an important lesson: the QDNs which arise from the harmonic expansion of ENMs are of a very special form, a form we are generally not guaranteed if we create QDNs in some other way. And indeed, coarse-graining an hENM via Eqn.~(\ref{cgHess3}) destroys that special form. 
In a nutshell, \emph{the functional form of the interactions obtained by exactly coarse-graining an hENM---a general quadratic form---is different from that obtained when linearising a CG ENM---a dyadic form}.

This technical point has an important consequence: the ultimate goal is to systematically construct a CG-ENM, exploiting the fact that the microscopic ENM can be expanded into a linearised hENM, for which one can perform an analytically closed bottom-up coarse-graining procedure; but the trouble is that the resulting coarse-grained QDN is no longer the harmonic expansion of a CG-ENM. However, we will now show how to make use of this discrepancy to identify the optimal subset of particles that will be retained from the fully atomistic ENM (that is, the set of $\{\VECr_i\}_A$). The idea is to minimise an appropriate measure quantifying the deviations between the coarse-grained hENM resulting from combining Eqn.~(\ref{enm1}) and (\ref{cgHess3}) and a true hENM satisfying the additional constraints 
(\ref{eq:Hessian-1-all},\ref{eq:Dkl}).

\subsection{Reconstructing an approximate CG-ENM from the CG-hENM}

Since the $3\times 3$ sub-blocks in the coarse-grained matrix $\MATH^{\rm CG}$ from Eqn.~(\ref{cgHess3}) are not dyads, an exact back-translation into an ENM is not possible. However, these blocks might be \emph{close} to dyads, in the sense that one of their eigenvalues strongly dominates the other two. To quantify this, let us consider the three eigenvalues of each $(K,L)$ sub-block of $\MATH^{\rm CG}_{KL}$. The form of Eqn.~(\ref{cgHess3}) makes it evident that the \emph{whole matrix} $\MATH^{\rm CG}$ is symmetric as long as $\MATH$ is; but this property does not extend to its $3\times 3$ sub-blocks, whose eigenvalues need not be real. Hence, we consider a symmetrised version of the matrix, defined as:
\begin{equation}
\MATS_{KL} =
\frac{1}{2}\Big(\MATH^{\rm CG}_{KL} + \MATH^{\rm CG}_{LK}\Big) \ ,
\end{equation}
which has real eigenvalues $\lambda_{KL}^{(i)}$ by construction. We then order these three eigenvalues of each $\MATS_{KL}$ by magnitude,
\begin{equation}
\lambda_{KL}^{(1)}\ge
\lambda_{KL}^{(2)}\ge
\lambda_{KL}^{(3)} \ ,
\end{equation}
and define the ratio $\rho_{KL}$ via
\begin{equation}
0 \le \rho_{KL} := \frac{\lambda_{KL}^{(2)}}{\lambda_{KL}^{(1)}} \le 1 \ .
\end{equation}

The case $\rho_{KL}=0$ corresponds to a real bond (the sub-block is indeed a dyad), while $\rho_{KL}=1$ deviates maximally from the ``desired'' form. From this information on individual pair-in\-ter\-ac\-ti\-ons, we will now define an intuitive metric for judging how the entire matrix fares. This is the average eigenvalue ratio, or AER for short, defined as:
\begin{equation}
\textrm{AER} := \frac{1}{N_b}\sum_{K<L} \rho_{KL}  \ ,
\end{equation}
where $N_b$ is the total number of bonds lying within the interaction cutoff. This is to say, only those bonds are considered that can be replaced by a potential of the form $\frac{1}{2} K_{IJ} (R_{IJ} - R^0_{IJ})^2$. Other interactions, which arise from the Boltzmann integration but connect sites farther away than the cutoff, will not be represented by the CG-ENM, and so they are not included in the computation of the AER. By construction, the AER lies in the range $[0,1]$, with $0$ being the best case scenario, and $1$ the worst case scenario. In the following, the AER will be presented in percent to ease the readability.

Together with this metric we also need to specify a prescription on how to define a CG-ENM from a CG-hENM that is not the expansion of any ENM. Essentially, we need to decide how to define an effective spring constant $K_{IJ}$ from a Hessian $\MATH_{IJ}^{\rm CG}$ whose sub-blocks do not describe springs. We choose to set
\begin{equation} \label{eq:trace_rule}
K_{IJ} = {\rm Tr}\Big(\MATH_{IJ}^{\rm CG}\Big) \ .
\end{equation}

This definition implements the assumption that the anisotropy of the system's response to the displacement of a bead can be (almost) completely ascribed to the functional form of the interaction, while the amplitude of the force is well approximated by the average over the three Cartesian directions. This assumption is in part consistent with other measures of a molecule's flexibility (e.g. b-factors), and has been employed in previous works \cite{phdPontiggia,Hinsen2000}.

\subsection{Optimising the selection of retained atoms in the CG-ENM}

We now employ the AER of a CG-ENM to guide us which atoms from the all-atom representation to retain upon coarse-graining. Fixing a trial set of CG sites, we exactly integrate out the other degrees of freedom (on the hENM level). The resulting AER serves as a cost function to be minimized when repeating this process over a large number of trial CG sites.

To perform the stochastic search in the space of all possible subsets of retained atoms we will use Monte Carlo (MC) simulated annealing \cite{KirkpatrickVecchi1983,Cerny1985}. Despite its  efficiency this process poses a potential bottleneck, because it requires inverting a $3N_B\times 3N_B$ matrix---see Eqn.~(\ref{cgHess3}). However, if we choose to employ MC moves that add and delete only a single site per step, the process can be significantly sped up, because due to the structure of $\MATH^{\rm CG}$ this change only affects those matrix elements directly connected with the removed or added sites. This allows calculation of the new matrix from the old one by a process that only needs to invert a significantly smaller matrix. The molecules examined in this work were small enough for this trick not to be critical, but it might be quite crucial for bigger ones, and so we outline its essence in the Supporting Information.

\begin{figure}[h]
\begin{center}
\includegraphics[width=\columnwidth]{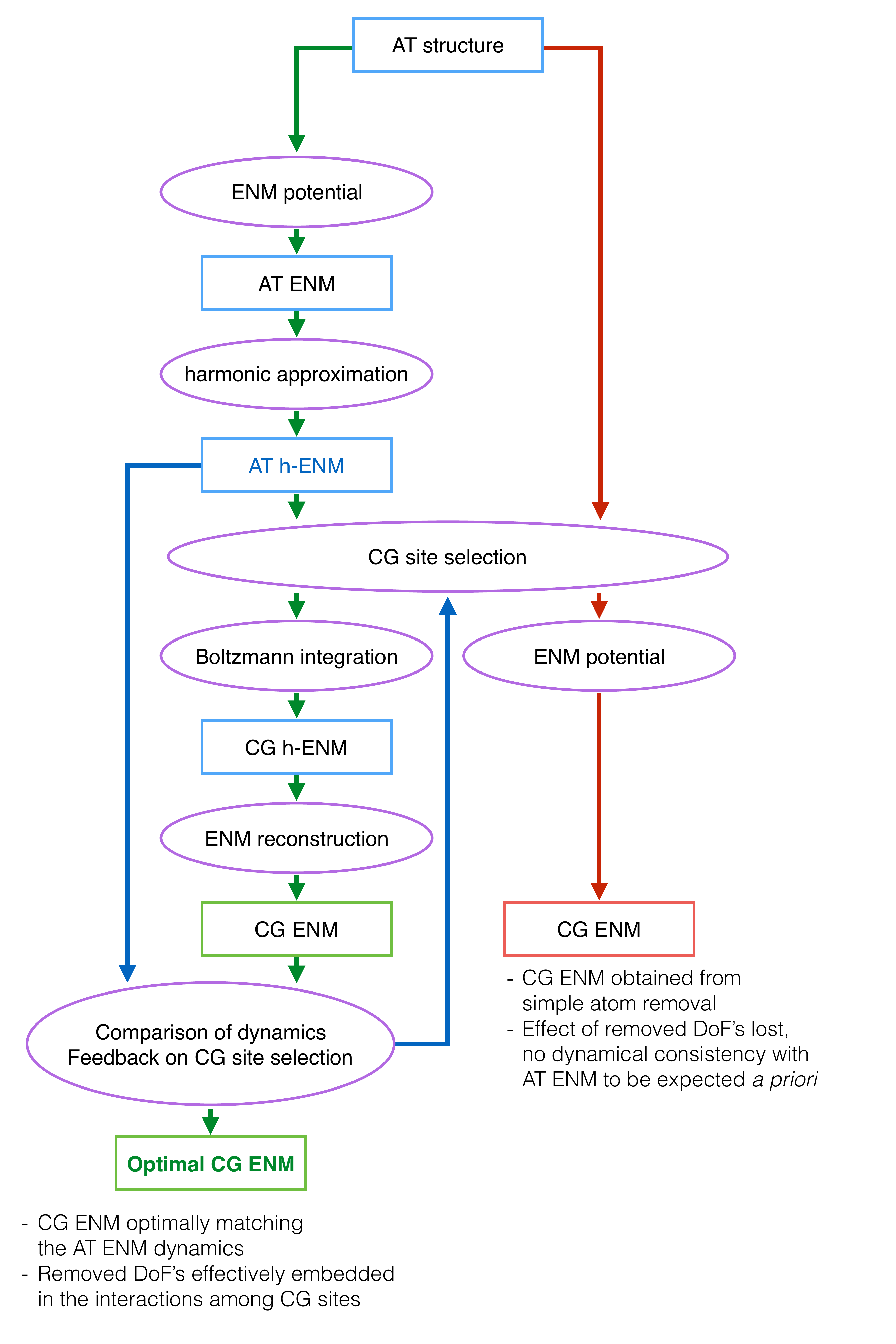}
\end{center}
\caption{Workflow of the method proposed here to construct a CG-ENM whose internal dynamics is maximally consistent with that of the reference AT ENM. Given a selection of atoms to play the role of CG sites, the input atomistic structure can be directly decimated to build an ENM with simple interactions among the surviving atoms, however with no {\it a priori} guarantee that the emerging dynamics will match the reference one (right half, red flow); alternatively, the harmonic expansion of the AT-ENM can be exactly integrated to leave out explicitly only the chosen CG sites, while the other ones are mapped onto the effective interactions (left half, green flow).}\label{fig:scheme}
\end{figure}

Let us now summarise the workflow of the proposed algorithm, presented schematically in Fig.~\ref{fig:scheme}. Starting from the fully atomistic structure, we equip it with ENM interactions to construct the reference model, i.e. the AT-ENM. A second order expansion of this model, as described in Eqn.~(\ref{eq:Hessian-1-all}), provides us with the exactly solvable harmonic ENM, or hENM, which still preserves the fully atomistic resolution but allows a simulation-free calculation of the essential dynamics. Once a subset of atoms has been selected as CG sites, the others are exactly integrated out, thereby renormalizing the interactions among the preserved sites. Up to this point, the model produces the same dynamics of the AT-hENM and, within the limits of the harmonic approximation, of the AT-ENM. This CG-hENM, however, \emph{cannot} be identified with the harmonic expansion of some CG-ENM, because it generally has a nonzero AER, and so it differs from a model obtained directly by removing the undesired atoms and building an ENM potential among them, as alternatively done in the right half of the workflow. Since for subsequent simulation we desire a full CG-ENM rather than a harmonic expansion, we employ the previously described criterion of AER minimization to guide a stochastic search for the best CG sites.

The parameters of the simulated annealing procedure are the same for both molecules. Specifically, we performed $10^4$ Monte Carlo steps: at each step one atom, currently being a CG site, is selected to be neglected (i.e. integrated out), while another atom which is not a CG site is promoted as such. The $i$-th move is accepted or rejected based on a Metropolis algorithm, with temperature decaying with an exponential law:
\begin{eqnarray}
&&T_i = T_0\ \textrm{e}^{-(i/n)^2}
\quad\textrm{with}\;\;
\left\{
\begin{array}{c}
T_0 = 0.2 \\
n = 50
\end{array}\right.
\end{eqnarray}

The outcome of this procedure is a model featuring the---ideally---smallest AER value. The problem at hand, however, bears the risk of being characterised by a multitude of (quasi-)degenerate minima, corresponding to different solutions with very close AER values. In order to avoid the risk of picking a suboptimal model stuck in such a minimum, and to get a qualitative idea of the free energy landscape structure, we have performed a two-layer set of parallel simulated annealing runs.

The first level consisted in running $18$ independent simulated annealing processes in parallel, and select as the optimal model the one with the lowest AER value among them. The second level is given by running 10 independent procedures as the aforementioned one, so as to have $10$ minimised AER values. Of these, only the model with the lowest AER is taken under examination, however the values of all $10$ ``local best'' values are considered to assess their dispersion and their optimality. The latter, in particular, is defined in terms of the separation between the lowest AER values and the random model AER distribution, as quantified by the Z-score:
\begin{equation}
Z = \frac{\textrm{AER}_{opt} - \mu}{\sigma}
\end{equation}
where $\mu$ and $\sigma$ are, respectively, the mean and standard deviation of the random model AER distribution. This measure is employed to determine if the model constructed through the simulated annealing is indeed better, in terms of the AER value, with respect to a random choice of CG sites, and how disperse the values obtained from independent optimisation runs are. The results of this analysis is reported in Fig. \ref{fig:AERdistrib} and Table \ref{table:summary}.

Once we have obtained the model maximising the consistency between CG interactions and the corresponding exact effective ones, we turn our attention to the dynamical properties of the CG-ENM. In particular, we first compare the harmonic expansion of the remapped CG-ENM to the CG-hENM from which it is reconstructed. This comparison is done in terms of the \emph{root weighted square inner product} (RWSIP), a measure of the overall consistency of different dynamical spaces. The RWSIP extends the concept of scalar product from single pairs of vectors to pairs of vector sets of equal dimension $s$ and number $Q$. Consider two sets of vectors, ${\bf u}_l$ and ${\bf v}_m$, with corresponding eigenvalues $\lambda^{u}_l$ and $\lambda^{v}_m$; in this context, they constitute a basis to describe the deformation of a molecule about a reference structure, and can be either obtained from an ENM or through principal component analysis of a molecular dynamics trajectory. Each ${\bf u}_l$ and ${\bf v}_m$ is a complete basis independent from the other, and as such they span the same vector space. On one extreme case, each vector of a basis could have a corresponding partner in the other one, albeit ranked in a different position; on the other extreme, no pair of vectors -each from one basis- could exist which point in the same direction. Depending on the strength of the corresponding eigenvalues, however, the essential spaces (i.e. the subsets of vectors with highest eigenvalues) of the two bases might overlap or not. The RWSIP quantifies this overlap by giving larger weight to the more collective modes. The RWSIP between subspaces composed of up to a number $Q$ of vectors is defined as:
\begin{eqnarray}
\textrm{RWSIP} &=& \sqrt{ \frac{\sum_{l, m =1}^Q \lambda^{u}_l \lambda^{v}_m | {\bf u}_l \cdot {\bf v}_m|^2}{\sum_{l =1}^Q \lambda^{u}_l \lambda^{v}_l}}\\ \nonumber
&=& \sqrt{\frac{\sum_{l, m =1}^Q \lambda^{u}_l \lambda^{v}_m \bigg|\sum_{i, j}^s u^i_l \cdot v^j_m \bigg|^2}{\sum_{l =1}^Q \lambda^{u}_l \lambda^{v}_l}} \ ,
\end{eqnarray}
and it lies by construction in the range $[0, 1]$. In the case of two sets of vectors representing the internal dynamics of a molecule composed by $N$ atoms, one has $s = Q = 3N$; correspondingly, the scalars $\lambda^u_i$ and $\lambda^v_i$ are the eigenvalues of the correlation matrix, that is, the inverse eigenvalues of the harmonic Hamiltonian. The measure of the RWSIP between the harmonic expansion of the CG-ENM and the exactly integrated CG-hENM provides a measure of how the properties of the latter are encoded into the former through the reconstruction procedure introduced in Eq.~(\ref{eq:trace_rule}).

Second, we consider the effectiveness of the various CG models in terms of the groups of atoms that are ascribed to specific CG sites, and of their internal dynamics. Specifically, we partition the atomistic structure of each molecule by means of a Voronoi tessellation, in which an atom is associated to the closest CG site (or, in case it is a CG site, to itself). We then perform a model dynamics exciting the eigenmodes of the AT-hENM, and compute how much of the dynamics, measured as the mean square fluctuation about the reference structure, can be ascribed to the motion of these groups of atoms {\it relative to each other}, and how much to the motion {\it internal to each group} \cite{Potestio2009,Polles2013}. The intra-block dynamics fraction (IBDF) is thus defined as follows.

Let each atom $i \in\{1, \ldots, N_\text{atoms}\}$ of the molecule be assigned to \emph{one and only one} Voronoi group $\mathcal{G}_I$ with $I \in\{1, \ldots, N_\text{groups}\}$, such that
\begin{equation}
\sum_{I=1}^{N_\text{groups}}\big|\mathcal{G}_I\big| =
\sum_{I=1}^{N_\text{groups}} \sum_{i \in \mathcal{G}_I} 1 = N_\text{atoms} \ .
\end{equation}
Furthermore, consider the two sets $\{{\bf r}_i\}_{i\in\mathcal{G}_I}$ and $\{{\bf r}_i^0\}_{i\in\mathcal{G}_I}$ of coordinates belonging to atoms $i\in\mathcal{G}_I$, in their present and reference configuration, respectively. We now define the mean square fluctuation $\sigma^2_I$ of these atoms with respect to their reference positions \textit{in the group} as the residual of a Kabsch alignment procedure \cite{Kabsch:a12999} carried out independently for each frame of the model dynamics. This procedure minimises the mean-square deviation between the sets $\{{\bf r}_i\}_{i\in\mathcal{G}_I}$ and $\{{\bf r}_i^0\}_{i\in\mathcal{G}_I}$ under all rotation-translation operations $\mathcal{K}$:
\begin{equation}
\sigma^2_I = \min_{\mathcal{K}}\left\langle \sum_{i \in \mathcal{G}_I} \Big[\mathcal K({\bf r}_i) - {\bf r}_i^0\Big]^2 \right\rangle \ .
\end{equation}
Similarly, one can define the residual mean square fluctuation for the whole molecule as:
\begin{equation}
\sigma^2_\text{full} = \min_{\mathcal{K}}\left\langle \sum_{i = 1}^{N_\text{atoms}} \Big[\mathcal K({\bf r}_i) - {\bf r}_i^0\Big]^2 \right\rangle \ .
\end{equation}
With these local and global fluctuation measures in place, we can now define the IBDF as
\begin{equation}
\text{IBDF} = \frac{\sum_{I=1}^{N_\text{groups}} \sigma^2_I}{\sigma^2_\text{full}} \ . \label{def_IBDF}
\end{equation}

Let us reiterate that the difference between the numerator and the denominator in Eq. \ref{def_IBDF} is that in the former the contribution from the relative motion \emph{among} the groups is absent.Hence, if the fluctuations \emph{within} each group are negligible, the IBDF is small, even if different groups move significantly with respect to each other. The IBDF thus provides a measure of the viability of these groups as quasi-rigid units in which the molecule can be decomposed. While these quasi-rigid units are formally similar to the ones customarily considered in the literature, they are different in spirit: the latter are in fact groups of \emph{amino acids} which provide a very coarse representation of the molecule in few large, function-oriented subunits; here, on the other hand, we consider groups of \emph{atoms} purveying a low-level coarse-graining  alternative to the conventional choice of one or two beads per amino acid.

Finally, we analyse the structure of the molecules in terms of various observables, namely: the structure of the interaction network; the distribution of local density of particles in proximity of an atom or CG site; and the size distribution of the Voronoi blocks associated to each CG site. Taken together, these properties offer a detailed, qualitative and quantitative, picture of the various models and their differences.

\begin{figure}[h]
\begin{center}
\includegraphics[width=\columnwidth]{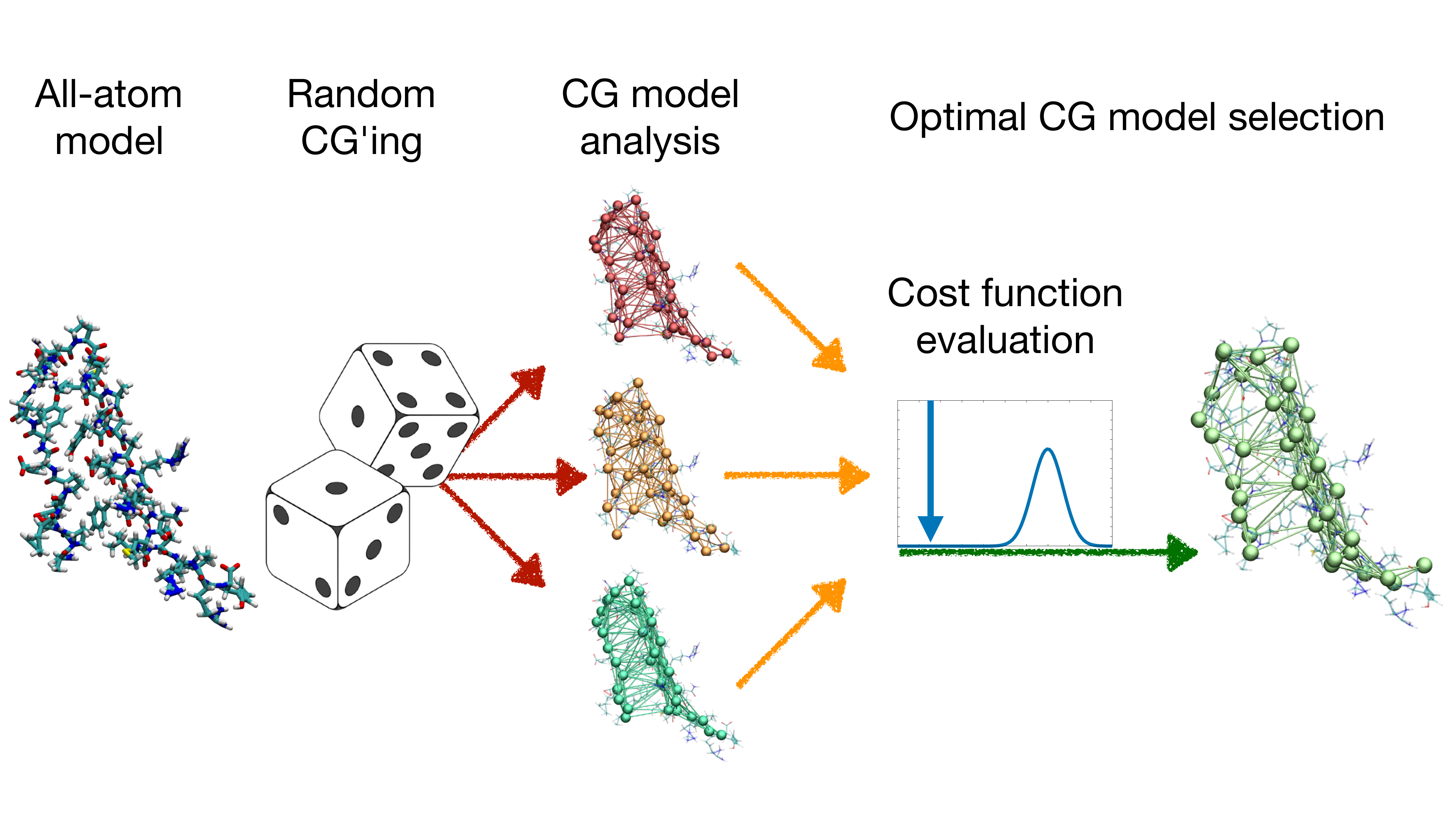}
\end{center}
\caption{Schematic of the main steps underlying the construction process of the coarse-grained model. Starting from a fully atomistic representation of the molecule, an atomistic elastic network model is constructed; from this, a selection of coarse-grained models is obtained by randomly choosing a set of coarse-grained degrees of freedom and exactly integrating out all the others; these models are assessed by a cost function that is optimized in a simulated annealing procedure. The CG model with the lowest value of the cost function is retained and used for all subsequent analyses.}\label{fig:scheme2}
\end{figure}

The main steps of the algorithm described above have been illustrated schematically in Fig. \ref{fig:scheme2}, which highlights the stochastic character of the coarse-grained model generation procedure and the selection based on an optimality criterion.

\begin{figure}[h]
\begin{center}
\includegraphics[width=0.9\columnwidth]{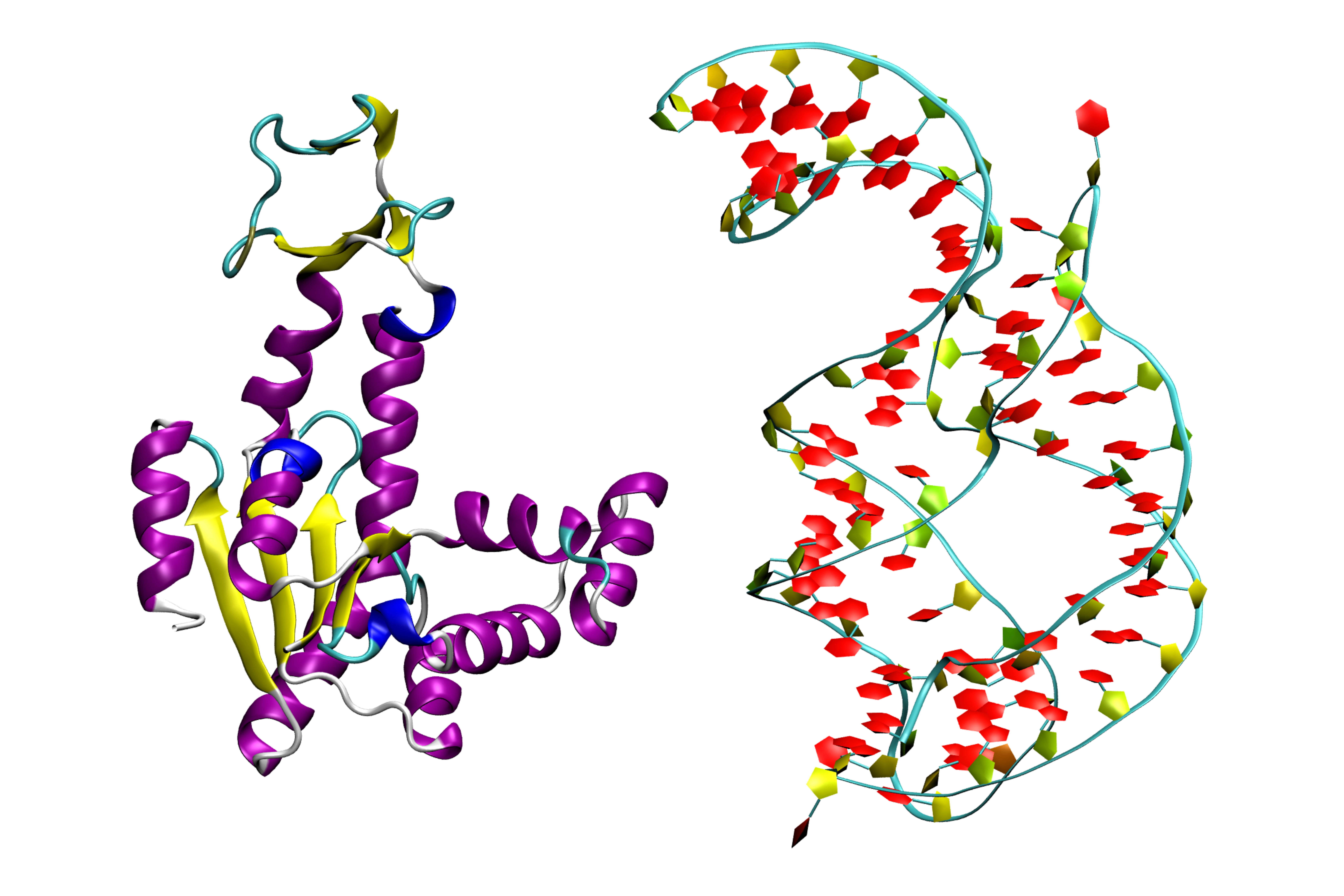}
\end{center}
\caption{The two molecules employed here to validate the proposed approach. Left: cartoon representation of Adenylate kinase (PDB code: 4AKE). Right: ribbon representation of adenine riboswitch (PDB code: 1Y26).}\label{fig:bothmolecules}
\end{figure}

In the following, we describe and discuss the results of applying our optimisation procedure to the two molecules depicted in Fig. \ref{fig:bothmolecules}, namely Adenylate kinase (\ake) \cite{4AKE} and the adenine riboswitch ({\it add}) \cite{1Y26}. These two molecules are similar in size ($\sim 1500$ atoms) and both undergo large-scale conformational rearrangements upon binding with their respective substrates. Their biological function thus largely relies on their internal, {\it collective} dynamics. Consequently, it is reasonable to expect that functional units can be identified in their structure, whose role and properties acquire meaning at an intermediate level between the atomic and the whole-protein ones. The process of coarse-graining should thus serve a twofold purpose: on the one hand, it should highlight the existence of these emergent structures; on the other hand, it would provide the ``language'' to express them, i.e. the interaction potentials among the coarse-grained constituents of the molecule. As it will subsequently become evident, this expectation may or may not be met---depending on specific intrinsic properties of the system under examination.

\section{Results and discussion}

\begin{table*}[htp]
\caption{Summary of data pertaining to the properties of the various models discussed in the text. For each CG model of both \ake\ and \add\ we report the number of coarse-grained sites employed; the value of the average eigenvalue ratio (AER, in percent); the $Z$-score of a given model with respect to the reference random distribution; the root weighted square inner product (RWSIP) between the exactly integrated CG model and the approximated model; and the fraction of intra-block dynamics not captured by the model (in percent).}
\begin{center}
\scalebox{1}{
\begin{tabular}{|c|c|c|c|c|c|c|c|c|}
\hline
 & AKE CA & AKE CB & AKE OPT & & RNA P & RNA C1$^\prime$ & RNA C2 & RNA OPT\\
\hline\hline
Number of CG sites & 214 & 194 & 214 & & 70 & 71 & 71 & 70 \\
\hline
AER (\%) & 43.549 & 47.737 & 34.564 &  & 57.895 & 53.692 & 55.075 & 37.646\\
\hline
Z-score & 4.493 & 2.599 & 19.705 & & 3.170 & 1.837 & 0.270 & 20.952\\
\hline
RWSIP CG ex--CG approx & 0.991 & 0.996 & 0.928 &  & 0.906 & 0.891 & 0.897 & 0.658\\
\hline
Fraction of intra-block dynamics (\%) & 3.00 &  3.05 & 2.30 &  & 88.52 & 88.28 & 88.24 & 87.02\\
\hline
\end{tabular}}
\end{center}
\label{table:summary}
\end{table*}%

Adenylate kinase, represented in Fig.~\ref{fig:bothmolecules} (left), is a globular protein of 214 amino acids (1656 atoms), responsible for the energy balance in the cell. Its relatively small size, biochemical relevance \cite{Pontiggia2008}, and flexible structure \cite{Potestio2009} make it a perfect candidate for the application of our approach. We investigated three different kinds of CG models: two ``standard'' ones, namely the one employing only the 214 C$_\alpha$ atoms, which are typically chosen as effective interaction centres in simplified models of polypeptides, and a model using only the 194 C$_\beta$ atoms; and the optimised model having 214 CG sites---as many as the $\alpha$ carbons. The interaction cutoff for all these models is set to $1$ nm, a typical value for protein ENM's \cite{Tirion1996,Micheletti2004,Bahar2005}.

\begin{figure*}
\begin{center}
\includegraphics[width=\columnwidth]{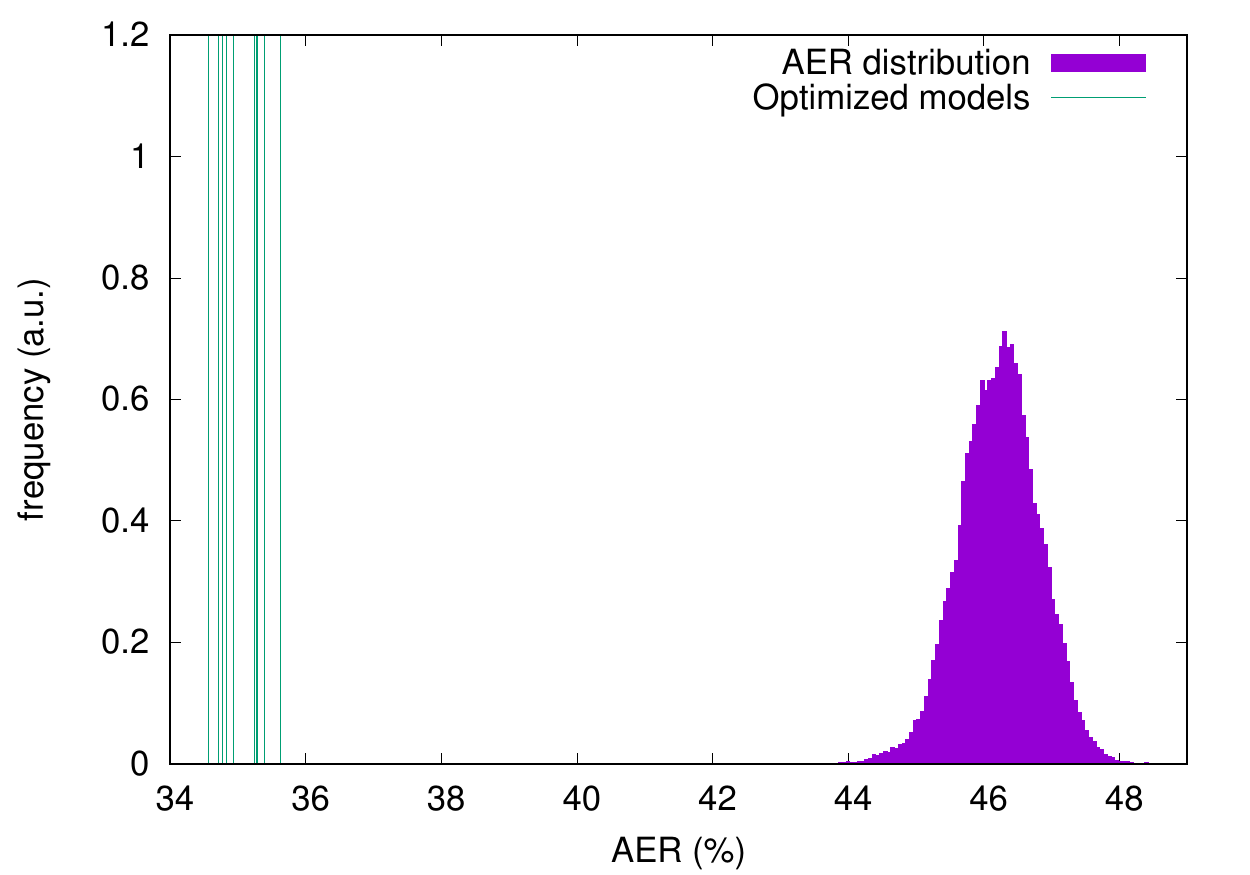}
\includegraphics[width=\columnwidth]{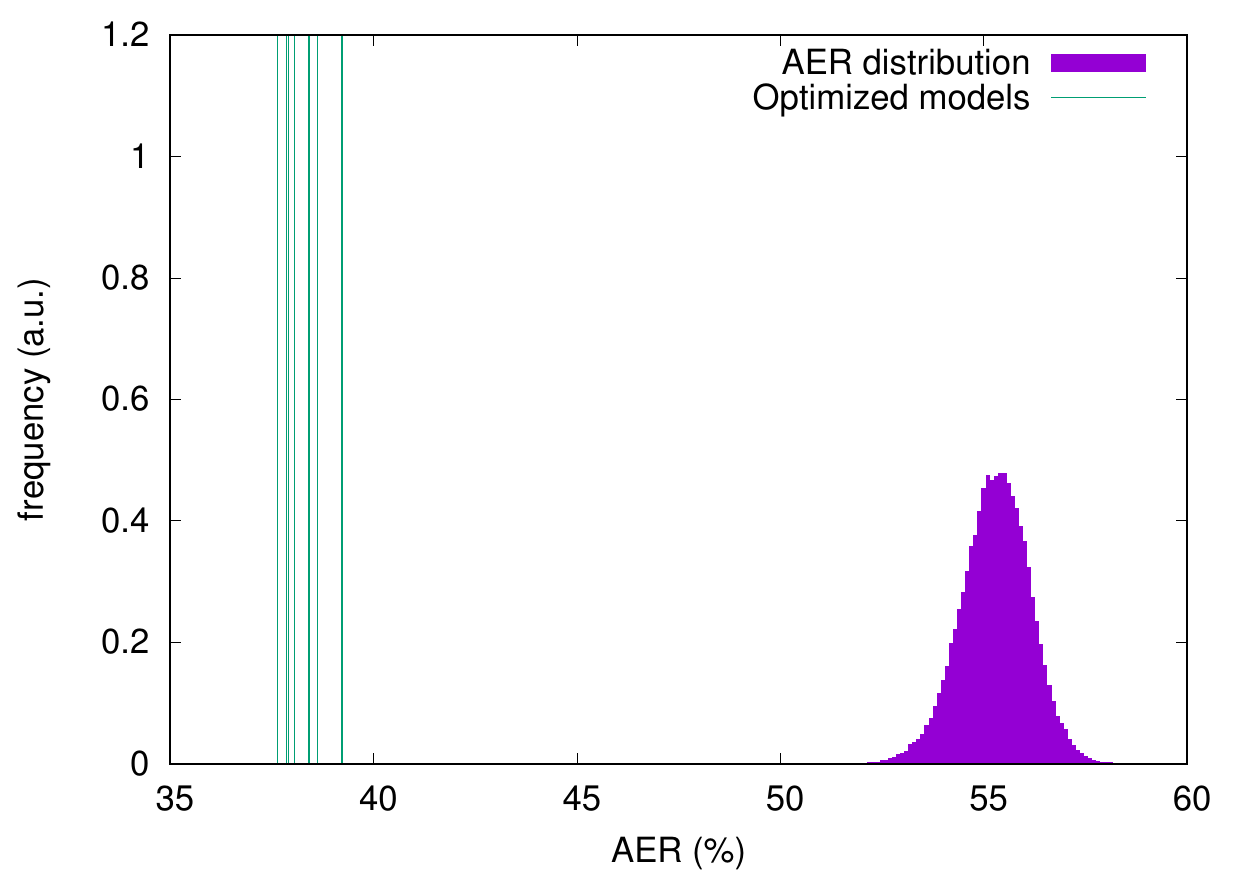}

\end{center}
\caption{AER values for randomly selected as well as optimised models. Left: \ake. Right: \add. The distribution, in purple, is obtained constructing $1.8 \times 10^5$ models with a fixed number of CG sites (214 for \ake, 71 for \add) randomly selected among all atoms. The green vertical lines indicate the positions of the AER values for each of the $10$ models obtained {\it via} simulated annealing optimisation. Of these, only the best --i.e., the one with the lowest AER value-- is further investigated.}\label{fig:AERdistrib}
\end{figure*}

In Fig. \ref{fig:AERdistrib} (left) we report the distribution of AER values for models of \ake\ having 214 CG sites. In these models the sites are selected at random; the resulting AER distribution is bell-shaped, with average and standard deviation being, respectively, $46.203$\% and $0.591$\%. The same figure also shows the AERs for the 10 independent simulated annealing minimisations. It is immediately evident that these values lie very far away from the average distribution: their average Z-score is $18.860$, while for the best one, which has an AER of $34.564$\%, the Z-score is as large as $19.705$. For comparison, standard CG models having only C$_\alpha$ or C$_\beta$ atoms feature Z-scores no larger than $4.5$, as reported in Table \ref{table:summary}.

We now turn our attention to the model with the lowest AER and its dynamical properties. From Table \ref{table:summary} we see that for the various models under examination the remapped CG-ENM shares a large dynamical consistency, as captured by the RWSIP, with the exactly integrated CG-hENM. The C$_\alpha$--only model has a value as high as $0.991$, while the C$_\beta$--only model is even slightly higher with $0.996$. The RWSIP between the reference CG-hENM and the harmonic expansion of the optimised model is not as high, however it is well above $0.9$; this result indicates that the criteria employed here to select the CG sites and to remap the interactions into a ``conventional'' CG-ENM guarantee a large overlap between the low-energy dynamical spaces of the model and the reference.

\begin{figure}
\begin{center}
\includegraphics[width=\columnwidth]{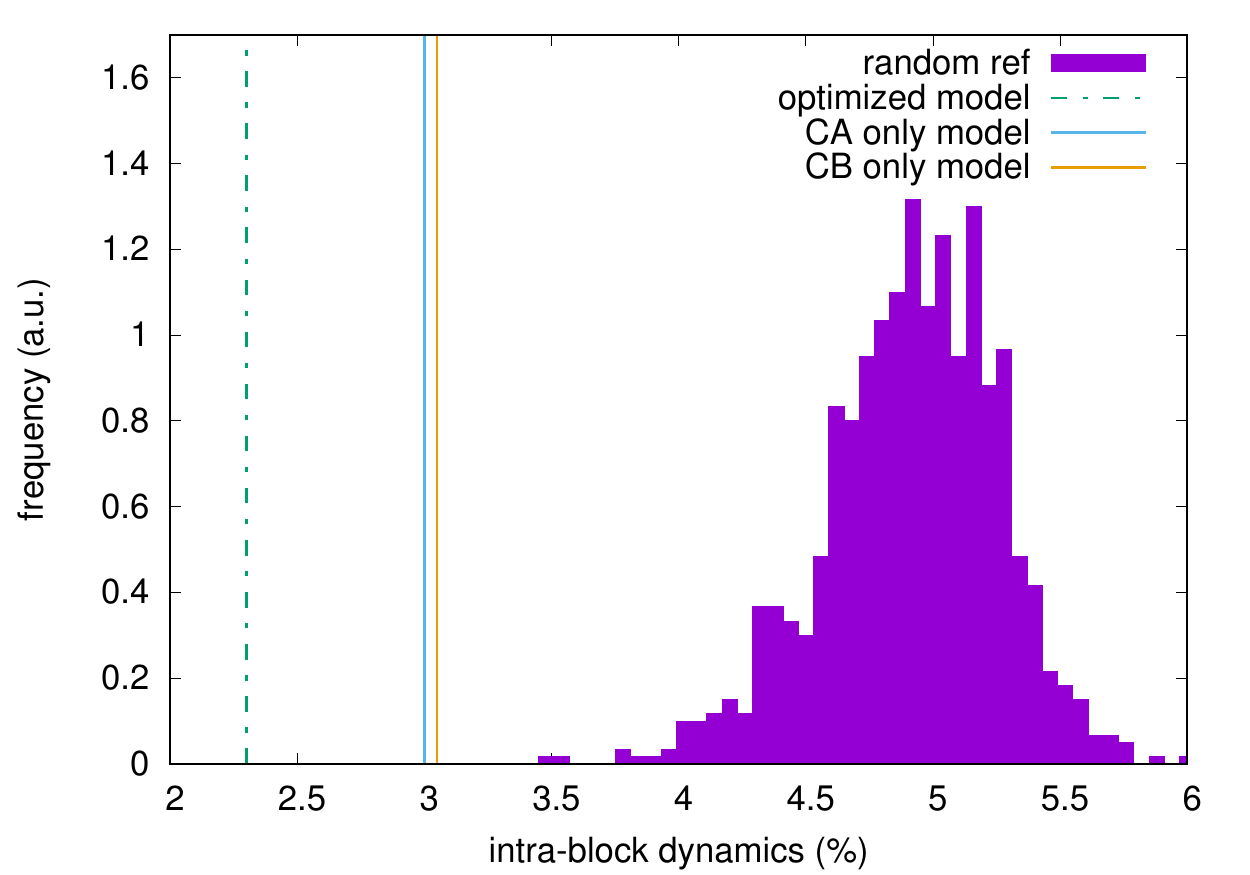}
\end{center}
\caption{Intra-block dynamics distribution for Adenylate kinase, obtained from 1000 models of \ake\ with 214 randomly-assigned CG sites. The vertical lines indicate the values of the intra-block dynamics fraction for the C$_\alpha$--only model (full blue line), the C$_\beta$--only model (full orange line), and the optimised model (dashed green line).}\label{fig:akeIBD}
\end{figure}

The second dynamical measure we employ is the fraction of dynamics that can be ascribed to the fluctuations {\it internal} to the Voronoi groups. Comparing the values reported in Table \ref{table:summary}, the model with the lowest AER also emerges as the one with the lowest IBDF value. In Fig. \ref{fig:akeIBD} we show the comparison of the IBDF of the various models with a reference distribution, obtained from 1000 models of \ake\ in which the 214 CG sites have been randomly assigned. All three CG models under examination feature an IBDF well below the average, with the C$_\alpha$--only and C$_\beta$--only models very close to each other; the optimised model, though, features an even lower value, highlighting its statistically relevant extremality.

This suggests that the CG site selection and remapping algorithm favours the construction of models in which the effective sites are representative of more rigid, i.e. more collectively fluctuating groups of atoms. This result is doubly interesting: on the one hand because it was not sought after nor encoded in the modelling strategy; on the other hand because it is at odds with the dynamical properties of the models as measured by the RWSIP. The picture that emerges thus hints at the (not entirely unsurprising) fact that which model performs best depends on the metric one choses to quantify performance.

\begin{figure}
\begin{center}
\includegraphics[width=\columnwidth]{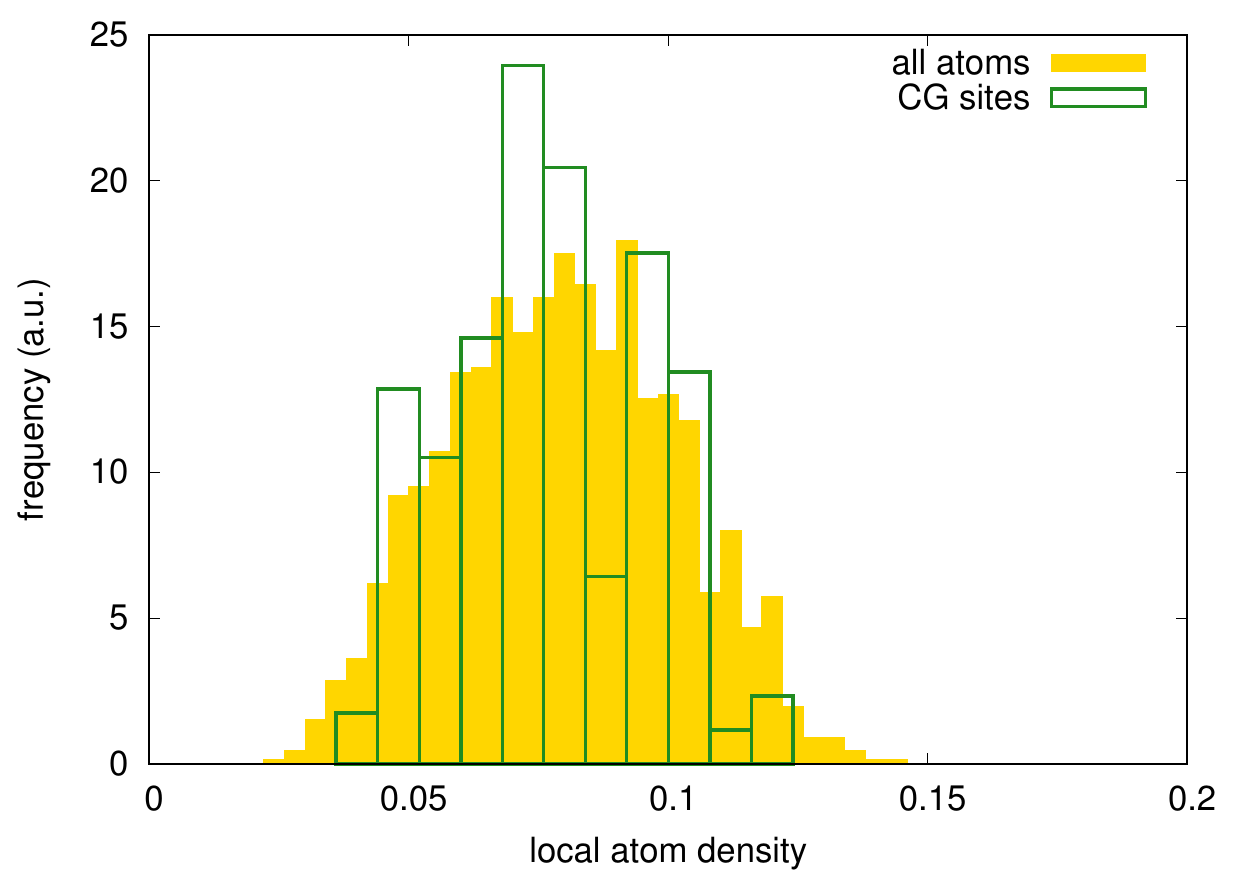}\\
\includegraphics[width=\columnwidth]{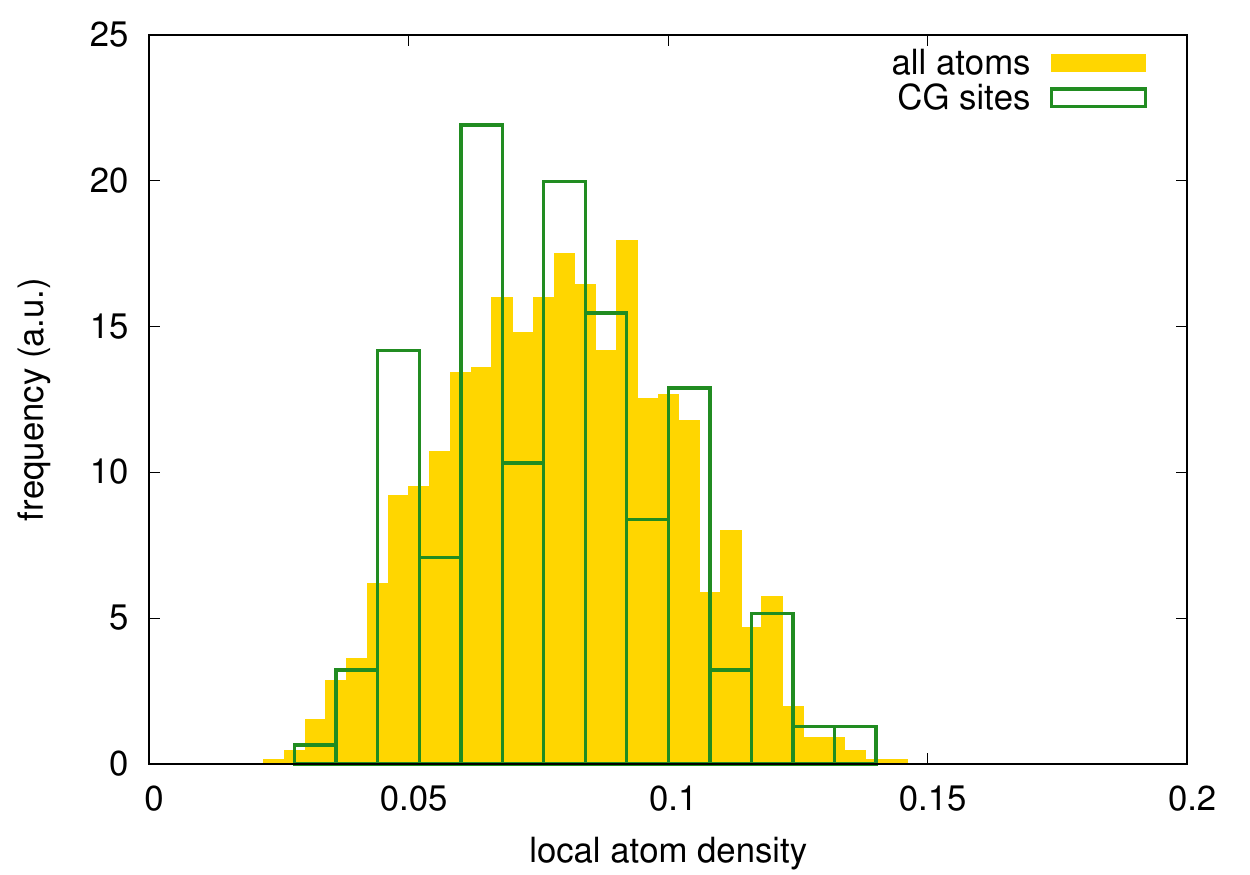}\\
\includegraphics[width=\columnwidth]{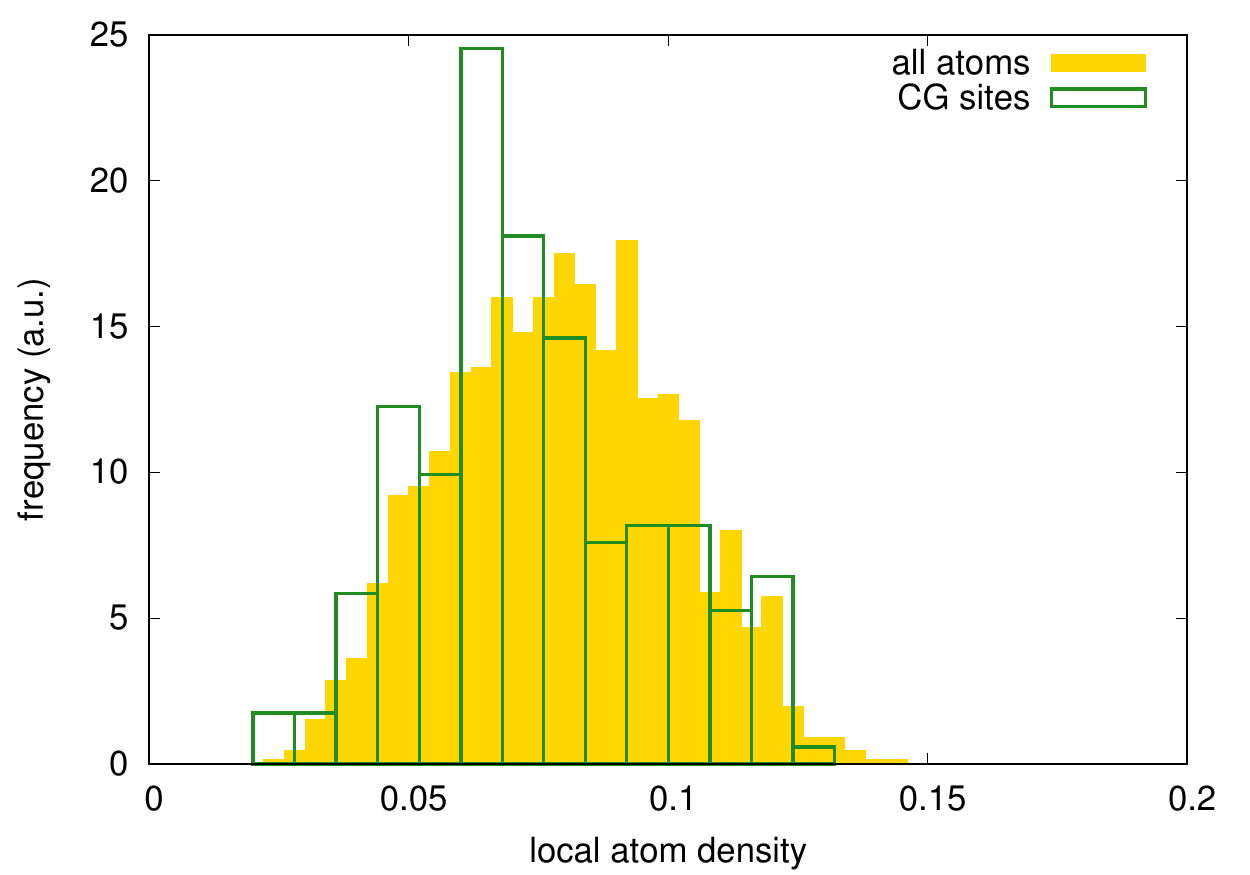}
\end{center}
\caption{Local normalised density distribution of particles in the all-atom model (yellow, filled boxes) and CG sites (green, empty boxes) for Adenylate kinase. The all-atom density distribution is the same in all cases; the CG density distribution is given for the various models as follows. Top: C$_\alpha$--only atoms; centre: C$_\beta$--only atoms; bottom: optimised model.}\label{fig:akeLocDens}
\end{figure}

\begin{figure*}[htp]
\begin{center}
\includegraphics[width=2\columnwidth]{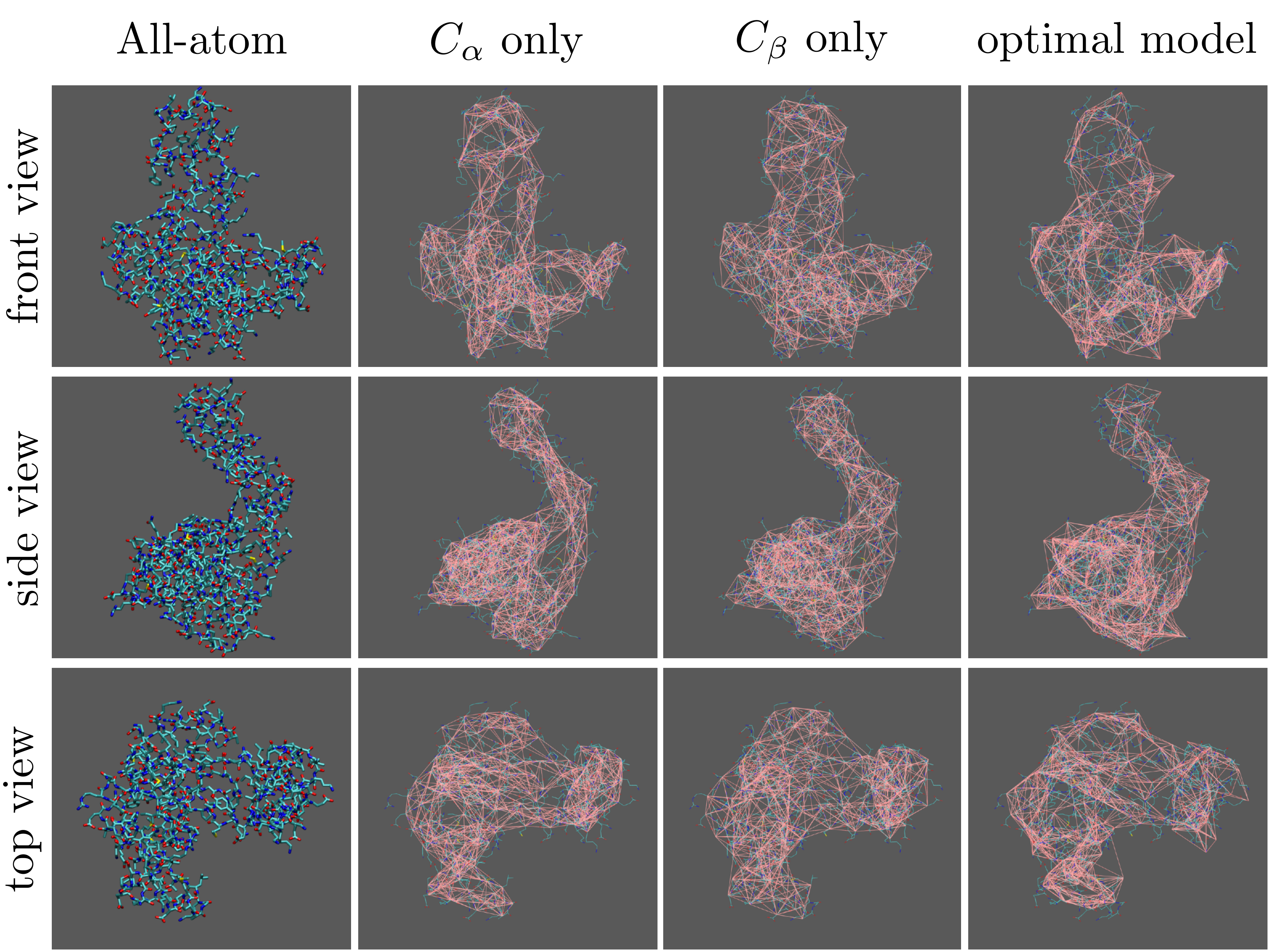}
\end{center}
\caption{The structure of Adenylate kinase (leftmost column, in licorice representation) from three orthogonal perspectives, compared to the atom selections discussed in the text. From left to right: all-atom representation; C$_\alpha$ atoms only; C$_\beta$ atoms only; all those atoms included in the optimal model by the simulated annealing approach. In all figures except the ones in the first column, the all-atom structure is provided as a faint lines representation in the background for the sake of comparison, while the network of ENM interactions among CG sites is shown in pink.}\label{fig:ProteinComparison}
\end{figure*}

How nontrivial the choice of CG sites is that results from the optimisation procedure can be illustrated by looking at the local density distribution, reported in Fig. \ref{fig:akeLocDens}. The local density is computed as the number of atoms within the interaction cutoff divided by the total number of particles: these are atoms in the all-atom model (yellow, filled histogram), and CG sites for all three CG models under examination (green, empty histogram); the former distribution does not depend on the CG model and is the same in all three plots. There appears to be no appreciable difference between the density distribution for the C$_\alpha$--only and C$_\beta$--only models; both are also fairly consistent with the background all-atom distribution, highlighting the uniformity of the assignment of these specific CG sites. This can also be seen from the networks reported in Fig. \ref{fig:ProteinComparison}: in particular, the network of the C$_\alpha$--only model strictly follows the peptide backbone, drawing a tube-like interaction pattern, while in the C$_\beta$--only model the network looks even more compact and uniform. The optimised model, on the other hand, favours a more inhomogeneous distribution, i.e. the occurrence of both ``dense clusters'' and ``voids''. This impression is consistent with the network shown in Fig. \ref{fig:ProteinComparison}, where fairly large ``holes'' in the interaction pattern can be seen especially in the protein's head; however, a more quantitative picture would be helpful.

Such a picture is once again provided by the Voronoi-like tessellation of the molecule, which allows for its decomposition in terms of groups of atoms each represented by the nearest CG site. We can then measure the distribution of the number of atoms included in such groups. A regular, homogeneous distribution of CG sites will be associated with a fairly peaked atom number distribution, indicating that each block contains roughly the same number of particles; on the other hand, if the CG sites are allocated in a less homogeneous manner, a broader distribution will emerge.

\begin{figure}[htp]
\begin{center}
\includegraphics[width=\columnwidth]{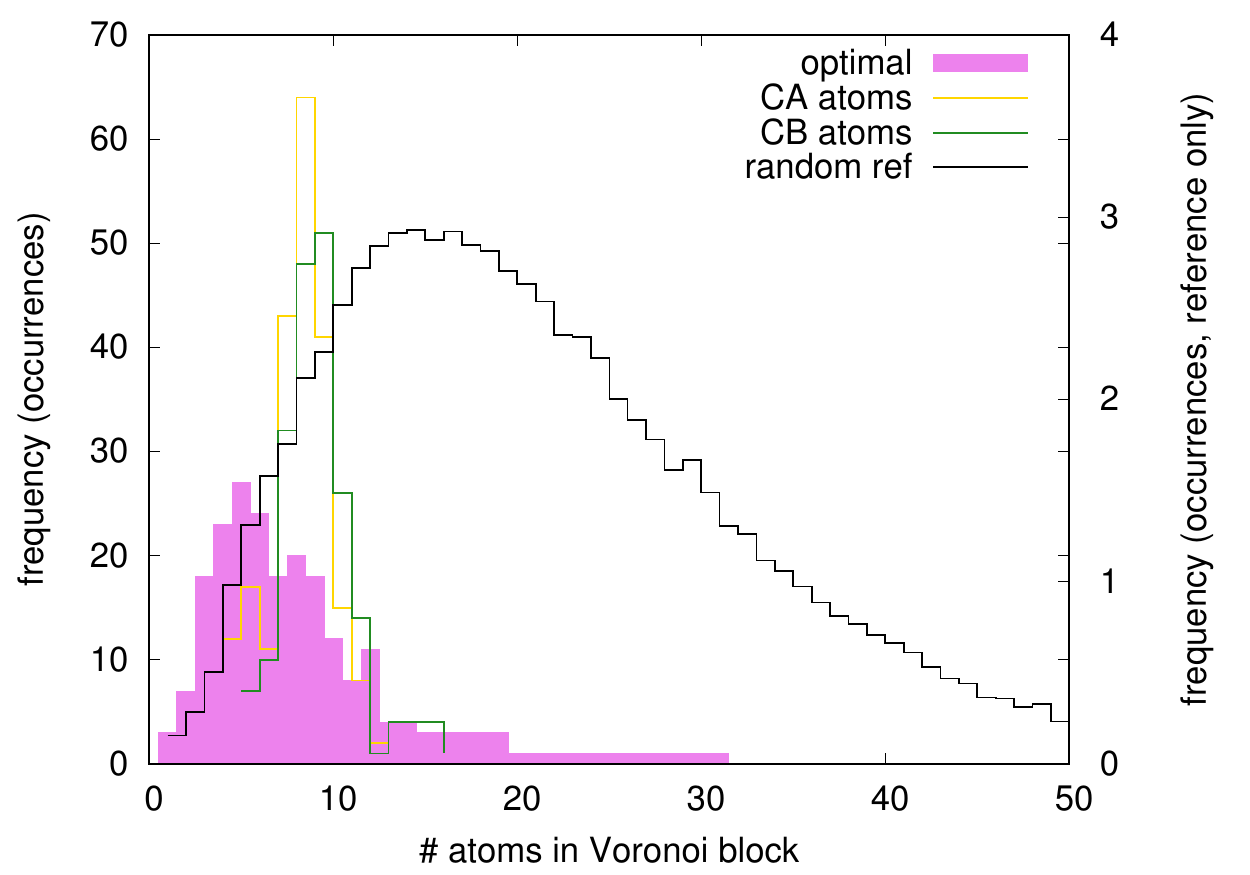}
\end{center}
\caption{Distribution of the number of atoms included in the Voronoi blocks for different models of Adenylate kinase: C$_\alpha$ atoms only (yellow empty line); C$_\beta$ atoms only (green empty line); random CG site assignment (black empty line); optimised model (full magenta line). The curves are normalised so that the average number of atoms, weighted by the distribution, equals the total number of atoms in the molecule (1656). Note that the right y-axis applies to the random reference curve only.}\label{fig:akeVoronoi}
\end{figure}

In Fig. \ref{fig:akeVoronoi} we report the distribution of atoms in the Voronoi blocks for the three models of \ake\ under examination. The C$_\alpha$-only and C$_\beta$-only models indeed exhibit peaked distributions, indicating that a CG site has typically $8$ neighbouring atoms, with deviations in the number of $\pm 4$ atoms. The optimised model, on the other hand, features a much broader distribution covering the whole range from a single neighbouring atom up to $30$, with a peak for $5$ atoms. This behaviour substantially departs from the standard cases as well as from a random assignment of CG sites: the latter, in fact, gives rise to the ``Maxwellian'' distribution reported in Fig. \ref{fig:akeVoronoi}, which is similar in shape to the optimised model distribution, however with substantially different average and width. The observed pattern is consistent with a nontrivial disposition of CG sites in the optimised CG model, where both rather ``high-resolution'' and ``low-resolution'' regions can be found. The most striking feature of this model can thus be identified in the non-uniform character of the CG site distribution across the structure.

Our second case study is the adenine riboswitch \add, pictured in Fig.~\ref{fig:bothmolecules} (right). This 71-bases-long RNA molecule, similar in size to \ake\ with 1499 heavy atoms, undergoes large-scale conformational changes upon binding to adenine. The internal dynamics of this class of molecules has been little investigated by means of ENM-like models, with a few notable exceptions \cite{setny2013,Zimmermann2014,Pinamonti2015,falconi2017}, thus it not only represents an interesting case study for our method, but also allows a direct comparison with pioneering studies in the field of RNA ENM-based modelling. As a reference, we consider models that employ the same atom from each base, specifically the phosphorus atom P, the C1$^\prime$ carbon atom, and the C2 carbon atom from the phosphate, sugar, and base moieties of the nucleic acid, respectively. The interaction cutoff is set to $2$ nm: this value was found in previous work \cite{Pinamonti2015} to provide the best results for P--only RNA hENM's; smaller optimal cutoff values were found for the other two model types, however, we decided to employ the largest among them for simplicity and to provide the most uniform and consistent set of parameters across different models. We point out that, in spite of a rather similar number of atoms between the two molecules under examination, the CG-sites-to-atoms ratio for \add\ (1:20) is almost three times smaller that of \ake. This is the case because the numbers of amino acids and nucleic bases in the two molecules differ in the same proportion. The aim of the present work is to perform a comparison among different models of the same system, while preserving the same overall level of coarse-graining within each case. This makes a direct comparison between \ake\ and \add\ necessarily unfair in terms of CG-sites-to-atoms ratio, however maintaining the rule of thumb {\it one atom per polymeric unit} valid for both.

The same dynamical analysis performed for \ake\ was carried out for \add, the results being reported in Table \ref{table:summary}. In this case we notice a qualitative behaviour consistent with the one previously described, however with a few remarkable differences. First, the RWSIP between the harmonic expansion of the remapped CG-ENM and its reference CG-hENM obtained {\it via} exact Boltzmann integration is substantially lower for the optimised model than for the standard one-atom choices for CG sites (P, C1$^\prime$, and C2 atoms): this is qualitatively the same trend observed for \ake, however the gap is wider. Furthermore, also for the ``standard'', better performing CG models --the best being the P-only model-- the RWSIP is $10\%$ lower than the best model of \ake, and they all have very similar values of RWSIP. The closeness of these values makes it difficult to rank the same-atom coarse-grained representations in terms of their representativeness of the reference, all-atom system. Previous work by Pinamonti {\it et al.} \cite{Pinamonti2015} has investigated these three models using a Hessian network model, and found that the C1$^\prime$-only model performed best at reproducing the fluctuations of all-atom reference simulations employing realistic force fields. This was followed by the C2-only model and, finally, by the P-only model. Similarly, Setny and Zacharias \cite{setny2013} have observed better performing ENMs when the effective interaction center was placed in the ribose ring rather than the phosphorus atom. If we look at the data in Table \ref{table:summary}, we find that the P-only model has the highest RWSIP; however, the small ($\sim 1 \%$) differences among the three conventional representations do not justify their ranking.

It also deserves to be noted that the model ranking proposed by Pinamonti {\it et al.} \cite{Pinamonti2015} is based on differences among the models' RWSIP that do not exceed $0.05-0.06$, thus consistent with the ones observed in this work and compatible with a substantial equivalence within deviations that can depend on several factors (model parameters, numerical accuracy, measure of dynamical consistency etc.). A second observable employed in \cite{Pinamonti2015}, comparing the dynamical properties of the ENM's to those of reference, all-atom simulations with accurate force fields, clearly indicates the P-only CG model as poorly performing, however the other two models are again quantitatively very close to each other.

\begin{figure}
\begin{center}
\includegraphics[width=\columnwidth]{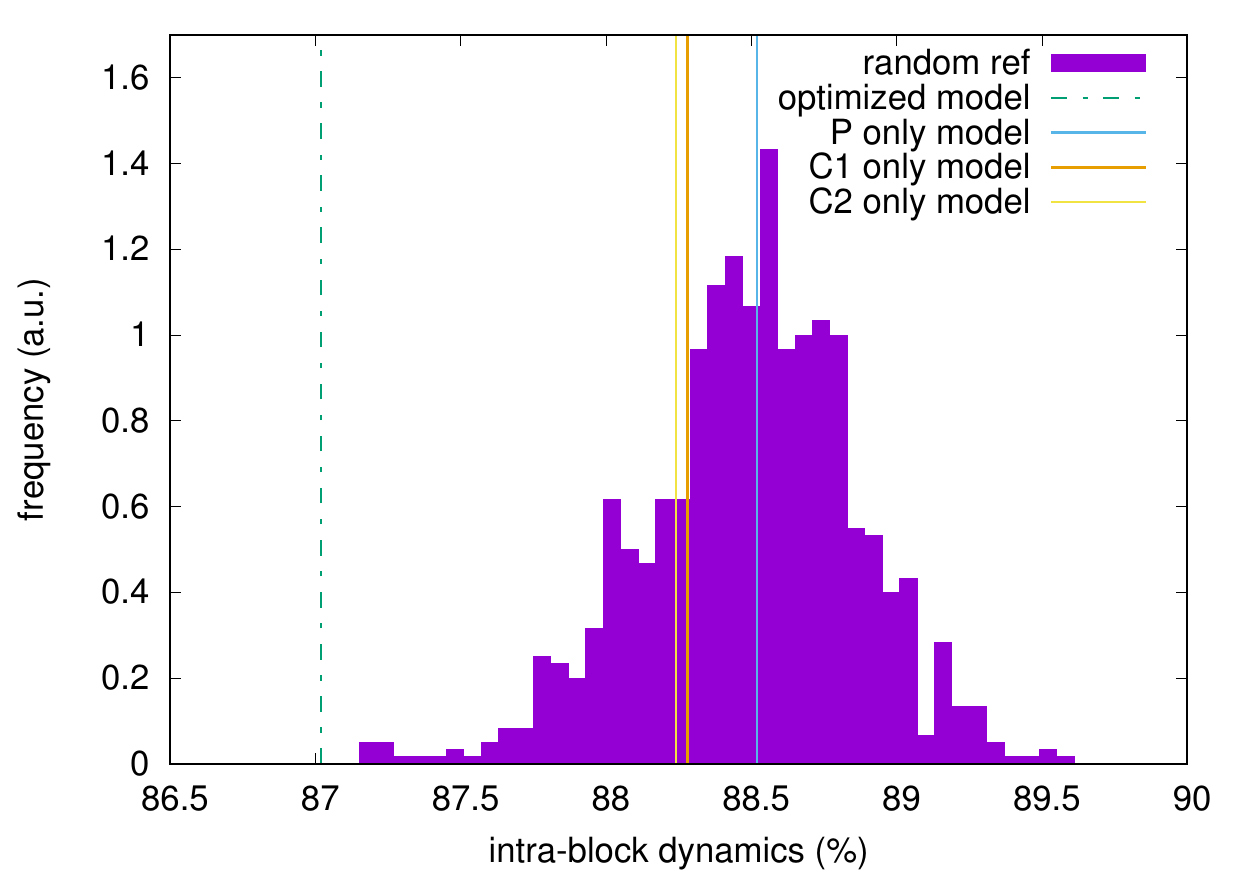}
\end{center}
\caption{Intra-block dynamics distribution for the adenine riboswitch, obtained from 1000 models of \add\ with 70 randomly-assigned CG sites. The vertical lines indicate the values of the intra-block dynamics fraction for the P--only model (full blue line), the C1$^\prime$--only model (full orange line), the C2--only model (full yellow line), and the optimised model (dashed green line).}\label{fig:rnaIBD}
\end{figure}

Second, we note that the fraction of motion internal to the Voronoi block is, for all models, much larger than what was observed for \ake, with all values in the range $87-88.5\%$. A high fraction of intra-block fluctuation is suggestive of a poorly collective dynamics: this behaviour is markedly at odds with Adenylate kinase, which on the contrary is thoroughly characterised by a highly modular, function-oriented dynamics \cite{Pontiggia2007,Potestio2009}. Indeed, \add\ also undergoes large-scale motions upon binding \cite{PRIYAKUMAR20101422,Allner01072013,DiPalma01112013,Pinamonti2015}, however these are qualitatively different from those of \ake, in that they largely consist of sequence rearrangements and base-pair breakage/formation; the large flexibility necessary to perform this dramatic structural rewiring is encoded, at least at a very basic level, into the contact network, and hence into a model as simple as an ENM. The large amount of molecular fluctuation {\it within} a compact group of atoms makes this lack of collectivity and directed dynamics manifest.

The typical intra-block dynamics fraction of \add, i.e. the amount of molecule dynamics that cannot be ascribed to the relative motion among the blocks, is much larger than for \ake, as it can be seen in Fig. \ref{fig:rnaIBD}. The average of the IBDF distribution, computed over 1000 random CG models, is in fact $\sim 88.5$\%. The standard, same-atom CG models feature values just at or slightly below the average, in any case well within the distribution. The optimised model, on the other hand, lies about three standard deviations below the average and just outside of the left tail of the distribution. While, on the one hand, the optimised model features a statistically significant improvement of the IBDF with respect to both random and standard CG models, this improvement is not, on the other hand, as important as in the case of Adenylate kinase.

\begin{figure*}[htp]
\begin{center}
\includegraphics[width=2\columnwidth]{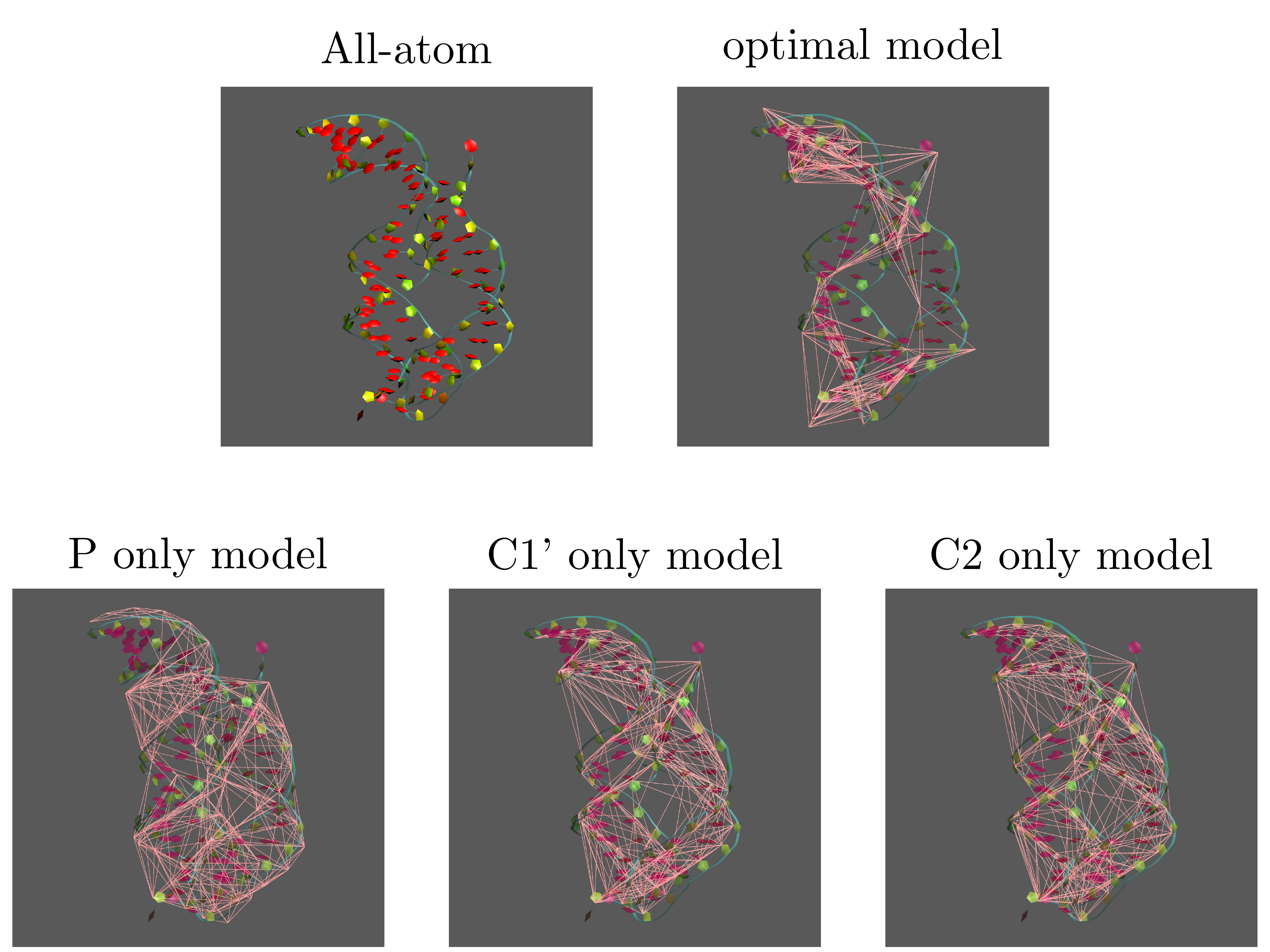}
\end{center}
\caption{Structure of the adenine riboswitch in the various models. Top row from left: ribbon representation of the all-atom structure; optimised model. Bottom row from left: P atoms only; C1$^\prime$ atoms only; C2 atoms only. In all figures except the first, the all-atom structure is provided as a faint ghost representation in the background for the sake of comparison, while the network of ENM interactions among CG sites is shown in pink.}\label{fig:RNAComparison}
\end{figure*}

\begin{figure}[htp]
\begin{center}
\includegraphics[width=\columnwidth]{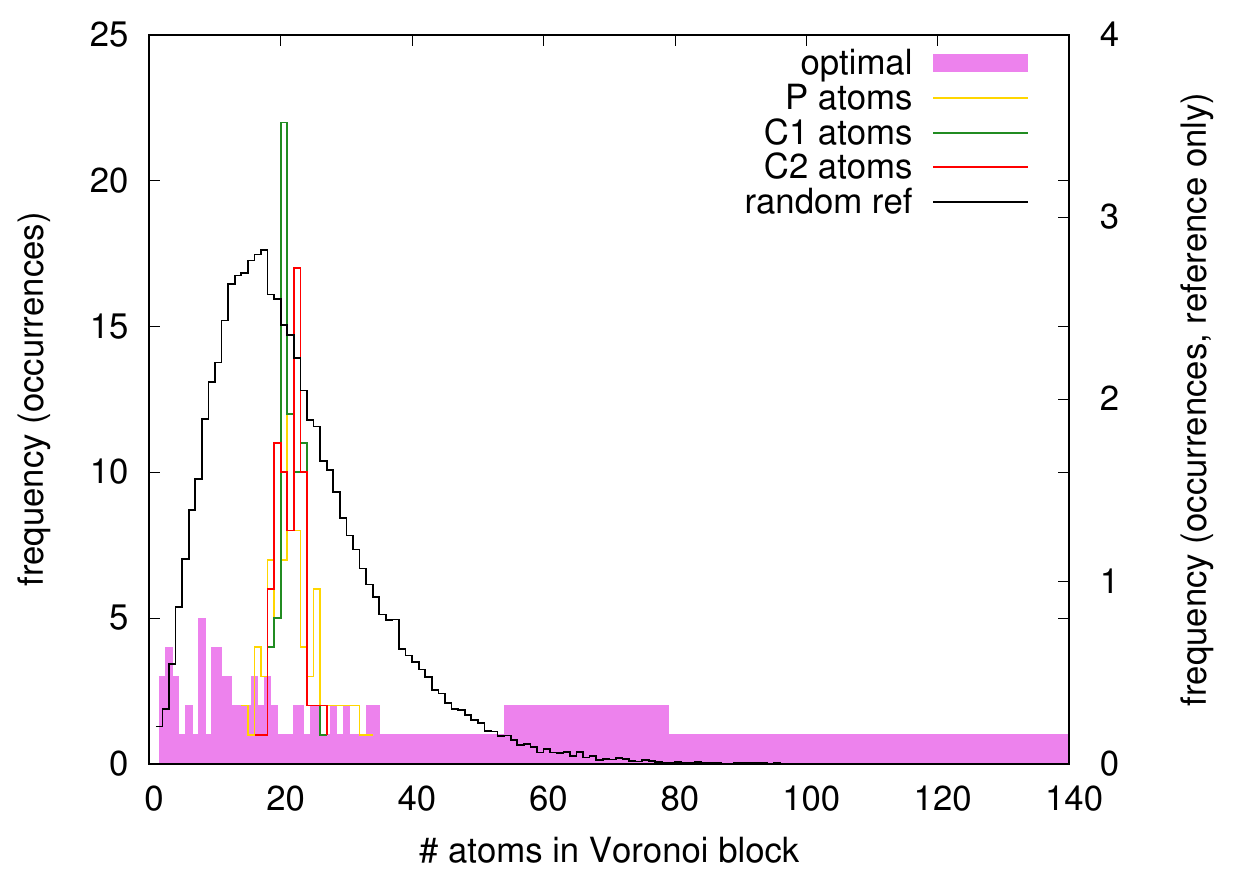}
\end{center}
\caption{Distribution of the number of atoms included in the Voronoi blocks for different models of adenine riboswitch: P atoms only (yellow empty line); C1$^\prime$ atoms only (green empty line); C2 atoms only (red empty line); random CG site assignment (black empty line); optimised model (full magenta line). Note that the random reference distribution is nonzero for values up to 148 atoms. The curves are normalised so that the average number of atoms, weighted by the distribution, equals the total number of atoms in the molecule (1499). Note that the right y-axis applies to the random reference curve only.}\label{fig:rnaVoronoi}
\end{figure}

\begin{figure*}[htp]
\begin{center}
\includegraphics[width=\columnwidth]{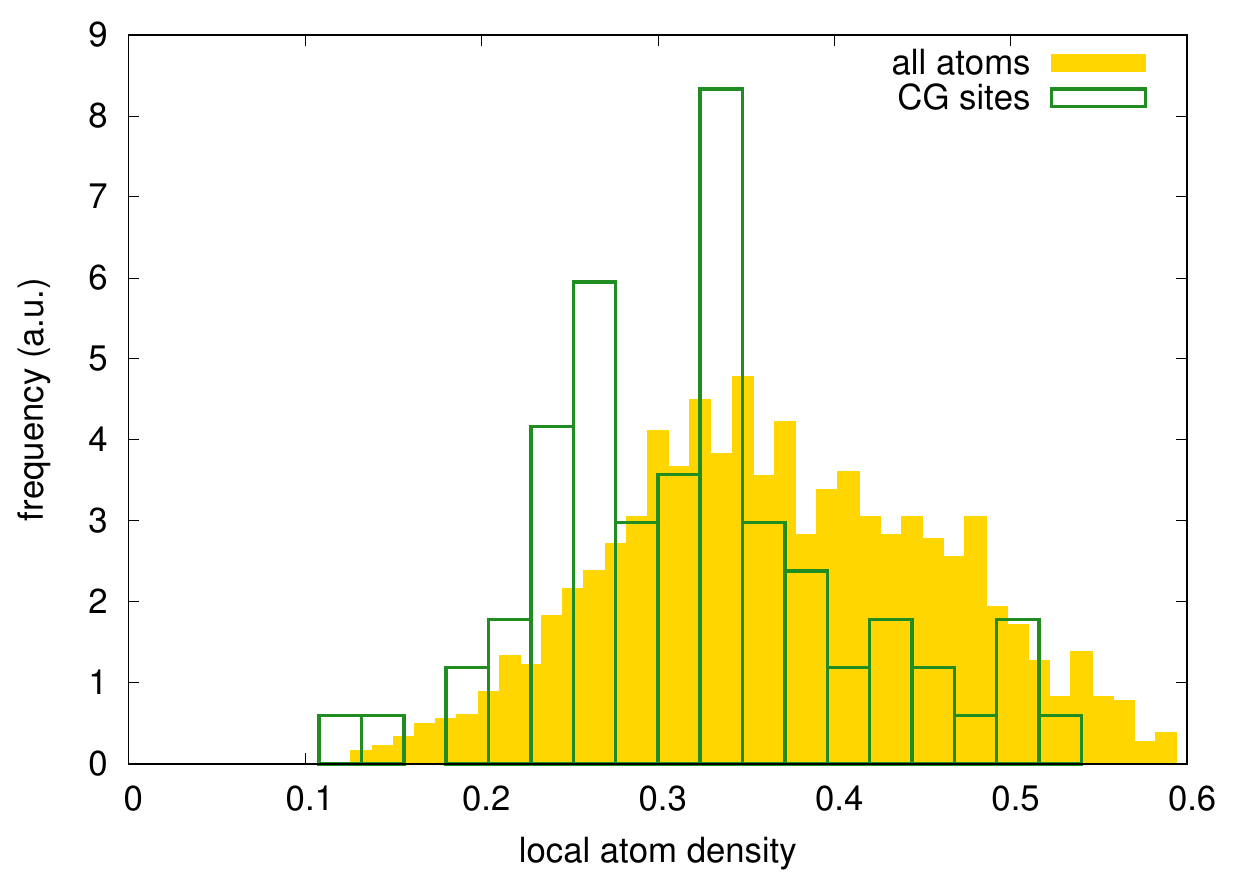}
\includegraphics[width=\columnwidth]{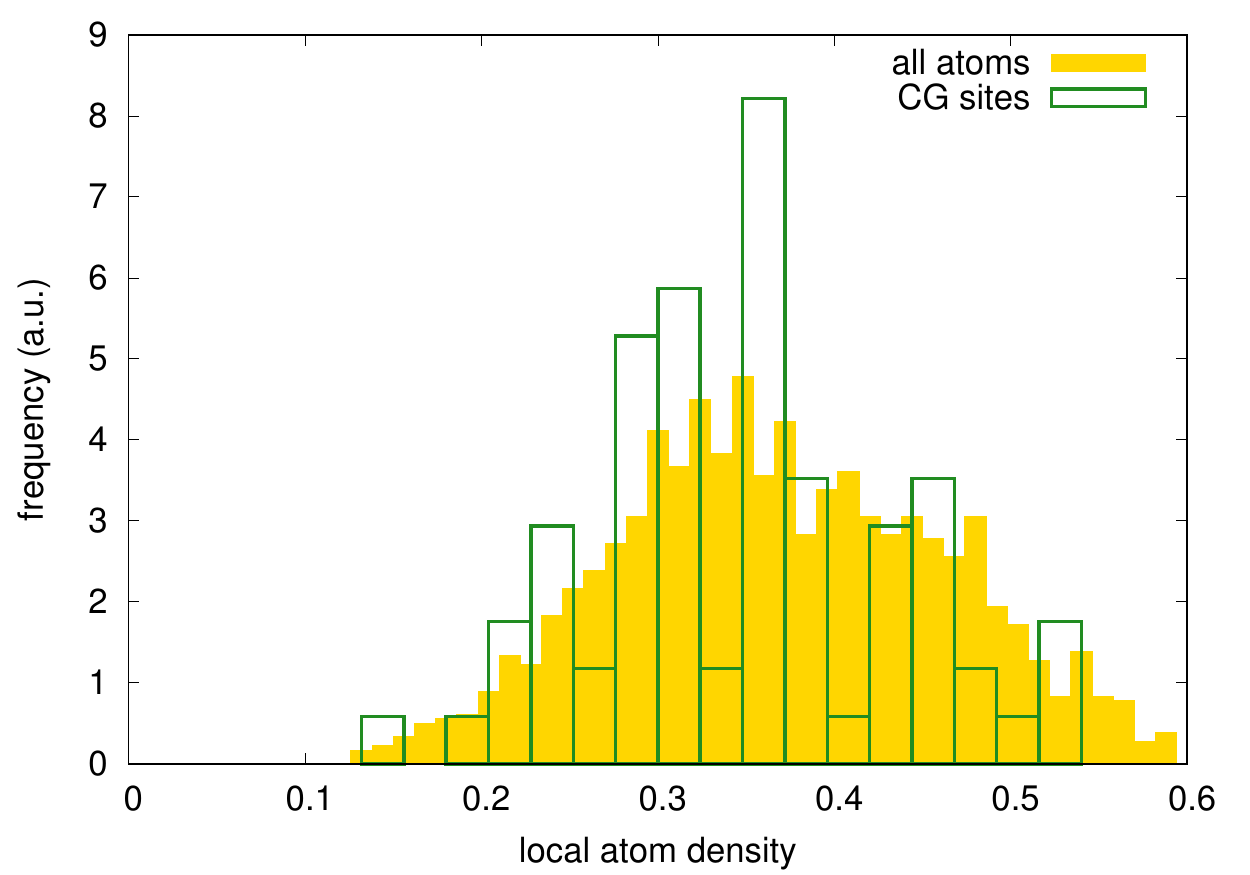}\\
\includegraphics[width=\columnwidth]{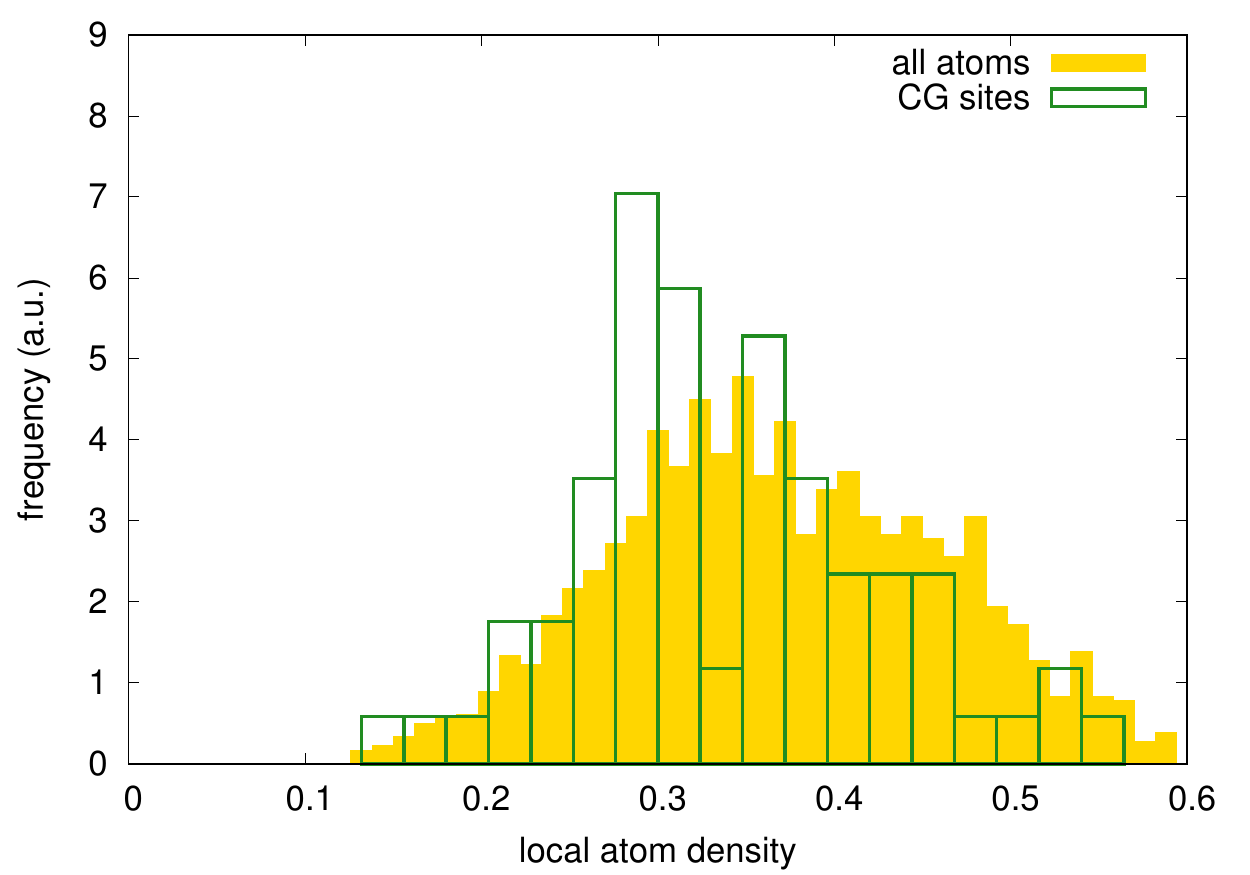}
\includegraphics[width=\columnwidth]{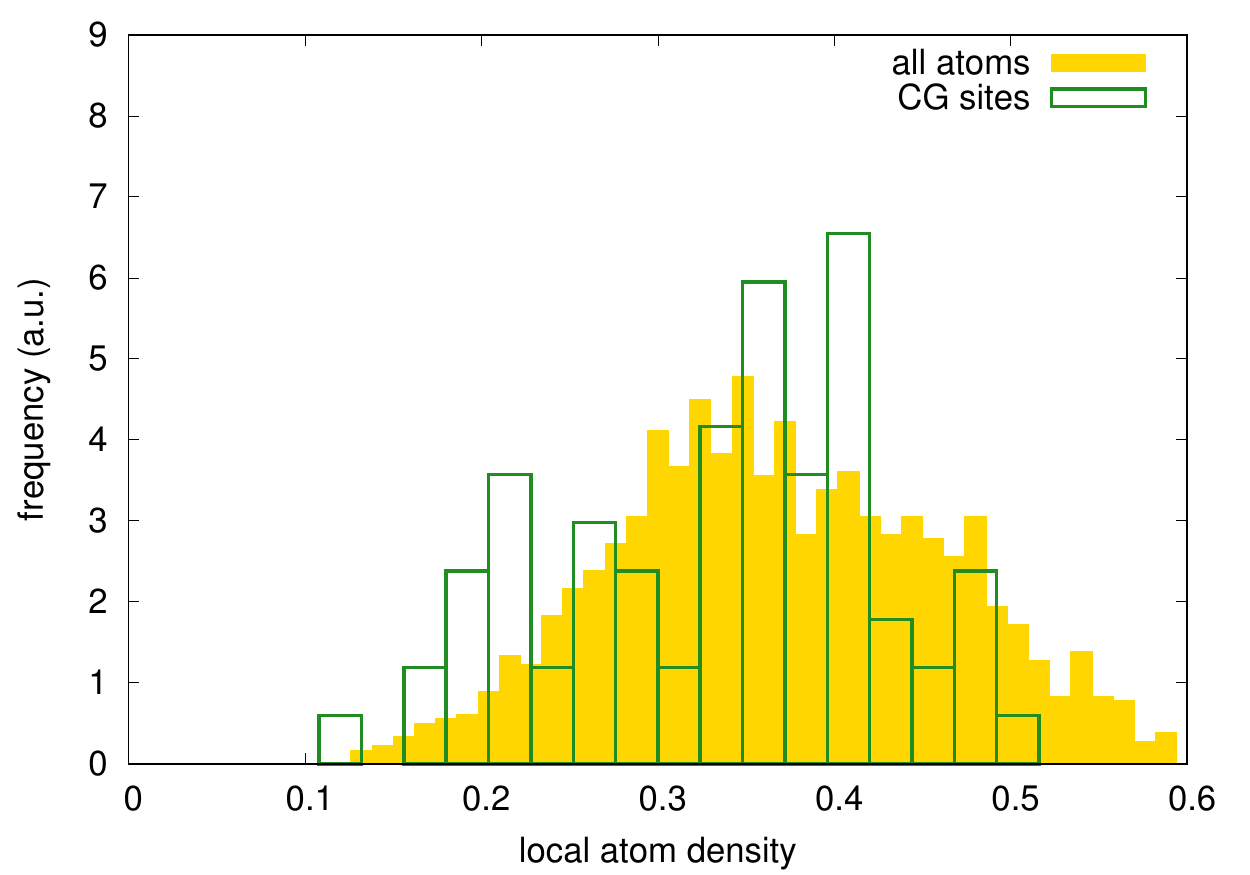}
\end{center}
\caption{Local normalised density distribution of particles in the all-atom model (yellow, filled boxes) and CG sites (green, empty boxes) for the Adenine riboswitch. The all-atom density distribution is the same in all cases; the CG density distribution is given for the various models as follows. Top left: P--only atoms; top right: C1$^\prime$--only atoms; bottom left: C2--only atoms; bottom right: optimised model.}\label{fig:rnaLocDens}
\end{figure*}

In summary, the optimised model shows a RWSIP between exactly integrated and remapped CG-hENM that is substantially lower than the ones observed in the other cases; in contrast, and consistently with the trends featured by \ake, the fraction of intra-block dynamics is lowest for the optimised representation, however by a small amount with respect to the other models. This is a nontrivial result, given the remarkable structural difference existing among the models. If, on the one hand, the CG models employing the same type of atoms have rather similar interaction network structures, as it can be seen in Fig. \ref{fig:RNAComparison}, the one of the optimised model deviates remarkably from this evenness: the distribution of CG sites is highly irregular, as it can be seen in the interaction network figure as well as in the Voronoi block size distributions, reported in Fig. \ref{fig:rnaVoronoi}. The intuitive structure of the RNA molecule is lost in favour of a  hollow web of interactions among the CG sites, each being representative of a group of atoms --the closest ones that have been integrated out-- whose number ranges from a few up to several tens. The distributions of local atom densities, shown in Fig. \ref{fig:rnaLocDens}, are consistent with this trend and in line with the one observed for \ake: that is, a relatively small deviation of the optimised model with respect to the other ones towards lower values, compatible with the more inhomogeneous structure of the CG site network.

\section{Conclusions}

Elastic network models represent a milestone in the computer-aided study of biomolecules, in that they enabled the fast, inexpensive and remarkably accurate characterisation of the equilibrium, function-oriented dynamics of these systems. Relying {\it a priori} on solid statistical mechanical arguments and {\it a posteriori} on thorough consistency checks and cross validations against independent data (experiments, atomistic MD simulations etc.), ENM's have been and still are at the heart of a wealth of methods that require fast access to the large-scale collective dynamics of proteins and other molecules.

In general, the effective interaction centres employed in an ENM are a specific subset of a molecule's atoms -- e.g the C$_\alpha$ atoms of a protein. In this work we have proposed and tested an algorithmic procedure to select these centres based on an extremality criterion. Starting from the harmonic approximation to an atomistic ENM, we have selected a subset of atoms to be retained as CG sites, thus generating a new harmonic ENM (hENM). The complementary subset of removed atoms is integrated out and embedded in effective interactions, whose functional form, albeit harmonic in the atoms' displacements, is not compatible with the straightforward harmonic expansion of an ENM. This difference can be used to generate the CG ENM whose harmonic expansion is the closest, according to a well-defined measure (the AER), to the integrated-out hENM. The optimal model is defined as the one minimising the distance between integrated hENM and ENM harmonic expansion over all possible removed atoms selections. This approach enables one to remove a given fraction of atoms from a structure without imposing a prescribed mapping, i.e.\ allowing each atom of the molecule to become a CG site.

The method has been tested on two case studies, namely Adenylate kinase and the adenine riboswitch. In the case of \ake, the dynamical consistency between the reference, integrated-out CG-hENM and the remapped coarse-grained ENM, as quantified by the RWSIP, turned out to be quite high with the traditional choices of mapping, namely selecting only C$_\alpha$ or C$_\beta$ atoms as CG sites, and only slightly lower when selecting the CG site subset based on an optimality criterion. When looking at the atom partition induced by the selection of CG sites, on the contrary, the optimised model proved more suited to represent the structure in terms of quasi-rigid groups of atoms, with small internal fluctuation and larger inter-domain dynamics. From the structural point of view, the optimal model was characterised by a higher degree of non-uniformity with respect to the conventional CG models, a property which can be expected to underlie the improved fraction of intra-block dynamics.

A qualitatively identical behaviour could be seen in the case of the adenine riboswitch, however with varying absolute numbers. The one-atom-type models showed a rather good dynamical consistency between the exactly integrated CG-hENM and the remapped one, while the optimised model featured a lower RWSIP; in both cases, the values were lower than the case of \ake, with a wider gap between conventional and optimised model. As for the fraction of intra-domain dynamics, the optimised model performed better than the others also in this case, in spite of generally larger amounts of the system's fluctuation within the blocks.

The construction of a simple, efficient, yet accurate coarse-grained representation of a macromolecule is a difficult task, whose intricacy does not only lie in the correct parametrisation of the interaction potentials. In fact, two crucial aspects have to be taken into account, namely the identification of the most appropriate interaction centres and the intrinsic viability of the coarse-graining procedure. As for the first aspect, the appropriateness of one choice of mapping over another largely depends on the {\it desiderata} of the model: these are the characteristics it is expected to entail and the physical properties it should reproduce. Indeed, a biased selection of the CG sites can produce a model which is optimal with respect to the quantity employed as a bias (the AER in this case), but whose performance is better or worse than average depending on the observable used to assess it. This behaviour is inherent in the process of optimisation, in that the search for the model that is optimal {\it in terms of a given property} necessarily drives the solution away from the optimality {\it in terms of other, orthogonal properties}. This situation is reminiscent of the coarse-grained modelling of liquids with approaches such as iterative Boltzmann inversion \cite{Reith2003,Noid2013b,Potestio2014}, where a model parametrised to reproduce exactly a single feature (the radial distribution function) performs well on some properties (compressibility) and poorly on others (pressure, three body correlation functions).

The second aspect is related to the extent to which the system under examination {\it can} be coarse-grained. In general, a model featuring a sensible but quite arbitrary mapping and interaction forces derived by the multi-body potential of mean force will satisfy all expectations one can have from a coarse-grained representation, since the MB-PMF reproduces the desired Boltzmann distribution {\it by construction}. However, the typical impossibility in calculating the PMF and, more importantly, the need to project it onto an efficient and computable basis set pose severe restrictions on the effectiveness of this strategy \cite{Foley2015}. Consistently, the (counter)example of the adenine riboswitch showed that, in spite of the optimisation procedure providing results in line with the trends observed in the case of Adenylate kinase, the performance of the model {\it in absolute terms} was not comparably good. The coarse-graining algorithm ``did its best'' to obtain a model with the lowest AER value -- succeeding indeed, however the result was quantitatively poorer than for \ake\ in terms of RWSIP and IBDF.

Even the simplest coarse-grained model, such as an ENM, entails a great amount of information about the properties of a system: this information is not only extracted through the {\it application} of the model, i.e. its usage in a calculation or simulation. Rather, useful insight can emerge from the study of how given properties depend on the strategy employed to construct the model. The approach discussed in this work, in which an algorithmic procedure was presented to identify the ideal CG sites in a macromolecule based on an optimality criterion, represents a first step in this direction.

\begin{acknowledgements}
We thank Gianluca Lattanzi and Pietro Faccioli for a critical reading of the manuscript and insightful discussion. MD acknowledges support by the NSF under grants CHE \#1464926 and CHE \#1764257. Simulations have been performed on Marconi at the CINECA supercomputing facility, under project INF18\_biophys. This project has received funding from the European Research Council (ERC) under the European Union's Horizon 2020 research and innovation programme (grant agreement No 758588).
\end{acknowledgements}

\section{Supporting information}
The Supporting Information file contains detailed information on the following topics:
\begin{itemize}
\item Description of an algorithm to implement an efficient inversion of the hENM Hamiltonian matrix
\item Plots of the mean square fluctuation of the atoms in all-atom and CG hENMs
\end{itemize}

\bibliography{main_revtex}

\begin{thebibliography}{80}%
\makeatletter
\providecommand \@ifxundefined [1]{%
 \@ifx{#1\undefined}
}%
\providecommand \@ifnum [1]{%
 \ifnum #1\expandafter \@firstoftwo
 \else \expandafter \@secondoftwo
 \fi
}%
\providecommand \@ifx [1]{%
 \ifx #1\expandafter \@firstoftwo
 \else \expandafter \@secondoftwo
 \fi
}%
\providecommand \natexlab [1]{#1}%
\providecommand \enquote  [1]{``#1''}%
\providecommand \bibnamefont  [1]{#1}%
\providecommand \bibfnamefont [1]{#1}%
\providecommand \citenamefont [1]{#1}%
\providecommand \href@noop [0]{\@secondoftwo}%
\providecommand \href [0]{\begingroup \@sanitize@url \@href}%
\providecommand \@href[1]{\@@startlink{#1}\@@href}%
\providecommand \@@href[1]{\endgroup#1\@@endlink}%
\providecommand \@sanitize@url [0]{\catcode `\\12\catcode `\$12\catcode
  `\&12\catcode `\#12\catcode `\^12\catcode `\_12\catcode `\%12\relax}%
\providecommand \@@startlink[1]{}%
\providecommand \@@endlink[0]{}%
\providecommand \url  [0]{\begingroup\@sanitize@url \@url }%
\providecommand \@url [1]{\endgroup\@href {#1}{\urlprefix }}%
\providecommand \urlprefix  [0]{URL }%
\providecommand \Eprint [0]{\href }%
\providecommand \doibase [0]{http://dx.doi.org/}%
\providecommand \selectlanguage [0]{\@gobble}%
\providecommand \bibinfo  [0]{\@secondoftwo}%
\providecommand \bibfield  [0]{\@secondoftwo}%
\providecommand \translation [1]{[#1]}%
\providecommand \BibitemOpen [0]{}%
\providecommand \bibitemStop [0]{}%
\providecommand \bibitemNoStop [0]{.\EOS\space}%
\providecommand \EOS [0]{\spacefactor3000\relax}%
\providecommand \BibitemShut  [1]{\csname bibitem#1\endcsname}%
\let\auto@bib@innerbib\@empty
\bibitem [{\citenamefont {Alder}\ and\ \citenamefont
  {Wainwright}(1957)}]{alder1957}%
  \BibitemOpen
  \bibfield  {author} {\bibinfo {author} {\bibfnamefont {B.}~\bibnamefont
  {Alder}}\ and\ \bibinfo {author} {\bibfnamefont {T.}~\bibnamefont
  {Wainwright}},\ }\href@noop {} {\bibfield  {journal} {\bibinfo  {journal} {J.
  Chem. Phys.}\ }\textbf {\bibinfo {volume} {5}} (\bibinfo {year}
  {1957})}\BibitemShut {NoStop}%
\bibitem [{\citenamefont {Levitt}\ and\ \citenamefont
  {Warshel}(1975)}]{Levitt1975}%
  \BibitemOpen
  \bibfield  {author} {\bibinfo {author} {\bibfnamefont {M.}~\bibnamefont
  {Levitt}}\ and\ \bibinfo {author} {\bibfnamefont {A.}~\bibnamefont
  {Warshel}},\ }\href@noop {} {\bibfield  {journal} {\bibinfo  {journal}
  {Nature}\ }\textbf {\bibinfo {volume} {253}},\ \bibinfo {pages} {694}
  (\bibinfo {year} {1975})}\BibitemShut {NoStop}%
\bibitem [{\citenamefont {Shaw}\ \emph {et~al.}(2009)\citenamefont {Shaw},
  \citenamefont {Dror}, \citenamefont {Salmon}, \citenamefont {Grossman},
  \citenamefont {Mackenzie}, \citenamefont {Bank}, \citenamefont {Young},
  \citenamefont {Deneroff}, \citenamefont {Batson}, \citenamefont {Bowers},
  \citenamefont {Chow}, \citenamefont {Eastwood}, \citenamefont {Ierardi},
  \citenamefont {Klepeis}, \citenamefont {Kuskin}, \citenamefont {Larson},
  \citenamefont {Lindorff-Larsen}, \citenamefont {Maragakis}, \citenamefont
  {Moraes}, \citenamefont {Piana}, \citenamefont {Shan},\ and\ \citenamefont
  {Towles}}]{Shaw2009}%
  \BibitemOpen
  \bibfield  {author} {\bibinfo {author} {\bibfnamefont {D.~E.}\ \bibnamefont
  {Shaw}}, \bibinfo {author} {\bibfnamefont {R.~O.}\ \bibnamefont {Dror}},
  \bibinfo {author} {\bibfnamefont {J.~K.}\ \bibnamefont {Salmon}}, \bibinfo
  {author} {\bibfnamefont {J.~P.}\ \bibnamefont {Grossman}}, \bibinfo {author}
  {\bibfnamefont {K.~M.}\ \bibnamefont {Mackenzie}}, \bibinfo {author}
  {\bibfnamefont {J.~A.}\ \bibnamefont {Bank}}, \bibinfo {author}
  {\bibfnamefont {C.}~\bibnamefont {Young}}, \bibinfo {author} {\bibfnamefont
  {M.~M.}\ \bibnamefont {Deneroff}}, \bibinfo {author} {\bibfnamefont
  {B.}~\bibnamefont {Batson}}, \bibinfo {author} {\bibfnamefont {K.~J.}\
  \bibnamefont {Bowers}}, \bibinfo {author} {\bibfnamefont {E.}~\bibnamefont
  {Chow}}, \bibinfo {author} {\bibfnamefont {M.~P.}\ \bibnamefont {Eastwood}},
  \bibinfo {author} {\bibfnamefont {D.~J.}\ \bibnamefont {Ierardi}}, \bibinfo
  {author} {\bibfnamefont {J.~L.}\ \bibnamefont {Klepeis}}, \bibinfo {author}
  {\bibfnamefont {J.~S.}\ \bibnamefont {Kuskin}}, \bibinfo {author}
  {\bibfnamefont {R.~H.}\ \bibnamefont {Larson}}, \bibinfo {author}
  {\bibfnamefont {K.}~\bibnamefont {Lindorff-Larsen}}, \bibinfo {author}
  {\bibfnamefont {P.}~\bibnamefont {Maragakis}}, \bibinfo {author}
  {\bibfnamefont {M.~A.}\ \bibnamefont {Moraes}}, \bibinfo {author}
  {\bibfnamefont {S.}~\bibnamefont {Piana}}, \bibinfo {author} {\bibfnamefont
  {Y.}~\bibnamefont {Shan}}, \ and\ \bibinfo {author} {\bibfnamefont
  {B.}~\bibnamefont {Towles}},\ }in\ \href@noop {} {\emph {\bibinfo {booktitle}
  {Proceedings of the Conference on High Performance Computing Networking,
  Storage and Analysis}}},\ \bibinfo {series and number} {SC '09}\ (\bibinfo
  {publisher} {ACM},\ \bibinfo {address} {New York, NY, USA},\ \bibinfo {year}
  {2009})\ pp.\ \bibinfo {pages} {65:1--65:11}\BibitemShut {NoStop}%
\bibitem [{\citenamefont {Piana}\ \emph {et~al.}(2012)\citenamefont {Piana},
  \citenamefont {Lindorff-Larsen},\ and\ \citenamefont {Shaw}}]{Piana2012}%
  \BibitemOpen
  \bibfield  {author} {\bibinfo {author} {\bibfnamefont {S.}~\bibnamefont
  {Piana}}, \bibinfo {author} {\bibfnamefont {K.}~\bibnamefont
  {Lindorff-Larsen}}, \ and\ \bibinfo {author} {\bibfnamefont {D.~E.}\
  \bibnamefont {Shaw}},\ }\href@noop {} {\bibfield  {journal} {\bibinfo
  {journal} {Proceedings of the National Academy of Sciences}\ }\textbf
  {\bibinfo {volume} {109}},\ \bibinfo {pages} {17845} (\bibinfo {year}
  {2012})}\BibitemShut {NoStop}%
\bibitem [{\citenamefont {Piana}\ \emph {et~al.}(2013)\citenamefont {Piana},
  \citenamefont {Lindorff-Larsen},\ and\ \citenamefont {Shaw}}]{Piana2013}%
  \BibitemOpen
  \bibfield  {author} {\bibinfo {author} {\bibfnamefont {S.}~\bibnamefont
  {Piana}}, \bibinfo {author} {\bibfnamefont {K.}~\bibnamefont
  {Lindorff-Larsen}}, \ and\ \bibinfo {author} {\bibfnamefont {D.~E.}\
  \bibnamefont {Shaw}},\ }\href@noop {} {\bibfield  {journal} {\bibinfo
  {journal} {Proceedings of the National Academy of Sciences}\ }\textbf
  {\bibinfo {volume} {110}},\ \bibinfo {pages} {5915} (\bibinfo {year}
  {2013})}\BibitemShut {NoStop}%
\bibitem [{\citenamefont {Freddolino}\ \emph {et~al.}(2006)\citenamefont
  {Freddolino}, \citenamefont {Arkhipov}, \citenamefont {Larson}, \citenamefont
  {McPherson},\ and\ \citenamefont {Schulten}}]{Freddolino2006}%
  \BibitemOpen
  \bibfield  {author} {\bibinfo {author} {\bibfnamefont {P.~L.}\ \bibnamefont
  {Freddolino}}, \bibinfo {author} {\bibfnamefont {A.~S.}\ \bibnamefont
  {Arkhipov}}, \bibinfo {author} {\bibfnamefont {S.~B.}\ \bibnamefont
  {Larson}}, \bibinfo {author} {\bibfnamefont {A.}~\bibnamefont {McPherson}}, \
  and\ \bibinfo {author} {\bibfnamefont {K.}~\bibnamefont {Schulten}},\
  }\href@noop {} {\bibfield  {journal} {\bibinfo  {journal} {Structure}\
  }\textbf {\bibinfo {volume} {14}},\ \bibinfo {pages} {437 } (\bibinfo {year}
  {2006})}\BibitemShut {NoStop}%
\bibitem [{\citenamefont {Bock}\ \emph {et~al.}(2013)\citenamefont {Bock},
  \citenamefont {Blau}, \citenamefont {Schr{\"o}der}, \citenamefont {Davydov},
  \citenamefont {Fischer}, \citenamefont {Stark}, \citenamefont {Rodnina},
  \citenamefont {Vaiana},\ and\ \citenamefont {Grubm{\"u}ller}}]{Bock2013}%
  \BibitemOpen
  \bibfield  {author} {\bibinfo {author} {\bibfnamefont {L.~V.}\ \bibnamefont
  {Bock}}, \bibinfo {author} {\bibfnamefont {C.}~\bibnamefont {Blau}}, \bibinfo
  {author} {\bibfnamefont {G.~F.}\ \bibnamefont {Schr{\"o}der}}, \bibinfo
  {author} {\bibfnamefont {I.~I.}\ \bibnamefont {Davydov}}, \bibinfo {author}
  {\bibfnamefont {N.}~\bibnamefont {Fischer}}, \bibinfo {author} {\bibfnamefont
  {H.}~\bibnamefont {Stark}}, \bibinfo {author} {\bibfnamefont {M.~V.}\
  \bibnamefont {Rodnina}}, \bibinfo {author} {\bibfnamefont {A.~C.}\
  \bibnamefont {Vaiana}}, \ and\ \bibinfo {author} {\bibfnamefont
  {H.}~\bibnamefont {Grubm{\"u}ller}},\ }\href@noop {} {\bibfield  {journal}
  {\bibinfo  {journal} {Nat Struct Mol Biol}\ }\textbf {\bibinfo {volume}
  {20}},\ \bibinfo {pages} {1390 } (\bibinfo {year} {2013})}\BibitemShut
  {NoStop}%
\bibitem [{\citenamefont {Borges}(2002)}]{borges2002exactitude}%
  \BibitemOpen
  \bibfield  {author} {\bibinfo {author} {\bibfnamefont {J.~L.}\ \bibnamefont
  {Borges}},\ }\href@noop {} {\bibfield  {journal} {\bibinfo  {journal}
  {Quaderns--Barcelona Collegi d'Arquitectes de Catalunya}\ ,\ \bibinfo {pages}
  {12}} (\bibinfo {year} {2002})}\BibitemShut {NoStop}%
\bibitem [{\citenamefont {Takada}(2012)}]{Takada2012}%
  \BibitemOpen
  \bibfield  {author} {\bibinfo {author} {\bibfnamefont {S.}~\bibnamefont
  {Takada}},\ }\href@noop {} {\bibfield  {journal} {\bibinfo  {journal} {Curr.
  Opin. Struct. Biol.}\ }\textbf {\bibinfo {volume} {22}},\ \bibinfo {pages}
  {130} (\bibinfo {year} {2012})}\BibitemShut {NoStop}%
\bibitem [{\citenamefont {Noid}(2013{\natexlab{a}})}]{Noid2013}%
  \BibitemOpen
  \bibfield  {author} {\bibinfo {author} {\bibfnamefont {W.~G.}\ \bibnamefont
  {Noid}},\ }\href@noop {} {\bibfield  {journal} {\bibinfo  {journal} {J. Chem.
  Phys.}\ }\textbf {\bibinfo {volume} {139}},\ \bibinfo {pages} {090901}
  (\bibinfo {year} {2013}{\natexlab{a}})}\BibitemShut {NoStop}%
\bibitem [{\citenamefont {Saunders}\ and\ \citenamefont
  {Voth}(2013)}]{Saunders2013}%
  \BibitemOpen
  \bibfield  {author} {\bibinfo {author} {\bibfnamefont {M.~G.}\ \bibnamefont
  {Saunders}}\ and\ \bibinfo {author} {\bibfnamefont {G.~A.}\ \bibnamefont
  {Voth}},\ }\href@noop {} {\bibfield  {journal} {\bibinfo  {journal} {Annu.
  Rev. Biophys.}\ }\textbf {\bibinfo {volume} {42}},\ \bibinfo {pages} {73}
  (\bibinfo {year} {2013})}\BibitemShut {NoStop}%
\bibitem [{\citenamefont {Potestio}\ \emph {et~al.}(2014)\citenamefont
  {Potestio}, \citenamefont {Peter},\ and\ \citenamefont
  {Kremer}}]{Potestio2014}%
  \BibitemOpen
  \bibfield  {author} {\bibinfo {author} {\bibfnamefont {R.}~\bibnamefont
  {Potestio}}, \bibinfo {author} {\bibfnamefont {C.}~\bibnamefont {Peter}}, \
  and\ \bibinfo {author} {\bibfnamefont {K.}~\bibnamefont {Kremer}},\
  }\href@noop {} {\bibfield  {journal} {\bibinfo  {journal} {Entropy}\ }\textbf
  {\bibinfo {volume} {16}},\ \bibinfo {pages} {4199} (\bibinfo {year}
  {2014})}\BibitemShut {NoStop}%
\bibitem [{\citenamefont {Ueda}\ \emph {et~al.}(1978)\citenamefont {Ueda},
  \citenamefont {Taketomi},\ and\ \citenamefont {Gō}}]{Ueda1978}%
  \BibitemOpen
  \bibfield  {author} {\bibinfo {author} {\bibfnamefont {Y.}~\bibnamefont
  {Ueda}}, \bibinfo {author} {\bibfnamefont {H.}~\bibnamefont {Taketomi}}, \
  and\ \bibinfo {author} {\bibfnamefont {N.}~\bibnamefont {Gō}},\ }\href@noop
  {} {\bibfield  {journal} {\bibinfo  {journal} {Biopolymers}\ }\textbf
  {\bibinfo {volume} {17}},\ \bibinfo {pages} {1531} (\bibinfo {year}
  {1978})}\BibitemShut {NoStop}%
\bibitem [{\citenamefont {Go}\ and\ \citenamefont {Taketomi}(1978)}]{Go1978}%
  \BibitemOpen
  \bibfield  {author} {\bibinfo {author} {\bibfnamefont {N.}~\bibnamefont
  {Go}}\ and\ \bibinfo {author} {\bibfnamefont {H.}~\bibnamefont {Taketomi}},\
  }\href@noop {} {\bibfield  {journal} {\bibinfo  {journal} {Proceedings of the
  National Academy of Sciences}\ }\textbf {\bibinfo {volume} {75}},\ \bibinfo
  {pages} {559} (\bibinfo {year} {1978})}\BibitemShut {NoStop}%
\bibitem [{\citenamefont {Golhlke}\ and\ \citenamefont
  {Thorpe}(2006)}]{Gohlke2006}%
  \BibitemOpen
  \bibfield  {author} {\bibinfo {author} {\bibfnamefont {H.}~\bibnamefont
  {Golhlke}}\ and\ \bibinfo {author} {\bibfnamefont {M.~F.}\ \bibnamefont
  {Thorpe}},\ }\href@noop {} {\bibfield  {journal} {\bibinfo  {journal}
  {Biophysical Journal}\ }\textbf {\bibinfo {volume} {91}},\ \bibinfo {pages}
  {2115} (\bibinfo {year} {2006})}\BibitemShut {NoStop}%
\bibitem [{\citenamefont {Potestio}\ \emph {et~al.}(2009)\citenamefont
  {Potestio}, \citenamefont {Pontiggia},\ and\ \citenamefont
  {Micheletti}}]{Potestio2009}%
  \BibitemOpen
  \bibfield  {author} {\bibinfo {author} {\bibfnamefont {R.}~\bibnamefont
  {Potestio}}, \bibinfo {author} {\bibfnamefont {F.}~\bibnamefont {Pontiggia}},
  \ and\ \bibinfo {author} {\bibfnamefont {C.}~\bibnamefont {Micheletti}},\
  }\href@noop {} {\bibfield  {journal} {\bibinfo  {journal} {Biophys J}\
  }\textbf {\bibinfo {volume} {96}} (\bibinfo {year} {2009})}\BibitemShut
  {NoStop}%
\bibitem [{\citenamefont {Tozzini}(2010)}]{Tozzini2010}%
  \BibitemOpen
  \bibfield  {author} {\bibinfo {author} {\bibfnamefont {V.}~\bibnamefont
  {Tozzini}},\ }\href@noop {} {\bibfield  {journal} {\bibinfo  {journal} {Acc.
  Chem. Res.}\ }\textbf {\bibinfo {volume} {43}},\ \bibinfo {pages} {220}
  (\bibinfo {year} {2010})}\BibitemShut {NoStop}%
\bibitem [{\citenamefont {Polles}\ \emph {et~al.}(2013)\citenamefont {Polles},
  \citenamefont {Indelicato}, \citenamefont {Potestio}, \citenamefont
  {Cermelli}, \citenamefont {Twarock},\ and\ \citenamefont
  {Micheletti}}]{Polles2013}%
  \BibitemOpen
  \bibfield  {author} {\bibinfo {author} {\bibfnamefont {G.}~\bibnamefont
  {Polles}}, \bibinfo {author} {\bibfnamefont {G.}~\bibnamefont {Indelicato}},
  \bibinfo {author} {\bibfnamefont {R.}~\bibnamefont {Potestio}}, \bibinfo
  {author} {\bibfnamefont {P.}~\bibnamefont {Cermelli}}, \bibinfo {author}
  {\bibfnamefont {R.}~\bibnamefont {Twarock}}, \ and\ \bibinfo {author}
  {\bibfnamefont {C.}~\bibnamefont {Micheletti}},\ }\href@noop {} {\bibfield
  {journal} {\bibinfo  {journal} {PLOS Computational Biology}\ }\textbf
  {\bibinfo {volume} {9}},\ \bibinfo {pages} {1} (\bibinfo {year}
  {2013})}\BibitemShut {NoStop}%
\bibitem [{\citenamefont {Najafi}\ and\ \citenamefont
  {Potestio}(2015)}]{Najafi2015}%
  \BibitemOpen
  \bibfield  {author} {\bibinfo {author} {\bibfnamefont {S.}~\bibnamefont
  {Najafi}}\ and\ \bibinfo {author} {\bibfnamefont {R.}~\bibnamefont
  {Potestio}},\ }\href@noop {} {\bibfield  {journal} {\bibinfo  {journal} {The
  Journal of Chemical Physics}\ }\textbf {\bibinfo {volume} {143}},\ \bibinfo
  {pages} {243121} (\bibinfo {year} {2015})}\BibitemShut {NoStop}%
\bibitem [{\citenamefont {Gibbons}\ and\ \citenamefont
  {Klug}(2007)}]{Gibbons2007}%
  \BibitemOpen
  \bibfield  {author} {\bibinfo {author} {\bibfnamefont {M.~M.}\ \bibnamefont
  {Gibbons}}\ and\ \bibinfo {author} {\bibfnamefont {W.~S.}\ \bibnamefont
  {Klug}},\ }\href@noop {} {\bibfield  {journal} {\bibinfo  {journal} {Journal
  of Materials Science}\ }\textbf {\bibinfo {volume} {42}},\ \bibinfo {pages}
  {8995} (\bibinfo {year} {2007})}\BibitemShut {NoStop}%
\bibitem [{\citenamefont {Gibbons}\ and\ \citenamefont
  {Klug}(2008)}]{Gibbons2008}%
  \BibitemOpen
  \bibfield  {author} {\bibinfo {author} {\bibfnamefont {M.~M.}\ \bibnamefont
  {Gibbons}}\ and\ \bibinfo {author} {\bibfnamefont {W.~S.}\ \bibnamefont
  {Klug}},\ }\href@noop {} {\bibfield  {journal} {\bibinfo  {journal}
  {Biophysical Journal}\ }\textbf {\bibinfo {volume} {95}},\ \bibinfo {pages}
  {3640 } (\bibinfo {year} {2008})}\BibitemShut {NoStop}%
\bibitem [{\citenamefont {Roos}\ \emph {et~al.}(2010)\citenamefont {Roos},
  \citenamefont {Gibbons}, \citenamefont {Arkhipov}, \citenamefont {Uetrecht},
  \citenamefont {Watts}, \citenamefont {Wingfield}, \citenamefont {Steven},
  \citenamefont {Heck}, \citenamefont {Schulten}, \citenamefont {Klug},\ and\
  \citenamefont {Wuite}}]{Roos2010}%
  \BibitemOpen
  \bibfield  {author} {\bibinfo {author} {\bibfnamefont {W.}~\bibnamefont
  {Roos}}, \bibinfo {author} {\bibfnamefont {M.}~\bibnamefont {Gibbons}},
  \bibinfo {author} {\bibfnamefont {A.}~\bibnamefont {Arkhipov}}, \bibinfo
  {author} {\bibfnamefont {C.}~\bibnamefont {Uetrecht}}, \bibinfo {author}
  {\bibfnamefont {N.}~\bibnamefont {Watts}}, \bibinfo {author} {\bibfnamefont
  {P.}~\bibnamefont {Wingfield}}, \bibinfo {author} {\bibfnamefont
  {A.}~\bibnamefont {Steven}}, \bibinfo {author} {\bibfnamefont
  {A.}~\bibnamefont {Heck}}, \bibinfo {author} {\bibfnamefont {K.}~\bibnamefont
  {Schulten}}, \bibinfo {author} {\bibfnamefont {W.}~\bibnamefont {Klug}}, \
  and\ \bibinfo {author} {\bibfnamefont {G.}~\bibnamefont {Wuite}},\
  }\href@noop {} {\bibfield  {journal} {\bibinfo  {journal} {Biophysical
  Journal}\ }\textbf {\bibinfo {volume} {99}},\ \bibinfo {pages} {1175 }
  (\bibinfo {year} {2010})}\BibitemShut {NoStop}%
\bibitem [{\citenamefont {Aggarwal}\ \emph {et~al.}(2016)\citenamefont
  {Aggarwal}, \citenamefont {May}, \citenamefont {Brooks},\ and\ \citenamefont
  {Klug}}]{Aggarwal2016}%
  \BibitemOpen
  \bibfield  {author} {\bibinfo {author} {\bibfnamefont {A.}~\bibnamefont
  {Aggarwal}}, \bibinfo {author} {\bibfnamefont {E.~R.}\ \bibnamefont {May}},
  \bibinfo {author} {\bibfnamefont {C.~L.}\ \bibnamefont {Brooks}}, \ and\
  \bibinfo {author} {\bibfnamefont {W.~S.}\ \bibnamefont {Klug}},\ }\href@noop
  {} {\bibfield  {journal} {\bibinfo  {journal} {Phys. Rev. E}\ }\textbf
  {\bibinfo {volume} {93}},\ \bibinfo {pages} {012417} (\bibinfo {year}
  {2016})}\BibitemShut {NoStop}%
\bibitem [{\citenamefont {Marrink}\ \emph {et~al.}(2004)\citenamefont
  {Marrink}, \citenamefont {De~Vries},\ and\ \citenamefont
  {Mark}}]{marrink2004coarse}%
  \BibitemOpen
  \bibfield  {author} {\bibinfo {author} {\bibfnamefont {S.~J.}\ \bibnamefont
  {Marrink}}, \bibinfo {author} {\bibfnamefont {A.~H.}\ \bibnamefont
  {De~Vries}}, \ and\ \bibinfo {author} {\bibfnamefont {A.~E.}\ \bibnamefont
  {Mark}},\ }\href@noop {} {\bibfield  {journal} {\bibinfo  {journal} {The
  Journal of Physical Chemistry B}\ }\textbf {\bibinfo {volume} {108}},\
  \bibinfo {pages} {750} (\bibinfo {year} {2004})}\BibitemShut {NoStop}%
\bibitem [{\citenamefont {Izvekov}\ and\ \citenamefont
  {Voth}(2005)}]{izvekov2005multiscale}%
  \BibitemOpen
  \bibfield  {author} {\bibinfo {author} {\bibfnamefont {S.}~\bibnamefont
  {Izvekov}}\ and\ \bibinfo {author} {\bibfnamefont {G.~A.}\ \bibnamefont
  {Voth}},\ }\href@noop {} {\bibfield  {journal} {\bibinfo  {journal} {The
  Journal of Physical Chemistry B}\ }\textbf {\bibinfo {volume} {109}},\
  \bibinfo {pages} {2469} (\bibinfo {year} {2005})}\BibitemShut {NoStop}%
\bibitem [{\citenamefont {Venturoli}\ \emph {et~al.}(2006)\citenamefont
  {Venturoli}, \citenamefont {Sperotto}, \citenamefont {Kranenburg},\ and\
  \citenamefont {Smit}}]{venturoli2006mesoscopic}%
  \BibitemOpen
  \bibfield  {author} {\bibinfo {author} {\bibfnamefont {M.}~\bibnamefont
  {Venturoli}}, \bibinfo {author} {\bibfnamefont {M.~M.}\ \bibnamefont
  {Sperotto}}, \bibinfo {author} {\bibfnamefont {M.}~\bibnamefont
  {Kranenburg}}, \ and\ \bibinfo {author} {\bibfnamefont {B.}~\bibnamefont
  {Smit}},\ }\href@noop {} {\bibfield  {journal} {\bibinfo  {journal} {Physics
  Reports}\ }\textbf {\bibinfo {volume} {437}},\ \bibinfo {pages} {1} (\bibinfo
  {year} {2006})}\BibitemShut {NoStop}%
\bibitem [{\citenamefont {M{\"u}ller}\ \emph {et~al.}(2006)\citenamefont
  {M{\"u}ller}, \citenamefont {Katsov},\ and\ \citenamefont
  {Schick}}]{muller2006biological}%
  \BibitemOpen
  \bibfield  {author} {\bibinfo {author} {\bibfnamefont {M.}~\bibnamefont
  {M{\"u}ller}}, \bibinfo {author} {\bibfnamefont {K.}~\bibnamefont {Katsov}},
  \ and\ \bibinfo {author} {\bibfnamefont {M.}~\bibnamefont {Schick}},\
  }\href@noop {} {\bibfield  {journal} {\bibinfo  {journal} {Physics Reports}\
  }\textbf {\bibinfo {volume} {434}},\ \bibinfo {pages} {113} (\bibinfo {year}
  {2006})}\BibitemShut {NoStop}%
\bibitem [{\citenamefont {Murtola}\ \emph {et~al.}(2009)\citenamefont
  {Murtola}, \citenamefont {Bunker}, \citenamefont {Vattulainen}, \citenamefont
  {Deserno},\ and\ \citenamefont {Karttunen}}]{murtola2009multiscale}%
  \BibitemOpen
  \bibfield  {author} {\bibinfo {author} {\bibfnamefont {T.}~\bibnamefont
  {Murtola}}, \bibinfo {author} {\bibfnamefont {A.}~\bibnamefont {Bunker}},
  \bibinfo {author} {\bibfnamefont {I.}~\bibnamefont {Vattulainen}}, \bibinfo
  {author} {\bibfnamefont {M.}~\bibnamefont {Deserno}}, \ and\ \bibinfo
  {author} {\bibfnamefont {M.}~\bibnamefont {Karttunen}},\ }\href@noop {}
  {\bibfield  {journal} {\bibinfo  {journal} {Physical Chemistry Chemical
  Physics}\ }\textbf {\bibinfo {volume} {11}},\ \bibinfo {pages} {1869}
  (\bibinfo {year} {2009})}\BibitemShut {NoStop}%
\bibitem [{\citenamefont {Deserno}(2009)}]{Deserno2009}%
  \BibitemOpen
  \bibfield  {author} {\bibinfo {author} {\bibfnamefont {M.}~\bibnamefont
  {Deserno}},\ }\href@noop {} {\bibfield  {journal} {\bibinfo  {journal}
  {Macromol. Rapid Commun.}\ }\textbf {\bibinfo {volume} {30}},\ \bibinfo
  {pages} {752} (\bibinfo {year} {2009})}\BibitemShut {NoStop}%
\bibitem [{\citenamefont {Alber}\ \emph {et~al.}(2003)\citenamefont {Alber},
  \citenamefont {Kiskowski}, \citenamefont {Glazier},\ and\ \citenamefont
  {Jiang}}]{alber2003cellular}%
  \BibitemOpen
  \bibfield  {author} {\bibinfo {author} {\bibfnamefont {M.~S.}\ \bibnamefont
  {Alber}}, \bibinfo {author} {\bibfnamefont {M.~A.}\ \bibnamefont
  {Kiskowski}}, \bibinfo {author} {\bibfnamefont {J.~A.}\ \bibnamefont
  {Glazier}}, \ and\ \bibinfo {author} {\bibfnamefont {Y.}~\bibnamefont
  {Jiang}},\ }in\ \href@noop {} {\emph {\bibinfo {booktitle} {Mathematical
  Systems Theory in Biology, Communications, Computation, and Finance}}}\
  (\bibinfo  {publisher} {Springer},\ \bibinfo {year} {2003})\ pp.\ \bibinfo
  {pages} {1--39}\BibitemShut {NoStop}%
\bibitem [{\citenamefont {Izaguirre}\ \emph {et~al.}(2004)\citenamefont
  {Izaguirre}, \citenamefont {Chaturvedi}, \citenamefont {Huang}, \citenamefont
  {Cickovski}, \citenamefont {Coffland}, \citenamefont {Thomas}, \citenamefont
  {Forgacs}, \citenamefont {Alber}, \citenamefont {Hentschel},\ and\
  \citenamefont {Newman}}]{izaguirre2004compucell}%
  \BibitemOpen
  \bibfield  {author} {\bibinfo {author} {\bibfnamefont {J.~A.}\ \bibnamefont
  {Izaguirre}}, \bibinfo {author} {\bibfnamefont {R.}~\bibnamefont
  {Chaturvedi}}, \bibinfo {author} {\bibfnamefont {C.}~\bibnamefont {Huang}},
  \bibinfo {author} {\bibfnamefont {T.}~\bibnamefont {Cickovski}}, \bibinfo
  {author} {\bibfnamefont {J.}~\bibnamefont {Coffland}}, \bibinfo {author}
  {\bibfnamefont {G.}~\bibnamefont {Thomas}}, \bibinfo {author} {\bibfnamefont
  {G.}~\bibnamefont {Forgacs}}, \bibinfo {author} {\bibfnamefont
  {M.}~\bibnamefont {Alber}}, \bibinfo {author} {\bibfnamefont
  {G.}~\bibnamefont {Hentschel}}, \ and\ \bibinfo {author} {\bibfnamefont
  {S.~A.}\ \bibnamefont {Newman}},\ }\href@noop {} {\bibfield  {journal}
  {\bibinfo  {journal} {Bioinformatics}\ }\textbf {\bibinfo {volume} {20}},\
  \bibinfo {pages} {1129} (\bibinfo {year} {2004})}\BibitemShut {NoStop}%
\bibitem [{\citenamefont {Shirinifard}\ \emph {et~al.}(2009)\citenamefont
  {Shirinifard}, \citenamefont {Gens}, \citenamefont {Zaitlen}, \citenamefont
  {Pop{\l}awski}, \citenamefont {Swat},\ and\ \citenamefont
  {Glazier}}]{shirinifard20093d}%
  \BibitemOpen
  \bibfield  {author} {\bibinfo {author} {\bibfnamefont {A.}~\bibnamefont
  {Shirinifard}}, \bibinfo {author} {\bibfnamefont {J.~S.}\ \bibnamefont
  {Gens}}, \bibinfo {author} {\bibfnamefont {B.~L.}\ \bibnamefont {Zaitlen}},
  \bibinfo {author} {\bibfnamefont {N.~J.}\ \bibnamefont {Pop{\l}awski}},
  \bibinfo {author} {\bibfnamefont {M.}~\bibnamefont {Swat}}, \ and\ \bibinfo
  {author} {\bibfnamefont {J.~A.}\ \bibnamefont {Glazier}},\ }\href@noop {}
  {\bibfield  {journal} {\bibinfo  {journal} {PloS one}\ }\textbf {\bibinfo
  {volume} {4}},\ \bibinfo {pages} {e7190} (\bibinfo {year}
  {2009})}\BibitemShut {NoStop}%
\bibitem [{\citenamefont {Tirion}(1996)}]{Tirion1996}%
  \BibitemOpen
  \bibfield  {author} {\bibinfo {author} {\bibfnamefont {M.~M.}\ \bibnamefont
  {Tirion}},\ }\href@noop {} {\bibfield  {journal} {\bibinfo  {journal} {Phys.
  Rev. Lett.}\ }\textbf {\bibinfo {volume} {77}},\ \bibinfo {pages} {1905}
  (\bibinfo {year} {1996})}\BibitemShut {NoStop}%
\bibitem [{\citenamefont {Bahar}\ \emph {et~al.}(1997)\citenamefont {Bahar},
  \citenamefont {Atilgan},\ and\ \citenamefont {Erman}}]{Bahar1997}%
  \BibitemOpen
  \bibfield  {author} {\bibinfo {author} {\bibfnamefont {I.}~\bibnamefont
  {Bahar}}, \bibinfo {author} {\bibfnamefont {A.~R.}\ \bibnamefont {Atilgan}},
  \ and\ \bibinfo {author} {\bibfnamefont {B.}~\bibnamefont {Erman}},\
  }\href@noop {} {\bibfield  {journal} {\bibinfo  {journal} {Fold. Des.}\
  }\textbf {\bibinfo {volume} {2}},\ \bibinfo {pages} {173} (\bibinfo {year}
  {1997})}\BibitemShut {NoStop}%
\bibitem [{\citenamefont {Hinsen}(1998)}]{Hinsen1998}%
  \BibitemOpen
  \bibfield  {author} {\bibinfo {author} {\bibfnamefont {K.}~\bibnamefont
  {Hinsen}},\ }\href@noop {} {\bibfield  {journal} {\bibinfo  {journal}
  {Proteins}\ }\textbf {\bibinfo {volume} {33}},\ \bibinfo {pages} {417}
  (\bibinfo {year} {1998})}\BibitemShut {NoStop}%
\bibitem [{\citenamefont {Atilgan}\ \emph {et~al.}(2001)\citenamefont
  {Atilgan}, \citenamefont {Durell}, \citenamefont {Jernigan}, \citenamefont
  {Demirel}, \citenamefont {Keskin},\ and\ \citenamefont
  {Bahar}}]{Atilgan2001}%
  \BibitemOpen
  \bibfield  {author} {\bibinfo {author} {\bibfnamefont {A.~R.}\ \bibnamefont
  {Atilgan}}, \bibinfo {author} {\bibfnamefont {S.~R.}\ \bibnamefont {Durell}},
  \bibinfo {author} {\bibfnamefont {R.~L.}\ \bibnamefont {Jernigan}}, \bibinfo
  {author} {\bibfnamefont {M.~C.}\ \bibnamefont {Demirel}}, \bibinfo {author}
  {\bibfnamefont {O.}~\bibnamefont {Keskin}}, \ and\ \bibinfo {author}
  {\bibfnamefont {I.}~\bibnamefont {Bahar}},\ }\href@noop {} {\bibfield
  {journal} {\bibinfo  {journal} {Biophys. J.}\ }\textbf {\bibinfo {volume}
  {80}},\ \bibinfo {pages} {505} (\bibinfo {year} {2001})}\BibitemShut
  {NoStop}%
\bibitem [{\citenamefont {Delarue}\ and\ \citenamefont
  {Sanejouand}(2002)}]{Delarue2002}%
  \BibitemOpen
  \bibfield  {author} {\bibinfo {author} {\bibfnamefont {M.}~\bibnamefont
  {Delarue}}\ and\ \bibinfo {author} {\bibfnamefont {Y.~H.}\ \bibnamefont
  {Sanejouand}},\ }\href@noop {} {\bibfield  {journal} {\bibinfo  {journal} {J
  Mol Biol}\ }\textbf {\bibinfo {volume} {320}},\ \bibinfo {pages} {1011}
  (\bibinfo {year} {2002})}\BibitemShut {NoStop}%
\bibitem [{\citenamefont {Micheletti}\ \emph {et~al.}(2004)\citenamefont
  {Micheletti}, \citenamefont {Carloni},\ and\ \citenamefont
  {Maritan}}]{Micheletti2004}%
  \BibitemOpen
  \bibfield  {author} {\bibinfo {author} {\bibfnamefont {C.}~\bibnamefont
  {Micheletti}}, \bibinfo {author} {\bibfnamefont {P.}~\bibnamefont {Carloni}},
  \ and\ \bibinfo {author} {\bibfnamefont {A.}~\bibnamefont {Maritan}},\
  }\href@noop {} {\bibfield  {journal} {\bibinfo  {journal} {Proteins}\
  }\textbf {\bibinfo {volume} {55}},\ \bibinfo {pages} {635} (\bibinfo {year}
  {2004})}\BibitemShut {NoStop}%
\bibitem [{\citenamefont {Amadei}\ \emph {et~al.}(1993)\citenamefont {Amadei},
  \citenamefont {Linssen},\ and\ \citenamefont {Berendsen}}]{Amadei1993}%
  \BibitemOpen
  \bibfield  {author} {\bibinfo {author} {\bibfnamefont {A.}~\bibnamefont
  {Amadei}}, \bibinfo {author} {\bibfnamefont {A.~B.~M.}\ \bibnamefont
  {Linssen}}, \ and\ \bibinfo {author} {\bibfnamefont {H.~J.~C.}\ \bibnamefont
  {Berendsen}},\ }\href@noop {} {\bibfield  {journal} {\bibinfo  {journal}
  {Proteins: Structure, Function, and Bioinformatics}\ }\textbf {\bibinfo
  {volume} {17}},\ \bibinfo {pages} {412} (\bibinfo {year} {1993})}\BibitemShut
  {NoStop}%
\bibitem [{\citenamefont {Pontiggia}\ \emph {et~al.}(2007)\citenamefont
  {Pontiggia}, \citenamefont {Colombo}, \citenamefont {Micheletti},\ and\
  \citenamefont {Orland}}]{Pontiggia2007}%
  \BibitemOpen
  \bibfield  {author} {\bibinfo {author} {\bibfnamefont {F.}~\bibnamefont
  {Pontiggia}}, \bibinfo {author} {\bibfnamefont {G.}~\bibnamefont {Colombo}},
  \bibinfo {author} {\bibfnamefont {C.}~\bibnamefont {Micheletti}}, \ and\
  \bibinfo {author} {\bibfnamefont {H.}~\bibnamefont {Orland}},\ }\href@noop {}
  {\bibfield  {journal} {\bibinfo  {journal} {Phys Rev Lett}\ }\textbf
  {\bibinfo {volume} {98}},\ \bibinfo {pages} {048102} (\bibinfo {year}
  {2007})}\BibitemShut {NoStop}%
\bibitem [{\citenamefont {Hensen}\ \emph {et~al.}(2012)\citenamefont {Hensen},
  \citenamefont {Meyer}, \citenamefont {Haas}, \citenamefont {Rex},
  \citenamefont {Vriend},\ and\ \citenamefont {Grubmüller}}]{Hensen2012}%
  \BibitemOpen
  \bibfield  {author} {\bibinfo {author} {\bibfnamefont {U.}~\bibnamefont
  {Hensen}}, \bibinfo {author} {\bibfnamefont {T.}~\bibnamefont {Meyer}},
  \bibinfo {author} {\bibfnamefont {J.}~\bibnamefont {Haas}}, \bibinfo {author}
  {\bibfnamefont {R.}~\bibnamefont {Rex}}, \bibinfo {author} {\bibfnamefont
  {G.}~\bibnamefont {Vriend}}, \ and\ \bibinfo {author} {\bibfnamefont
  {H.}~\bibnamefont {Grubmüller}},\ }\href@noop {} {\bibfield  {journal}
  {\bibinfo  {journal} {PLOS ONE}\ }\textbf {\bibinfo {volume} {7}},\ \bibinfo
  {pages} {1} (\bibinfo {year} {2012})}\BibitemShut {NoStop}%
\bibitem [{\citenamefont {Nussinov}\ and\ \citenamefont
  {Wolynes}(2014)}]{Nussinov2014}%
  \BibitemOpen
  \bibfield  {author} {\bibinfo {author} {\bibfnamefont {R.}~\bibnamefont
  {Nussinov}}\ and\ \bibinfo {author} {\bibfnamefont {P.~G.}\ \bibnamefont
  {Wolynes}},\ }\href@noop {} {\bibfield  {journal} {\bibinfo  {journal} {Phys.
  Chem. Chem. Phys.}\ }\textbf {\bibinfo {volume} {16}},\ \bibinfo {pages}
  {6321} (\bibinfo {year} {2014})}\BibitemShut {NoStop}%
\bibitem [{\citenamefont {Wei}\ \emph {et~al.}(2016)\citenamefont {Wei},
  \citenamefont {Xi}, \citenamefont {Nussinov},\ and\ \citenamefont
  {Ma}}]{Wei2016}%
  \BibitemOpen
  \bibfield  {author} {\bibinfo {author} {\bibfnamefont {G.}~\bibnamefont
  {Wei}}, \bibinfo {author} {\bibfnamefont {W.}~\bibnamefont {Xi}}, \bibinfo
  {author} {\bibfnamefont {R.}~\bibnamefont {Nussinov}}, \ and\ \bibinfo
  {author} {\bibfnamefont {B.}~\bibnamefont {Ma}},\ }\href@noop {} {\bibfield
  {journal} {\bibinfo  {journal} {Chemical Reviews}\ }\textbf {\bibinfo
  {volume} {116}},\ \bibinfo {pages} {6516} (\bibinfo {year} {2016})},\
  \bibinfo {note} {pMID: 26807783}\BibitemShut {NoStop}%
\bibitem [{\citenamefont {Zhang}\ \emph {et~al.}(2008)\citenamefont {Zhang},
  \citenamefont {Lu}, \citenamefont {Noid}, \citenamefont {Krishna},
  \citenamefont {Pfaendtner},\ and\ \citenamefont {Voth}}]{ZHANG_BJ_2008}%
  \BibitemOpen
  \bibfield  {author} {\bibinfo {author} {\bibfnamefont {Z.}~\bibnamefont
  {Zhang}}, \bibinfo {author} {\bibfnamefont {L.}~\bibnamefont {Lu}}, \bibinfo
  {author} {\bibfnamefont {W.~G.}\ \bibnamefont {Noid}}, \bibinfo {author}
  {\bibfnamefont {V.}~\bibnamefont {Krishna}}, \bibinfo {author} {\bibfnamefont
  {J.}~\bibnamefont {Pfaendtner}}, \ and\ \bibinfo {author} {\bibfnamefont
  {G.~A.}\ \bibnamefont {Voth}},\ }\href@noop {} {\bibfield  {journal}
  {\bibinfo  {journal} {Biophysical Journal}\ }\textbf {\bibinfo {volume}
  {95}},\ \bibinfo {pages} {5073 } (\bibinfo {year} {2008})}\BibitemShut
  {NoStop}%
\bibitem [{\citenamefont {Zhang}\ \emph {et~al.}(2009)\citenamefont {Zhang},
  \citenamefont {Pfaendtner}, \citenamefont {Grafmüller},\ and\ \citenamefont
  {Voth}}]{ZHANG_BJ_2009}%
  \BibitemOpen
  \bibfield  {author} {\bibinfo {author} {\bibfnamefont {Z.}~\bibnamefont
  {Zhang}}, \bibinfo {author} {\bibfnamefont {J.}~\bibnamefont {Pfaendtner}},
  \bibinfo {author} {\bibfnamefont {A.}~\bibnamefont {Grafmüller}}, \ and\
  \bibinfo {author} {\bibfnamefont {G.~A.}\ \bibnamefont {Voth}},\ }\href@noop
  {} {\bibfield  {journal} {\bibinfo  {journal} {Biophysical Journal}\ }\textbf
  {\bibinfo {volume} {97}},\ \bibinfo {pages} {2327 } (\bibinfo {year}
  {2009})}\BibitemShut {NoStop}%
\bibitem [{\citenamefont {Zhang}\ and\ \citenamefont
  {Voth}(2010)}]{ZHANG_JCTC_2010}%
  \BibitemOpen
  \bibfield  {author} {\bibinfo {author} {\bibfnamefont {Z.}~\bibnamefont
  {Zhang}}\ and\ \bibinfo {author} {\bibfnamefont {G.~A.}\ \bibnamefont
  {Voth}},\ }\href@noop {} {\bibfield  {journal} {\bibinfo  {journal} {Journal
  of Chemical Theory and Computation}\ }\textbf {\bibinfo {volume} {6}},\
  \bibinfo {pages} {2990} (\bibinfo {year} {2010})}\BibitemShut {NoStop}%
\bibitem [{\citenamefont {Aleksiev}\ \emph {et~al.}(2009)\citenamefont
  {Aleksiev}, \citenamefont {Potestio}, \citenamefont {Pontiggia},
  \citenamefont {Cozzini},\ and\ \citenamefont {Micheletti}}]{aleksiev2009}%
  \BibitemOpen
  \bibfield  {author} {\bibinfo {author} {\bibfnamefont {T.}~\bibnamefont
  {Aleksiev}}, \bibinfo {author} {\bibfnamefont {R.}~\bibnamefont {Potestio}},
  \bibinfo {author} {\bibfnamefont {F.}~\bibnamefont {Pontiggia}}, \bibinfo
  {author} {\bibfnamefont {S.}~\bibnamefont {Cozzini}}, \ and\ \bibinfo
  {author} {\bibfnamefont {C.}~\bibnamefont {Micheletti}},\ }\href@noop {}
  {\bibfield  {journal} {\bibinfo  {journal} {Bioinformatics}\ }\textbf
  {\bibinfo {volume} {25}},\ \bibinfo {pages} {2743} (\bibinfo {year}
  {2009})}\BibitemShut {NoStop}%
\bibitem [{\citenamefont {Sinitskiy}\ \emph {et~al.}(2012)\citenamefont
  {Sinitskiy}, \citenamefont {Saunders},\ and\ \citenamefont
  {Voth}}]{Sinitskiy2012}%
  \BibitemOpen
  \bibfield  {author} {\bibinfo {author} {\bibfnamefont {A.~V.}\ \bibnamefont
  {Sinitskiy}}, \bibinfo {author} {\bibfnamefont {M.~G.}\ \bibnamefont
  {Saunders}}, \ and\ \bibinfo {author} {\bibfnamefont {G.~A.}\ \bibnamefont
  {Voth}},\ }\href@noop {} {\bibfield  {journal} {\bibinfo  {journal} {The
  Journal of Physical Chemistry B}\ }\textbf {\bibinfo {volume} {116}},\
  \bibinfo {pages} {8363} (\bibinfo {year} {2012})},\ \bibinfo {note} {pMID:
  22276676}\BibitemShut {NoStop}%
\bibitem [{\citenamefont {Foley}\ \emph {et~al.}(2015)\citenamefont {Foley},
  \citenamefont {Shell},\ and\ \citenamefont {Noid}}]{Foley2015}%
  \BibitemOpen
  \bibfield  {author} {\bibinfo {author} {\bibfnamefont {T.~T.}\ \bibnamefont
  {Foley}}, \bibinfo {author} {\bibfnamefont {M.~S.}\ \bibnamefont {Shell}}, \
  and\ \bibinfo {author} {\bibfnamefont {W.~G.}\ \bibnamefont {Noid}},\
  }\href@noop {} {\bibfield  {journal} {\bibinfo  {journal} {The Journal of
  Chemical Physics}\ }\textbf {\bibinfo {volume} {143}},\ \bibinfo {pages}
  {243104} (\bibinfo {year} {2015})}\BibitemShut {NoStop}%
\bibitem [{\citenamefont {Shell}(2008)}]{Shell2008}%
  \BibitemOpen
  \bibfield  {author} {\bibinfo {author} {\bibfnamefont {M.~S.}\ \bibnamefont
  {Shell}},\ }\href@noop {} {\bibfield  {journal} {\bibinfo  {journal} {J.
  Chem. Phys.}\ }\textbf {\bibinfo {volume} {129}},\ \bibinfo {pages} {144108}
  (\bibinfo {year} {2008})}\BibitemShut {NoStop}%
\bibitem [{\citenamefont {Hinsen}\ \emph {et~al.}(2000)\citenamefont {Hinsen},
  \citenamefont {Petrescu}, \citenamefont {Dellerue}, \citenamefont
  {Bellissent-Funel},\ and\ \citenamefont {Kneller}}]{Hinsen2000}%
  \BibitemOpen
  \bibfield  {author} {\bibinfo {author} {\bibfnamefont {K.}~\bibnamefont
  {Hinsen}}, \bibinfo {author} {\bibfnamefont {A.}~\bibnamefont {Petrescu}},
  \bibinfo {author} {\bibfnamefont {S.}~\bibnamefont {Dellerue}}, \bibinfo
  {author} {\bibfnamefont {M.}~\bibnamefont {Bellissent-Funel}}, \ and\
  \bibinfo {author} {\bibfnamefont {G.}~\bibnamefont {Kneller}},\ }\href@noop
  {} {\bibfield  {journal} {\bibinfo  {journal} {Chem. Phys.}\ }\textbf
  {\bibinfo {volume} {261}},\ \bibinfo {pages} {25} (\bibinfo {year}
  {2000})}\BibitemShut {NoStop}%
\bibitem [{\citenamefont {Carnevale}\ \emph {et~al.}(2007)\citenamefont
  {Carnevale}, \citenamefont {Pontiggia},\ and\ \citenamefont
  {Micheletti}}]{Carnevale2007}%
  \BibitemOpen
  \bibfield  {author} {\bibinfo {author} {\bibfnamefont {V.}~\bibnamefont
  {Carnevale}}, \bibinfo {author} {\bibfnamefont {F.}~\bibnamefont
  {Pontiggia}}, \ and\ \bibinfo {author} {\bibfnamefont {C.}~\bibnamefont
  {Micheletti}},\ }\href@noop {} {\bibfield  {journal} {\bibinfo  {journal}
  {Journal of Physics: Condensed Matter}\ }\textbf {\bibinfo {volume} {19}},\
  \bibinfo {pages} {285206+} (\bibinfo {year} {2007})}\BibitemShut {NoStop}%
\bibitem [{\citenamefont {Zen}\ \emph {et~al.}(2008)\citenamefont {Zen},
  \citenamefont {Carnevale}, \citenamefont {Lesk},\ and\ \citenamefont
  {Micheletti}}]{Zen2007}%
  \BibitemOpen
  \bibfield  {author} {\bibinfo {author} {\bibfnamefont {A.}~\bibnamefont
  {Zen}}, \bibinfo {author} {\bibfnamefont {V.}~\bibnamefont {Carnevale}},
  \bibinfo {author} {\bibfnamefont {A.~M.}\ \bibnamefont {Lesk}}, \ and\
  \bibinfo {author} {\bibfnamefont {C.}~\bibnamefont {Micheletti}},\
  }\href@noop {} {\bibfield  {journal} {\bibinfo  {journal} {Protein Sci.}\
  }\textbf {\bibinfo {volume} {17}},\ \bibinfo {pages} {918} (\bibinfo {year}
  {2008})}\BibitemShut {NoStop}%
\bibitem [{\citenamefont {Doruker}\ \emph {et~al.}(2000)\citenamefont
  {Doruker}, \citenamefont {Atilgan},\ and\ \citenamefont
  {Bahar}}]{Doruker2000}%
  \BibitemOpen
  \bibfield  {author} {\bibinfo {author} {\bibfnamefont {P.}~\bibnamefont
  {Doruker}}, \bibinfo {author} {\bibfnamefont {A.~R.}\ \bibnamefont
  {Atilgan}}, \ and\ \bibinfo {author} {\bibfnamefont {I.}~\bibnamefont
  {Bahar}},\ }\href@noop {} {\bibfield  {journal} {\bibinfo  {journal}
  {Proteins: Structure, Function, and Bioinformatics}\ }\textbf {\bibinfo
  {volume} {40}},\ \bibinfo {pages} {512} (\bibinfo {year} {2000})}\BibitemShut
  {NoStop}%
\bibitem [{\citenamefont {Cui}\ and\ \citenamefont {Bahar}(2005)}]{Cui2005}%
  \BibitemOpen
  \bibfield  {author} {\bibinfo {author} {\bibfnamefont {Q.}~\bibnamefont
  {Cui}}\ and\ \bibinfo {author} {\bibfnamefont {I.}~\bibnamefont {Bahar}}\
  }(\bibinfo  {publisher} {Chapman and Hall/CRC},\ \bibinfo {address} {United
  Kingdom},\ \bibinfo {year} {2005})\BibitemShut {NoStop}%
\bibitem [{\citenamefont {M{\"{u}}ller-Plathe}(2002)}]{Muller-Plathe2002}%
  \BibitemOpen
  \bibfield  {author} {\bibinfo {author} {\bibfnamefont {F.}~\bibnamefont
  {M{\"{u}}ller-Plathe}},\ }\href@noop {} {\bibfield  {journal} {\bibinfo
  {journal} {ChemPhysChem}\ }\textbf {\bibinfo {volume} {3}},\ \bibinfo {pages}
  {754} (\bibinfo {year} {2002})}\BibitemShut {NoStop}%
\bibitem [{\citenamefont {Reith}\ \emph {et~al.}(2003)\citenamefont {Reith},
  \citenamefont {P{\"{u}}tz},\ and\ \citenamefont
  {M{\"{u}}ller-Plathe}}]{Reith2003}%
  \BibitemOpen
  \bibfield  {author} {\bibinfo {author} {\bibfnamefont {D.}~\bibnamefont
  {Reith}}, \bibinfo {author} {\bibfnamefont {M.}~\bibnamefont {P{\"{u}}tz}}, \
  and\ \bibinfo {author} {\bibfnamefont {F.}~\bibnamefont
  {M{\"{u}}ller-Plathe}},\ }\href@noop {} {\bibfield  {journal} {\bibinfo
  {journal} {J. Comput. Chem.}\ }\textbf {\bibinfo {volume} {24}},\ \bibinfo
  {pages} {1624} (\bibinfo {year} {2003})}\BibitemShut {NoStop}%
\bibitem [{\citenamefont {Izvekov}\ and\ \citenamefont
  {Voth}(2006)}]{Izvekov2006}%
  \BibitemOpen
  \bibfield  {author} {\bibinfo {author} {\bibfnamefont {S.}~\bibnamefont
  {Izvekov}}\ and\ \bibinfo {author} {\bibfnamefont {G.~A.}\ \bibnamefont
  {Voth}},\ }\href@noop {} {\bibfield  {journal} {\bibinfo  {journal} {The
  Journal of Chemical Physics}\ }\textbf {\bibinfo {volume} {125}},\ \bibinfo
  {pages} {151101} (\bibinfo {year} {2006})}\BibitemShut {NoStop}%
\bibitem [{\citenamefont {Spyriouni}\ \emph {et~al.}(2007)\citenamefont
  {Spyriouni}, \citenamefont {Tzoumanekas}, \citenamefont {Theodorou},
  \citenamefont {M{\"{u}}ller-Plathe},\ and\ \citenamefont
  {Milano}}]{Spyriouni2007}%
  \BibitemOpen
  \bibfield  {author} {\bibinfo {author} {\bibfnamefont {T.}~\bibnamefont
  {Spyriouni}}, \bibinfo {author} {\bibfnamefont {C.}~\bibnamefont
  {Tzoumanekas}}, \bibinfo {author} {\bibfnamefont {D.}~\bibnamefont
  {Theodorou}}, \bibinfo {author} {\bibfnamefont {F.}~\bibnamefont
  {M{\"{u}}ller-Plathe}}, \ and\ \bibinfo {author} {\bibfnamefont
  {G.}~\bibnamefont {Milano}},\ }\href@noop {} {\bibfield  {journal} {\bibinfo
  {journal} {Macromolecules}\ }\textbf {\bibinfo {volume} {40}},\ \bibinfo
  {pages} {3876} (\bibinfo {year} {2007})}\BibitemShut {NoStop}%
\bibitem [{\citenamefont {Noid}\ \emph {et~al.}(2008)\citenamefont {Noid},
  \citenamefont {Chu}, \citenamefont {Ayton}, \citenamefont {Krishna},
  \citenamefont {Izvekov}, \citenamefont {Voth}, \citenamefont {Das},\ and\
  \citenamefont {Andersen}}]{Noid2008}%
  \BibitemOpen
  \bibfield  {author} {\bibinfo {author} {\bibfnamefont {W.~G.}\ \bibnamefont
  {Noid}}, \bibinfo {author} {\bibfnamefont {J.-W.}\ \bibnamefont {Chu}},
  \bibinfo {author} {\bibfnamefont {G.~S.}\ \bibnamefont {Ayton}}, \bibinfo
  {author} {\bibfnamefont {V.}~\bibnamefont {Krishna}}, \bibinfo {author}
  {\bibfnamefont {S.}~\bibnamefont {Izvekov}}, \bibinfo {author} {\bibfnamefont
  {G.~A.}\ \bibnamefont {Voth}}, \bibinfo {author} {\bibfnamefont
  {A.}~\bibnamefont {Das}}, \ and\ \bibinfo {author} {\bibfnamefont {H.~C.}\
  \bibnamefont {Andersen}},\ }\href@noop {} {\bibfield  {journal} {\bibinfo
  {journal} {The Journal of Chemical Physics}\ }\textbf {\bibinfo {volume}
  {128}},\ \bibinfo {pages} {244114} (\bibinfo {year} {2008})}\BibitemShut
  {NoStop}%
\bibitem [{\citenamefont {Shell}(2012)}]{Shell2012}%
  \BibitemOpen
  \bibfield  {author} {\bibinfo {author} {\bibfnamefont {M.~S.}\ \bibnamefont
  {Shell}},\ }\href@noop {} {\bibfield  {journal} {\bibinfo  {journal} {J.
  Chem. Phys.}\ }\textbf {\bibinfo {volume} {137}},\ \bibinfo {pages} {084503}
  (\bibinfo {year} {2012})}\BibitemShut {NoStop}%
\bibitem [{\citenamefont {Noid}(2013{\natexlab{b}})}]{Noid2013b}%
  \BibitemOpen
  \bibfield  {author} {\bibinfo {author} {\bibfnamefont {W.~G.}\ \bibnamefont
  {Noid}},\ }\enquote {\bibinfo {title} {Systematic methods for structurally
  consistent coarse-grained models},}\ in\ \href@noop {} {\emph {\bibinfo
  {booktitle} {Biomolecular Simulations: Methods and Protocols}}},\ \bibinfo
  {editor} {edited by\ \bibinfo {editor} {\bibfnamefont {L.}~\bibnamefont
  {Monticelli}}\ and\ \bibinfo {editor} {\bibfnamefont {E.}~\bibnamefont
  {Salonen}}}\ (\bibinfo  {publisher} {Humana Press},\ \bibinfo {address}
  {Totowa, NJ},\ \bibinfo {year} {2013})\ pp.\ \bibinfo {pages}
  {487--531}\BibitemShut {NoStop}%
\bibitem [{\citenamefont {Pontiggia}(2008)}]{phdPontiggia}%
  \BibitemOpen
  \bibfield  {author} {\bibinfo {author} {\bibfnamefont {F.}~\bibnamefont
  {Pontiggia}},\ }\emph {\bibinfo {title} {Protein Structure and
  Functionally-oriented Dynamics: From Atomistic to Coarse-grained Models}},\
  \href@noop {} {Ph.D. thesis},\ \bibinfo  {school} {SISSA/ISAS - International
  School for Advanced Studies} (\bibinfo {year} {2008})\BibitemShut {NoStop}%
\bibitem [{\citenamefont {Globisch}\ \emph {et~al.}(2013)\citenamefont
  {Globisch}, \citenamefont {Krishnamani}, \citenamefont {Deserno},\ and\
  \citenamefont {Peter}}]{Globisch2013}%
  \BibitemOpen
  \bibfield  {author} {\bibinfo {author} {\bibfnamefont {C.}~\bibnamefont
  {Globisch}}, \bibinfo {author} {\bibfnamefont {V.}~\bibnamefont
  {Krishnamani}}, \bibinfo {author} {\bibfnamefont {M.}~\bibnamefont
  {Deserno}}, \ and\ \bibinfo {author} {\bibfnamefont {C.}~\bibnamefont
  {Peter}},\ }\href@noop {} {\bibfield  {journal} {\bibinfo  {journal} {PLOS
  ONE}\ }\textbf {\bibinfo {volume} {8}},\ \bibinfo {pages} {e60582} (\bibinfo
  {year} {2013})}\BibitemShut {NoStop}%
\bibitem [{\citenamefont {Carnevale}\ \emph {et~al.}(2006)\citenamefont
  {Carnevale}, \citenamefont {Raugei}, \citenamefont {Micheletti},\ and\
  \citenamefont {Carloni}}]{Carnevale2006}%
  \BibitemOpen
  \bibfield  {author} {\bibinfo {author} {\bibfnamefont {V.}~\bibnamefont
  {Carnevale}}, \bibinfo {author} {\bibfnamefont {S.}~\bibnamefont {Raugei}},
  \bibinfo {author} {\bibfnamefont {C.}~\bibnamefont {Micheletti}}, \ and\
  \bibinfo {author} {\bibfnamefont {P.}~\bibnamefont {Carloni}},\ }\href@noop
  {} {\bibfield  {journal} {\bibinfo  {journal} {J. Am. Chem. Soc.}\ }\textbf
  {\bibinfo {volume} {2}},\ \bibinfo {pages} {173} (\bibinfo {year}
  {2006})}\BibitemShut {NoStop}%
\bibitem [{\citenamefont {Zhou}\ and\ \citenamefont
  {Siegelbaum}(2008)}]{Zhou2008}%
  \BibitemOpen
  \bibfield  {author} {\bibinfo {author} {\bibfnamefont {L.}~\bibnamefont
  {Zhou}}\ and\ \bibinfo {author} {\bibfnamefont {S.~A.}\ \bibnamefont
  {Siegelbaum}},\ }\href@noop {} {\bibfield  {journal} {\bibinfo  {journal}
  {Biophys. J.}\ }\textbf {\bibinfo {volume} {94}},\ \bibinfo {pages} {3461}
  (\bibinfo {year} {2008})}\BibitemShut {NoStop}%
\bibitem [{\citenamefont {Kirkpatrick}\ \emph {et~al.}(1983)\citenamefont
  {Kirkpatrick}, \citenamefont {Gelatt~Jr},\ and\ \citenamefont
  {Vecchi}}]{KirkpatrickVecchi1983}%
  \BibitemOpen
  \bibfield  {author} {\bibinfo {author} {\bibfnamefont {S.}~\bibnamefont
  {Kirkpatrick}}, \bibinfo {author} {\bibfnamefont {C.~D.}\ \bibnamefont
  {Gelatt~Jr}}, \ and\ \bibinfo {author} {\bibfnamefont {M.~P.}\ \bibnamefont
  {Vecchi}},\ }\href@noop {} {\bibfield  {journal} {\bibinfo  {journal}
  {Science}\ }\textbf {\bibinfo {volume} {220}},\ \bibinfo {pages} {674}
  (\bibinfo {year} {1983})}\BibitemShut {NoStop}%
\bibitem [{\citenamefont {\v{C}ern{\'y}}(1985)}]{Cerny1985}%
  \BibitemOpen
  \bibfield  {author} {\bibinfo {author} {\bibfnamefont {V.}~\bibnamefont
  {\v{C}ern{\'y}}},\ }\href@noop {} {\bibfield  {journal} {\bibinfo  {journal}
  {J. Optim. Theory Appl.}\ }\textbf {\bibinfo {volume} {45}},\ \bibinfo
  {pages} {41} (\bibinfo {year} {1985})}\BibitemShut {NoStop}%
\bibitem [{\citenamefont {Kabsch}(1976)}]{Kabsch:a12999}%
  \BibitemOpen
  \bibfield  {author} {\bibinfo {author} {\bibfnamefont {W.}~\bibnamefont
  {Kabsch}},\ }\href@noop {} {\bibfield  {journal} {\bibinfo  {journal} {Acta
  Crystallographica Section A}\ }\textbf {\bibinfo {volume} {32}},\ \bibinfo
  {pages} {922} (\bibinfo {year} {1976})}\BibitemShut {NoStop}%
\bibitem [{\citenamefont {M{\"u}ller}\ \emph {et~al.}(1996)\citenamefont
  {M{\"u}ller}, \citenamefont {Schlauderer}, \citenamefont {Reinstein},\ and\
  \citenamefont {Schulz}}]{4AKE}%
  \BibitemOpen
  \bibfield  {author} {\bibinfo {author} {\bibfnamefont {C.~W.}\ \bibnamefont
  {M{\"u}ller}}, \bibinfo {author} {\bibfnamefont {G.~J.}\ \bibnamefont
  {Schlauderer}}, \bibinfo {author} {\bibfnamefont {J.}~\bibnamefont
  {Reinstein}}, \ and\ \bibinfo {author} {\bibfnamefont {G.~E.}\ \bibnamefont
  {Schulz}},\ }\href@noop {} {\bibfield  {journal} {\bibinfo  {journal}
  {Structure}\ }\textbf {\bibinfo {volume} {4}},\ \bibinfo {pages} {147}
  (\bibinfo {year} {1996})}\BibitemShut {NoStop}%
\bibitem [{\citenamefont {Serganov}\ \emph {et~al.}(2004)\citenamefont
  {Serganov}, \citenamefont {Yuan}, \citenamefont {Pikovskaya}, \citenamefont
  {Polonskaia}, \citenamefont {Malinina}, \citenamefont {Phan}, \citenamefont
  {Hobartner}, \citenamefont {Micura}, \citenamefont {Breaker},\ and\
  \citenamefont {Patel}}]{1Y26}%
  \BibitemOpen
  \bibfield  {author} {\bibinfo {author} {\bibfnamefont {A.}~\bibnamefont
  {Serganov}}, \bibinfo {author} {\bibfnamefont {Y.~R.}\ \bibnamefont {Yuan}},
  \bibinfo {author} {\bibfnamefont {O.}~\bibnamefont {Pikovskaya}}, \bibinfo
  {author} {\bibfnamefont {A.}~\bibnamefont {Polonskaia}}, \bibinfo {author}
  {\bibfnamefont {L.}~\bibnamefont {Malinina}}, \bibinfo {author}
  {\bibfnamefont {A.~T.}\ \bibnamefont {Phan}}, \bibinfo {author}
  {\bibfnamefont {C.}~\bibnamefont {Hobartner}}, \bibinfo {author}
  {\bibfnamefont {R.}~\bibnamefont {Micura}}, \bibinfo {author} {\bibfnamefont
  {R.~R.}\ \bibnamefont {Breaker}}, \ and\ \bibinfo {author} {\bibfnamefont
  {D.~J.}\ \bibnamefont {Patel}},\ }\href@noop {} {\bibfield  {journal}
  {\bibinfo  {journal} {Chem. Biol.}\ }\textbf {\bibinfo {volume} {11}},\
  \bibinfo {pages} {1729} (\bibinfo {year} {2004})}\BibitemShut {NoStop}%
\bibitem [{\citenamefont {Pontiggia}\ \emph {et~al.}(2008)\citenamefont
  {Pontiggia}, \citenamefont {Zen},\ and\ \citenamefont
  {Micheletti}}]{Pontiggia2008}%
  \BibitemOpen
  \bibfield  {author} {\bibinfo {author} {\bibfnamefont {F.}~\bibnamefont
  {Pontiggia}}, \bibinfo {author} {\bibfnamefont {A.}~\bibnamefont {Zen}}, \
  and\ \bibinfo {author} {\bibfnamefont {C.}~\bibnamefont {Micheletti}},\
  }\href@noop {} {\bibfield  {journal} {\bibinfo  {journal} {Biophys J}\
  }\textbf {\bibinfo {volume} {95}},\ \bibinfo {pages} {5901} (\bibinfo {year}
  {2008})}\BibitemShut {NoStop}%
\bibitem [{\citenamefont {Bahar}\ and\ \citenamefont
  {Rader}(2005)}]{Bahar2005}%
  \BibitemOpen
  \bibfield  {author} {\bibinfo {author} {\bibfnamefont {I.}~\bibnamefont
  {Bahar}}\ and\ \bibinfo {author} {\bibfnamefont {A.~J.}\ \bibnamefont
  {Rader}},\ }\href@noop {} {\bibfield  {journal} {\bibinfo  {journal} {Curr.
  Opin. Struct. Biol.}\ }\textbf {\bibinfo {volume} {15}},\ \bibinfo {pages}
  {586} (\bibinfo {year} {2005})}\BibitemShut {NoStop}%
\bibitem [{\citenamefont {Setny}\ and\ \citenamefont
  {Zacharias}(2013)}]{setny2013}%
  \BibitemOpen
  \bibfield  {author} {\bibinfo {author} {\bibfnamefont {P.}~\bibnamefont
  {Setny}}\ and\ \bibinfo {author} {\bibfnamefont {M.}~\bibnamefont
  {Zacharias}},\ }\href@noop {} {\bibfield  {journal} {\bibinfo  {journal}
  {Journal of Chemical Theory and Computation}\ }\textbf {\bibinfo {volume}
  {9}},\ \bibinfo {pages} {5460} (\bibinfo {year} {2013})},\ \bibinfo {note}
  {pMID: 26592282}\BibitemShut {NoStop}%
\bibitem [{\citenamefont {Zimmermann}\ and\ \citenamefont
  {Jernigan}(2014)}]{Zimmermann2014}%
  \BibitemOpen
  \bibfield  {author} {\bibinfo {author} {\bibfnamefont {M.~T.}\ \bibnamefont
  {Zimmermann}}\ and\ \bibinfo {author} {\bibfnamefont {R.~L.}\ \bibnamefont
  {Jernigan}},\ }\href@noop {} {\bibfield  {journal} {\bibinfo  {journal}
  {RNA}\ }\textbf {\bibinfo {volume} {20}},\ \bibinfo {pages} {792} (\bibinfo
  {year} {2014})}\BibitemShut {NoStop}%
\bibitem [{\citenamefont {Pinamonti}\ \emph {et~al.}(2015)\citenamefont
  {Pinamonti}, \citenamefont {Bottaro}, \citenamefont {Micheletti},\ and\
  \citenamefont {Bussi}}]{Pinamonti2015}%
  \BibitemOpen
  \bibfield  {author} {\bibinfo {author} {\bibfnamefont {G.}~\bibnamefont
  {Pinamonti}}, \bibinfo {author} {\bibfnamefont {S.}~\bibnamefont {Bottaro}},
  \bibinfo {author} {\bibfnamefont {C.}~\bibnamefont {Micheletti}}, \ and\
  \bibinfo {author} {\bibfnamefont {G.}~\bibnamefont {Bussi}},\ }\href@noop {}
  {\bibfield  {journal} {\bibinfo  {journal} {Nucl. Acids Res.}\ }\textbf
  {\bibinfo {volume} {43}},\ \bibinfo {pages} {7260} (\bibinfo {year}
  {2015})}\BibitemShut {NoStop}%
\bibitem [{\citenamefont {Hu}\ \emph {et~al.}(2017)\citenamefont {Hu},
  \citenamefont {He}, \citenamefont {Iacovelli},\ and\ \citenamefont
  {Falconi}}]{falconi2017}%
  \BibitemOpen
  \bibfield  {author} {\bibinfo {author} {\bibfnamefont {G.}~\bibnamefont
  {Hu}}, \bibinfo {author} {\bibfnamefont {L.}~\bibnamefont {He}}, \bibinfo
  {author} {\bibfnamefont {F.}~\bibnamefont {Iacovelli}}, \ and\ \bibinfo
  {author} {\bibfnamefont {M.}~\bibnamefont {Falconi}},\ }\href@noop {}
  {\bibfield  {journal} {\bibinfo  {journal} {Molecules}\ }\textbf {\bibinfo
  {volume} {22}} (\bibinfo {year} {2017})}\BibitemShut {NoStop}%
\bibitem [{\citenamefont {Priyakumar}\ and\ \citenamefont
  {MacKerell}(2010)}]{PRIYAKUMAR20101422}%
  \BibitemOpen
  \bibfield  {author} {\bibinfo {author} {\bibfnamefont {U.~D.}\ \bibnamefont
  {Priyakumar}}\ and\ \bibinfo {author} {\bibfnamefont {A.~D.}\ \bibnamefont
  {MacKerell}},\ }\href@noop {} {\bibfield  {journal} {\bibinfo  {journal}
  {Journal of Molecular Biology}\ }\textbf {\bibinfo {volume} {396}},\ \bibinfo
  {pages} {1422 } (\bibinfo {year} {2010})}\BibitemShut {NoStop}%
\bibitem [{\citenamefont {Alln{\'e}r}\ \emph {et~al.}(2013)\citenamefont
  {Alln{\'e}r}, \citenamefont {Nilsson},\ and\ \citenamefont
  {Villa}}]{Allner01072013}%
  \BibitemOpen
  \bibfield  {author} {\bibinfo {author} {\bibfnamefont {O.}~\bibnamefont
  {Alln{\'e}r}}, \bibinfo {author} {\bibfnamefont {L.}~\bibnamefont {Nilsson}},
  \ and\ \bibinfo {author} {\bibfnamefont {A.}~\bibnamefont {Villa}},\ }\href
  {\doibase 10.1261/rna.037549.112} {\bibfield  {journal} {\bibinfo  {journal}
  {RNA}\ }\textbf {\bibinfo {volume} {19}},\ \bibinfo {pages} {916} (\bibinfo
  {year} {2013})}\BibitemShut {NoStop}%
\bibitem [{\citenamefont {Di~Palma}\ \emph {et~al.}(2013)\citenamefont
  {Di~Palma}, \citenamefont {Colizzi},\ and\ \citenamefont
  {Bussi}}]{DiPalma01112013}%
  \BibitemOpen
  \bibfield  {author} {\bibinfo {author} {\bibfnamefont {F.}~\bibnamefont
  {Di~Palma}}, \bibinfo {author} {\bibfnamefont {F.}~\bibnamefont {Colizzi}}, \
  and\ \bibinfo {author} {\bibfnamefont {G.}~\bibnamefont {Bussi}},\ }\href
  {\doibase 10.1261/rna.040493.113} {\bibfield  {journal} {\bibinfo  {journal}
  {RNA}\ }\textbf {\bibinfo {volume} {19}},\ \bibinfo {pages} {1517} (\bibinfo
  {year} {2013})}\BibitemShut {NoStop}%
\end{thebibliography}%

\end{document}